%% file: Wyczalkowski_JMBBM_v2.tex
\documentclass[authoryear,3p,11pt]{elsarticle}   

\makeatletter
\def\ps@pprintTitle{%
 \let\@oddhead\@empty
 \let\@evenhead\@empty
 \def\@oddfoot{}%
 \let\@evenfoot\@oddfoot}
\makeatother

\input{Header.tex} % all usepackage and newcommand stuff below.
\input{Figures.tex} % all figures are defined in the file below, called here with \figureXXX command
\input{Tables.tex}  % tables defined here
\input{Title.tex}

\begin{document}
\input{0_Abstract}

\maketitle

\clearpage
%\clearpage

\section{Introduction}

\input{1_Introduction}

\section{Background}
\input{2_Background}

\section{Experimental Methods}

\input{3_ExperimentalMethods}

\section{Experimental Results} \label{sec:ExperimentalResults}
\input{4_ExperimentalResults}

\section{Computational Model for Wound Healing} \label{sec:theory_methods}
\input{5_ModelMethods}

\section{Model Results} \label{sec:ModelingResults}
\input{6_ModelResults}

\section{Discussion}
\input{7_Discussion}

%\clearpage
\hrulefill

\appendix
\renewcommand{\thefigure}{\arabic{figure}}
\section{Contractile Ring Geometry} \label{sec:coordinates}
\input{A_Coordinate_Systems}

\section{Comparison of Single and Multiple Fiber Models} \label{sec:SingleMultiFiber}
\input{B_SingleMultiFiber}

\section{Mechanical Properties of Blastoderm} \label{sec:varner}
\input{C_Properties}

% from http://support.river-valley.com/wiki/index.php?title=Ans._M7
\section*{Acknowledgments}
\input{X_Acknowledgements}

\clearpage

\section*{References}
\bibliographystyle{abbrvnat}  
\bibliography{Wyczalkowski_JMBBM_v2}

\clearpage

\setcounter{section}{18}  % http://texblog.org/2007/07/25/counters-in-latex/
\section{Supplemental Material} \label{sec:supplemental}
% Reset figure counter, make figure names start with S.
% http://stackoverflow.com/questions/3391540/renumbering-figure-in-latex

\makeatletter 
\setcounter{figure}{0}
\renewcommand{\thefigure}{S\@arabic\c@figure} 
\setcounter{table}{0}
\renewcommand{\thetable}{S\@arabic\c@table}
\makeatother

\input{S_Supplemental}
\end{document}

%% file: Header.tex
\usepackage{amssymb,amsmath}    % Handy: ftp://ftp.ams.org/ams/doc/amsmath/short-math-guide.pdf
\usepackage{array}    % Change row height, http://nepsweb.co.uk/docs/tableTricks.pdf
\usepackage{bm}
\usepackage[nodots]{numcompress}
\usepackage{graphicx}
\usepackage{subfig}
\usepackage{gensymb}
\usepackage{verbatim}
\usepackage[normalem]{ulem}   % for development.  Usage: \sout{foo} to strike out
\usepackage{comment}
\usepackage{xcolor}
\usepackage{setspace}
\usepackage[percent]{overpic} % so superpose (a), etc. on images.  http://www.ctex.org/documents/packages/graphics/opic-abs.pdf
\usepackage{caption}

\newcommand{\bolde}{{\bf e}}

\newcommand{\boldC}{{\bf C}}
\newcommand{\boldF}{{\bf F}}
\newcommand{\boldG}{{\bf G}}
\newcommand{\boldI}{{\bf I}}

\newcommand{\boldP}{{\bf P}}
\newcommand{\boldR}{{\bf R}}
\newcommand{\boldQ}{{\bf Q}}

\newcommand{\boldsigma}{\boldsymbol \sigma}
\newcommand{\boldrho}{\boldsymbol \rho}

\newcommand{\del}{\boldsymbol \nabla}
\newcommand{\tr}{\mbox{tr\,}}

\newcommand{\bF}{{\bf F}}
\newcommand{\bG}{{\bf G}}

\newcommand{\Ca}{Ca$^{2+}$}  

\newcommand{\phidot}{\mathring{\phi}}

\newcommand{\psubref}{\protect\subref}  

%% file: Figures.tex
\newcommand{\figureWoundImage}{  % Fig 1
\begin{figure}[htbp]
  \centering
  \subfloat{\label{fig:circular_img}
    \begin{overpic}[width=0.40\textwidth]{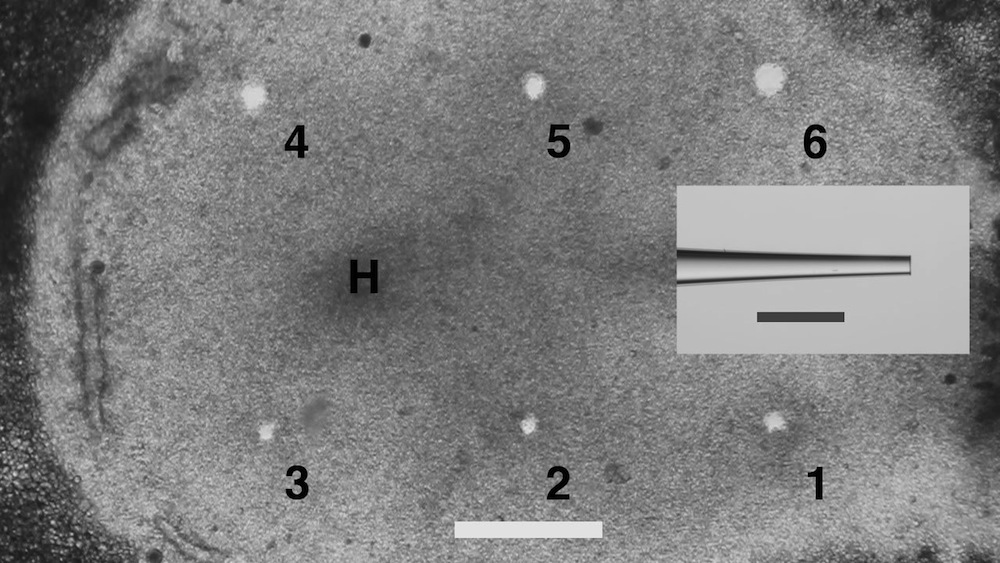}
      \put(1,3){\color{white}{\large (a)}} \end{overpic}}
  \subfloat{\label{fig:elliptical_img}
    \begin{overpic}[width=0.40\textwidth]{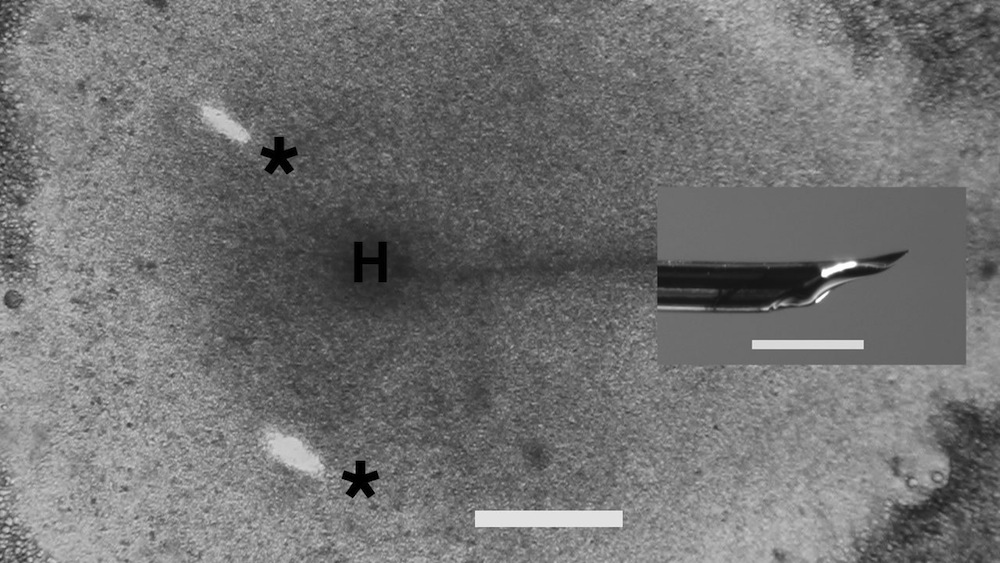}
      \put(1,3){\color{black}{\large (b)}} \end{overpic}}
 \caption
{
Early stage (HH4) chick embryos with circular and elliptical wounds, ventral
side up and anterior to the left.  Hensen's node marked by ``H''.  Scale bars
500$\mu m$. 
\psubref{fig:circular_img} Circular wounds (numbered 1-6) several seconds after being created with a circular punch (inset).  
\psubref{fig:elliptical_img}  Two elliptical wounds (indicated by *) created with microscalpel (inset). 
} \label{fig:woundImage}
\end{figure}
}

\newcommand{\figureCircularAreaFit}{  % Fig 2
\begin{figure}[htbp]
\begin{minipage}{\textwidth}
\centering
    \subfloat{\label{fig:Circular}
        \begin{overpic}[width=0.40\textwidth]{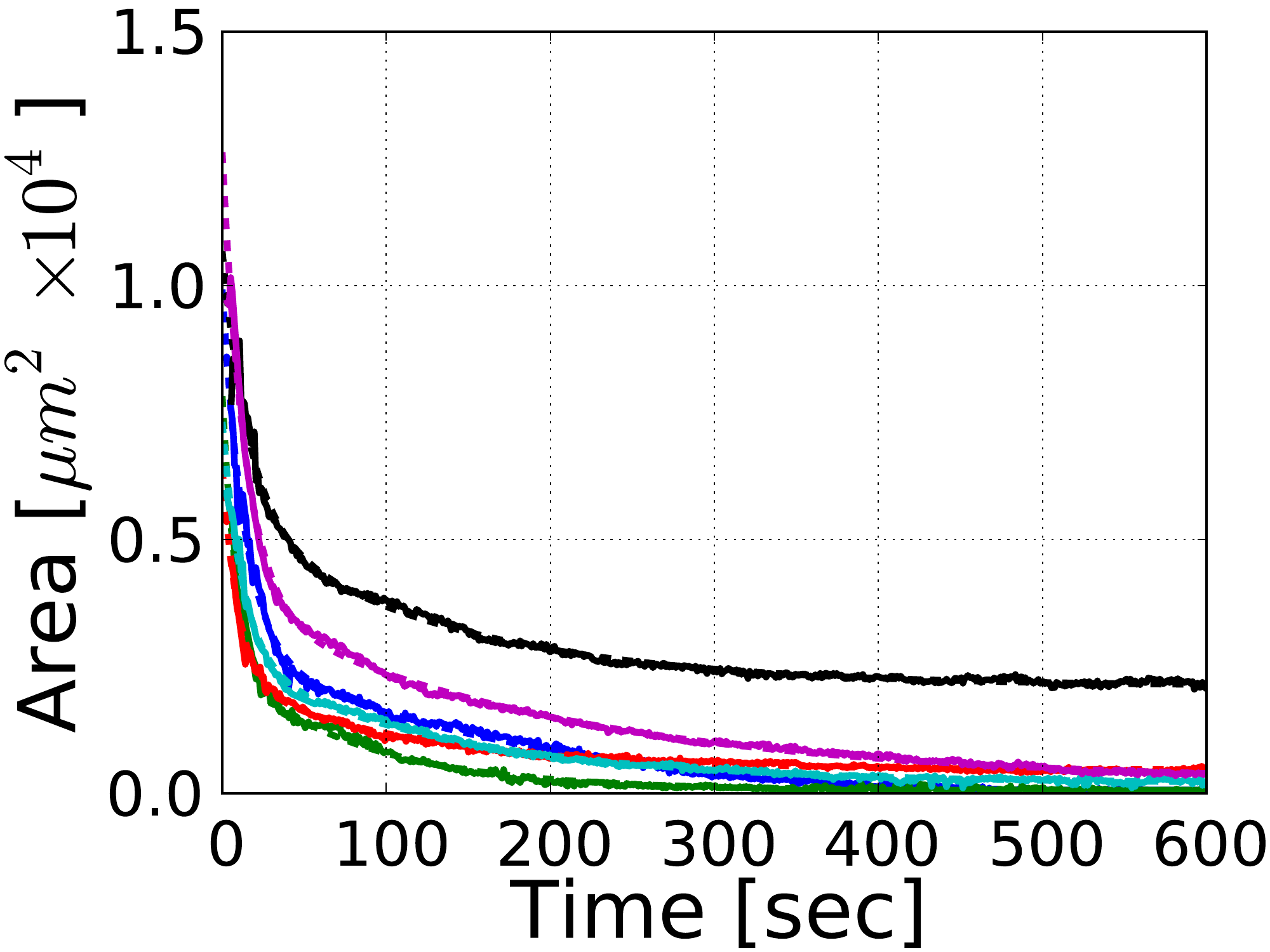}
      \put(0,0){\color{black}{\large (a)}} \end{overpic}}
    \subfloat{\label{fig:Circular1min}
        \begin{overpic}[width=0.40\textwidth]{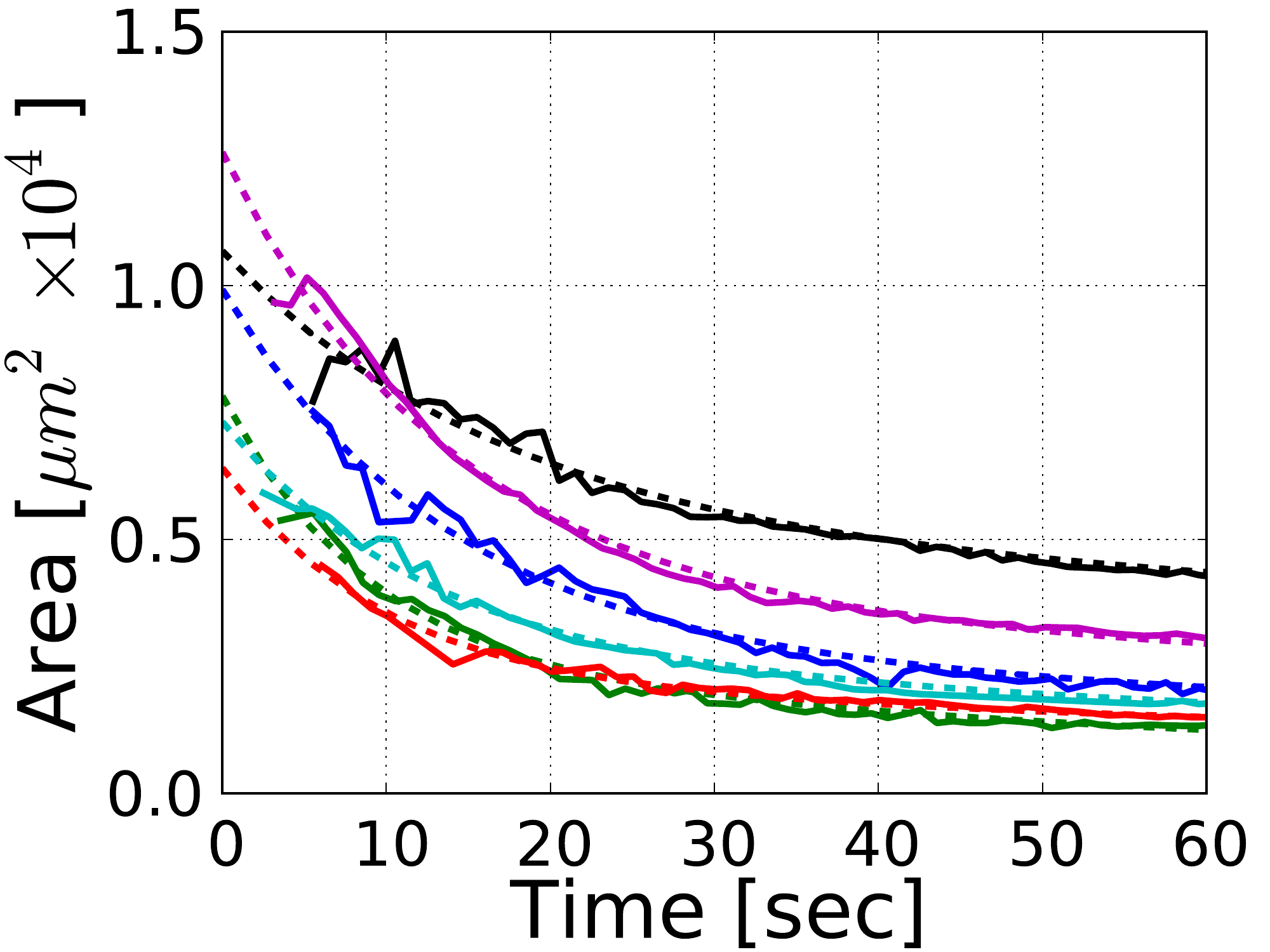}
      \put(0,0){\color{black}{\large (b)}} \end{overpic}}
\end{minipage}
\caption
{
Experimental wound area vs. time for circular wounds.  Experimental data (solid
curves) are for six wounds from a single representative embryo (pictured in
Fig.~\ref{fig:circular_img}).  Dashed curves indicate best fit of
Eq.~\eqref{eqn:fit} for each wound (parameter values in
Table~\ref{tab:WoundFitAll}).  Panels \psubref*{fig:Circular},
\psubref*{fig:Circular1min} show the first 10 minutes and 1 minute of wound healing,
respectively.  Good fit to data obscures dashed curves in panel a.
}
\label{fig:CircularAreaFit}
\end{figure}
}

\newcommand{\figureEllipseAreaAR}{  % Fig 3
\begin{figure}[htbp]
\begin{minipage}{\textwidth}
\centering
    \subfloat{\label{fig:EllipseArea}
        \begin{overpic}[width=0.40\textwidth]{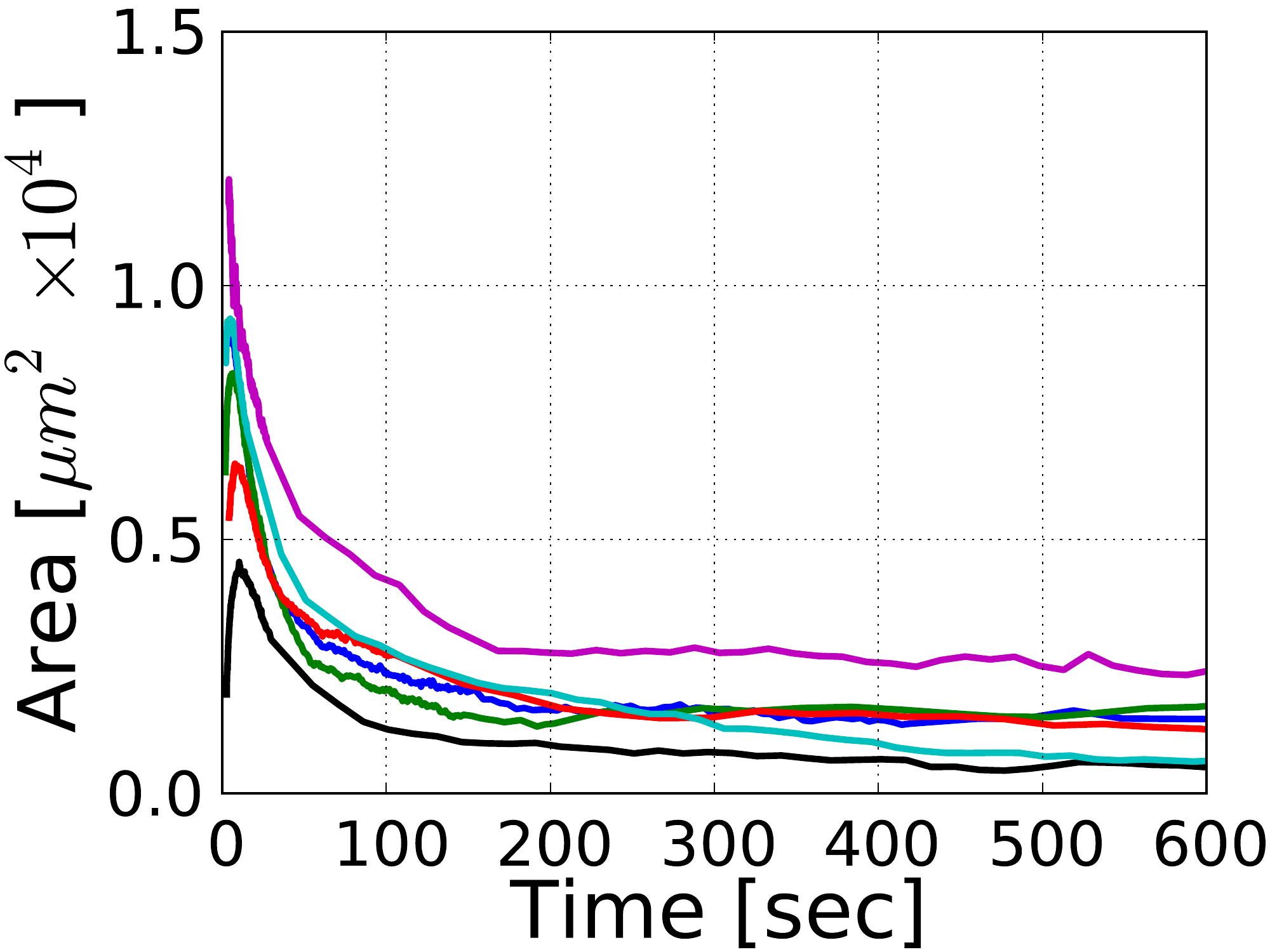}
        \put(0,0){\color{black}{\large(a)}} \end{overpic}}
    \subfloat{\label{fig:EllipseAR}
        \begin{overpic}[width=0.40\textwidth]{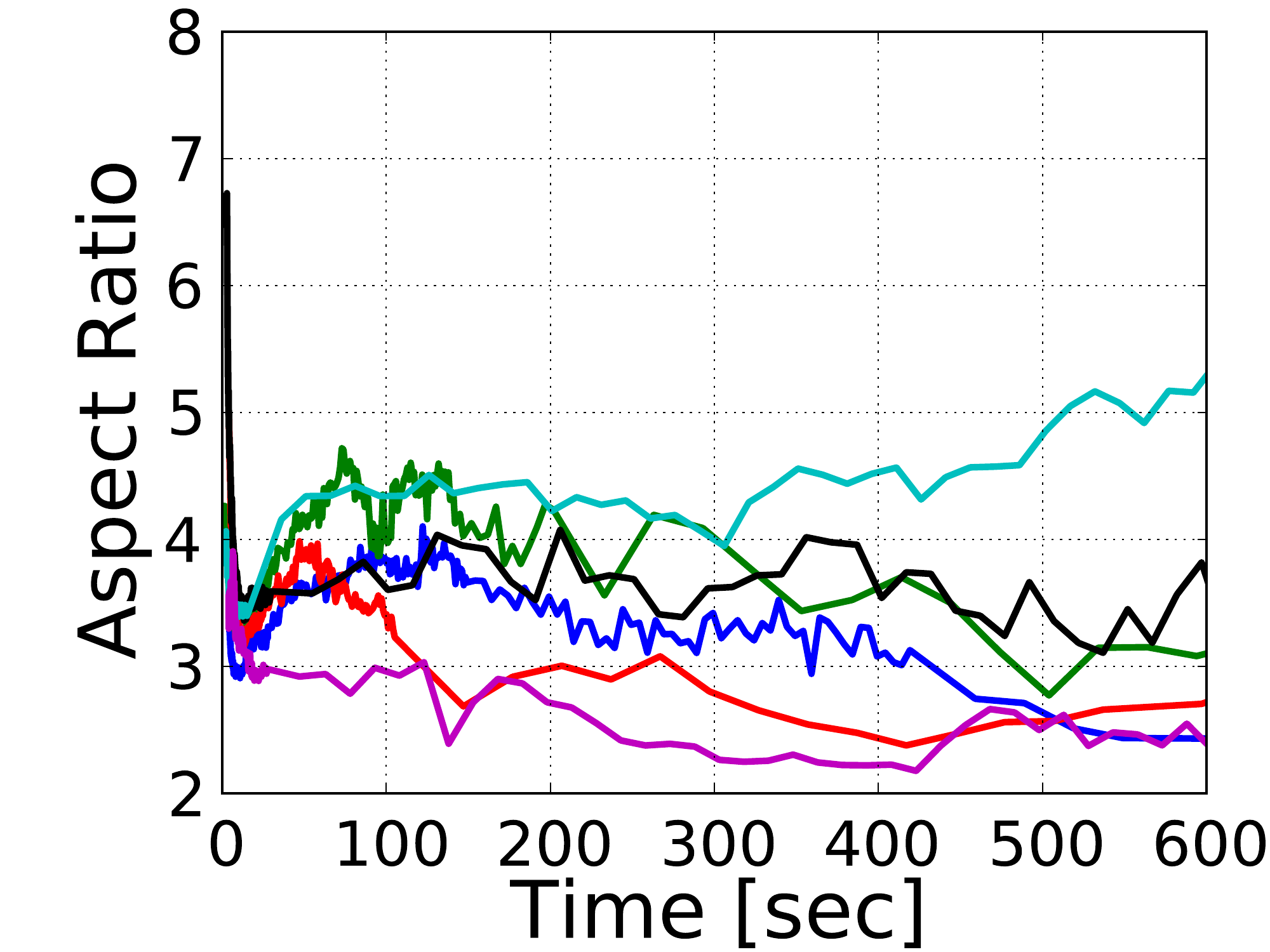}
        \put(0,0){\color{black}{\large(b)}} \end{overpic}}
\end{minipage}
\begin{minipage}{\textwidth}
\centering
    \subfloat{\label{fig:EllipseArea1min}
        \begin{overpic}[width=0.40\textwidth]{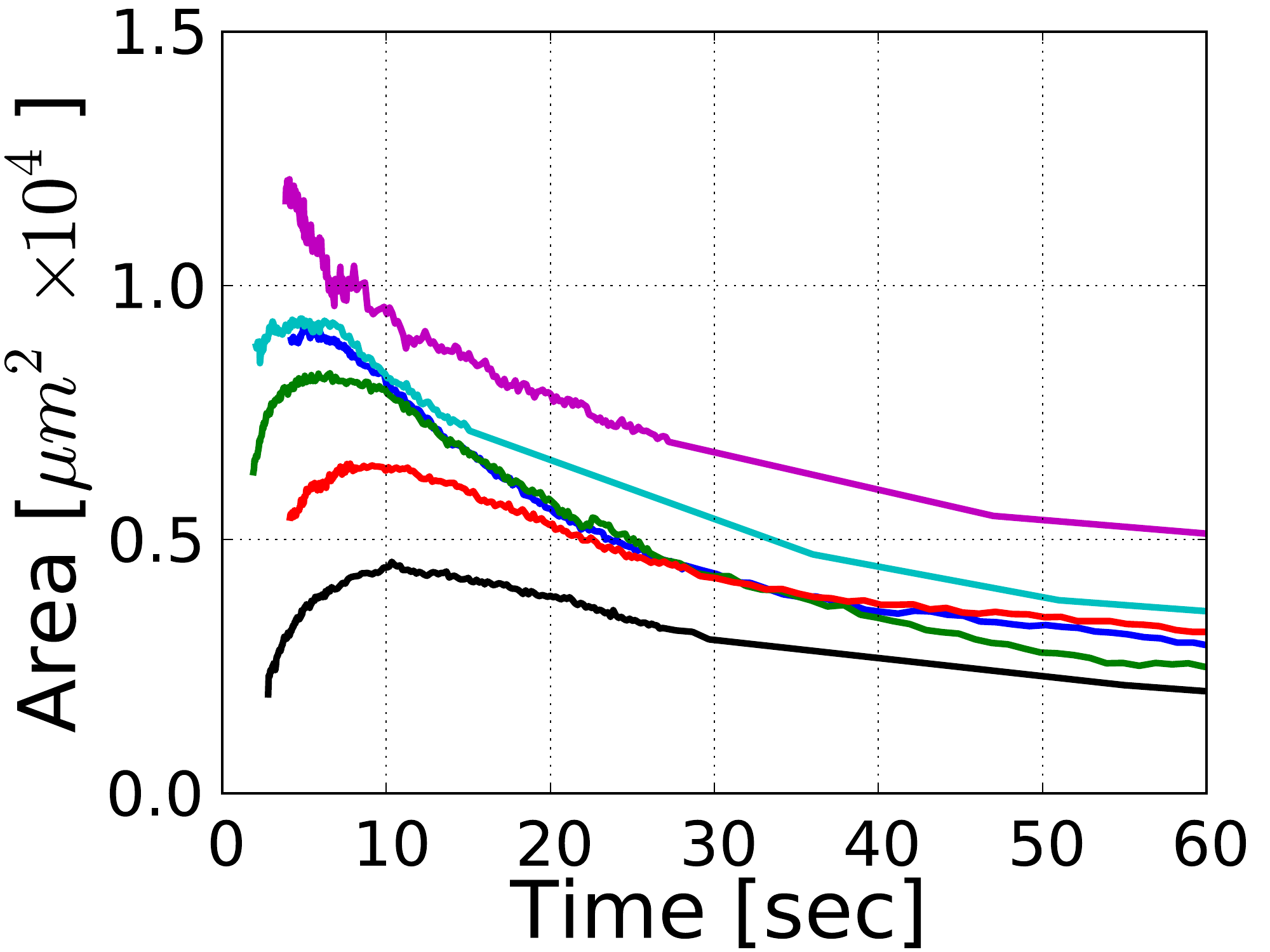}
        \put(0,0){\color{black}{\large (c)}} \end{overpic}}
    \subfloat{\label{fig:EllipseAR1min}
        \begin{overpic}[width=0.40\textwidth]{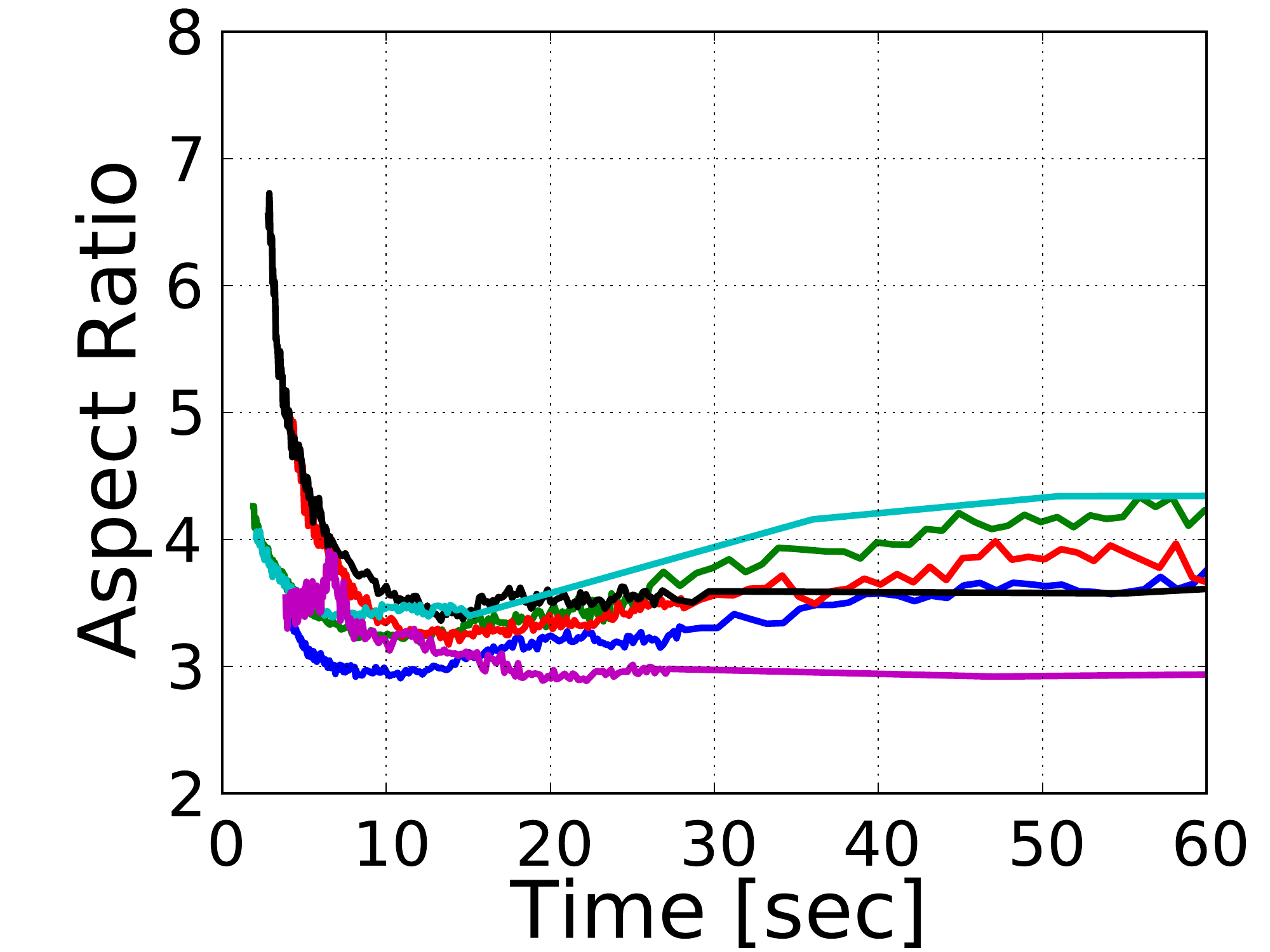}
        \put(0,0){\color{black}{\large (d)}} \end{overpic}}
\end{minipage}
\caption
{
Experimental wound area and aspect ratio (AR) vs. time for elliptical wounds.
Each wound is from a different embryo. (\psubref*{fig:EllipseArea},
\psubref*{fig:EllipseAR}) Area and AR, respectively, for ten
minutes post wounding. (\psubref*{fig:EllipseArea1min},
\psubref*{fig:EllipseAR1min}) The same quantities for the first minute.  
Sampling rate changes correspond to transition from video to still image acquisition
(see Sec.~\ref{sec:Imaging}).
}
\label{fig:EllipseAreaAR}
\end{figure}
}

\newcommand{\figureWoundActin}{  % Fig 4
\begin{figure}[htbp]
  \begin{minipage}{\textwidth}
  \centering
  \subfloat{\label{fig:ecto_20sec}
    \begin{overpic}[width=0.30\textwidth]{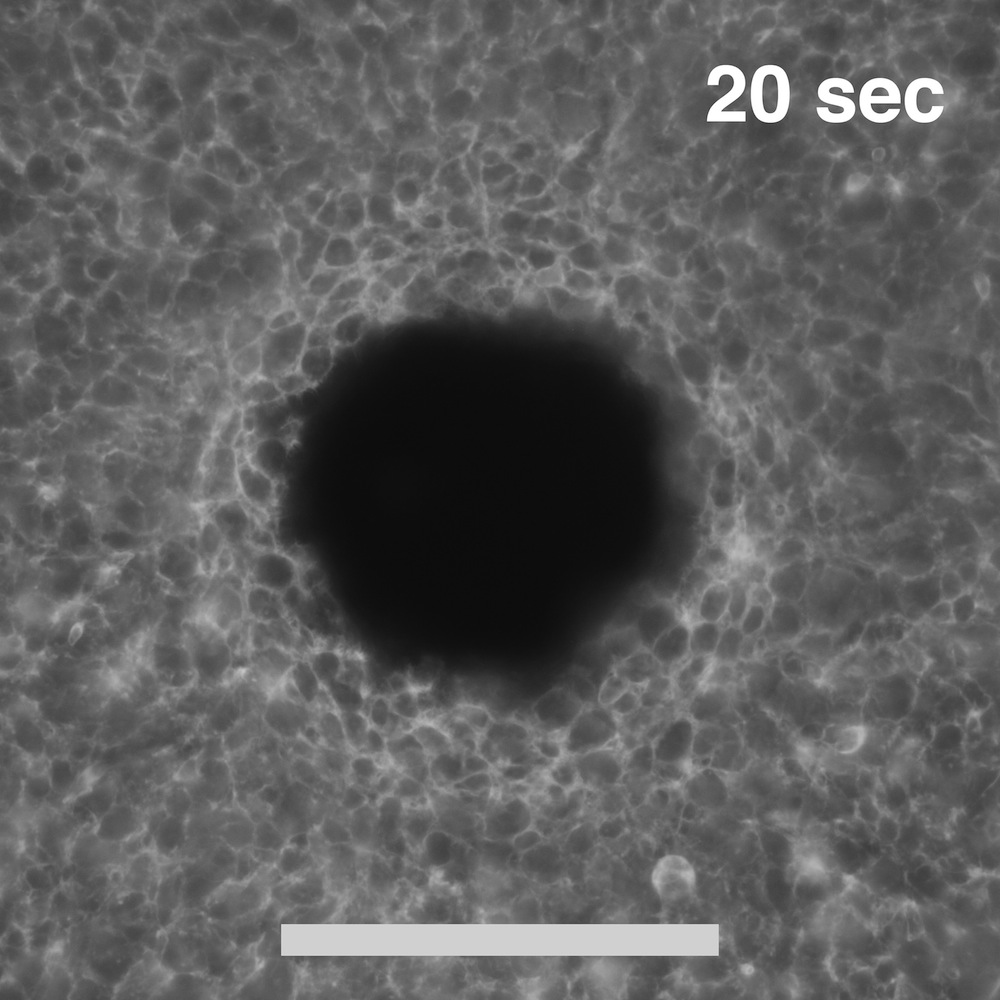}
        \put(1,3){\color{white}{\large (a)}} \end{overpic}}
  \subfloat{\label{fig:ecto_1min}
    \begin{overpic}[width=0.30\textwidth]{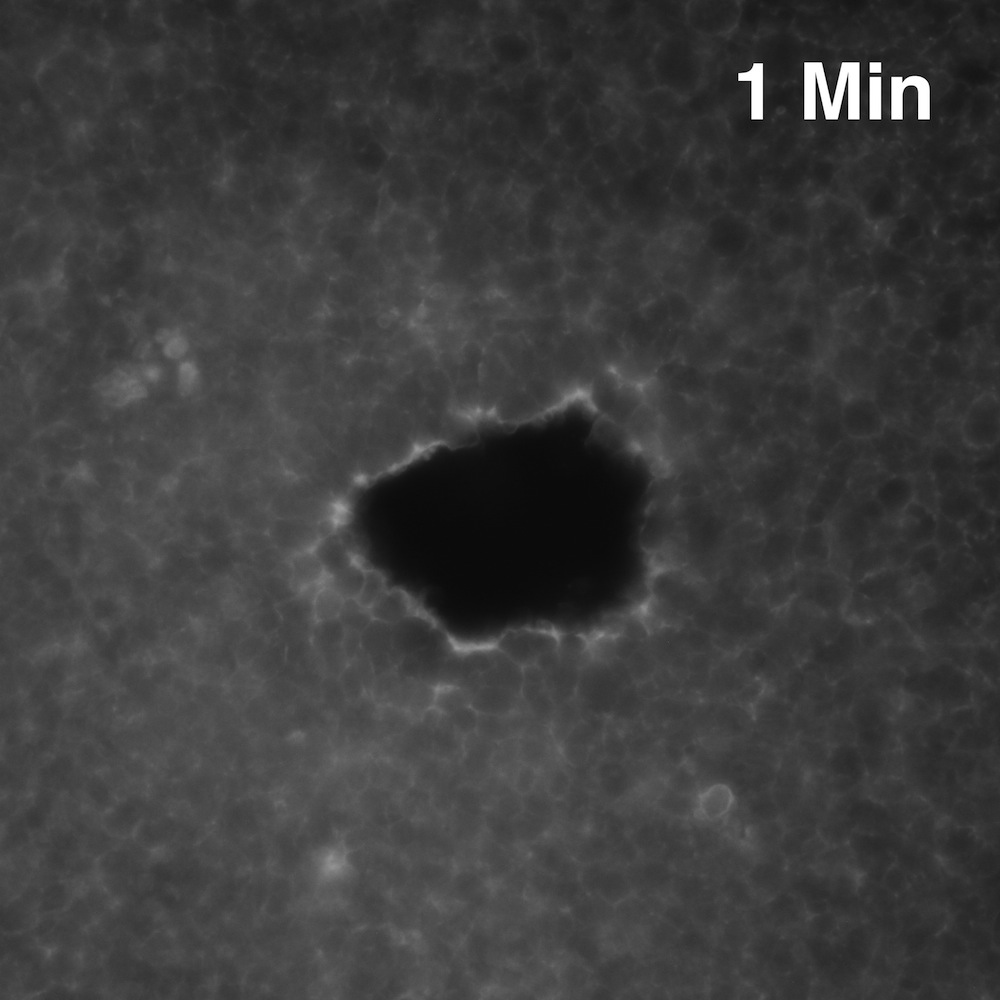}
        \put(1,3){\color{white}{\large (b)}} \end{overpic}}
  \subfloat{\label{fig:ecto_11min}
    \begin{overpic}[width=0.30\textwidth]{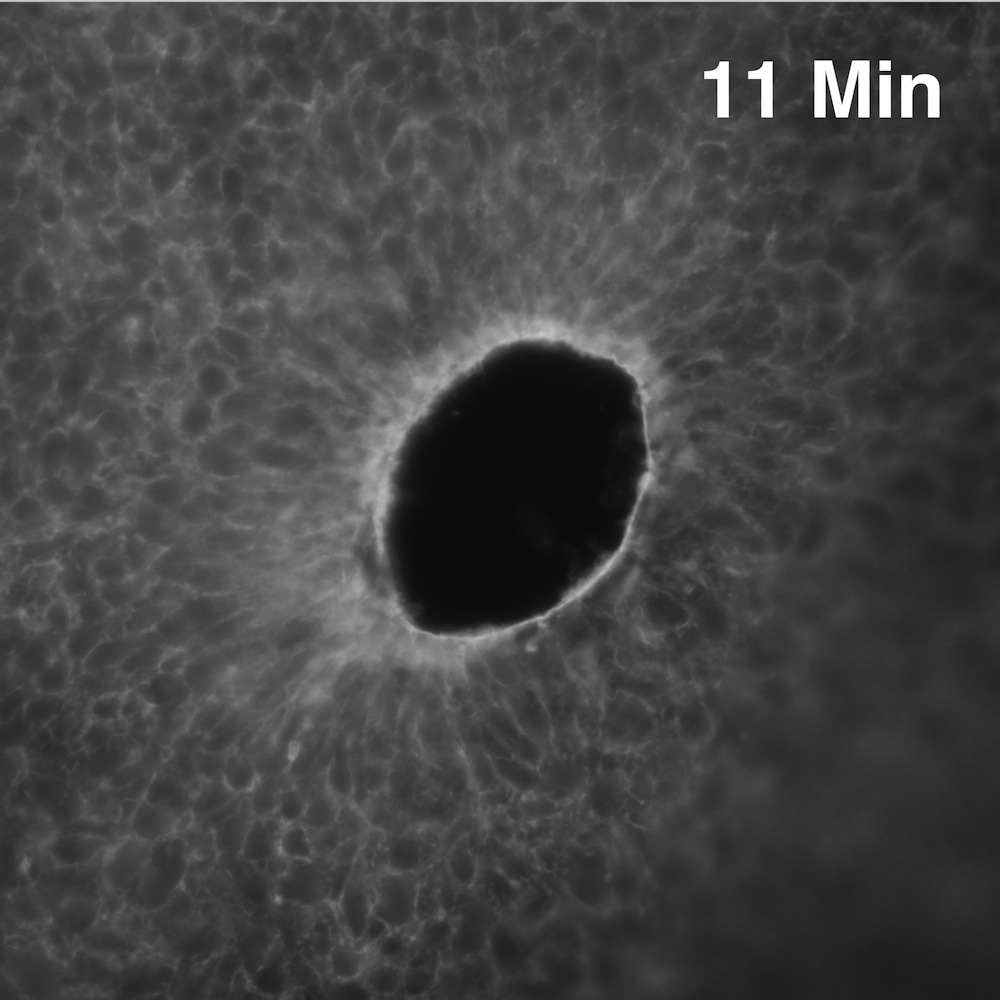}
        \put(1,3){\color{white}{\large (c)}} \end{overpic}}
  \end{minipage}
  \begin{minipage}{\textwidth}
  \centering
  \subfloat{\label{fig:endo_15sec}
    \begin{overpic}[width=0.30\textwidth]{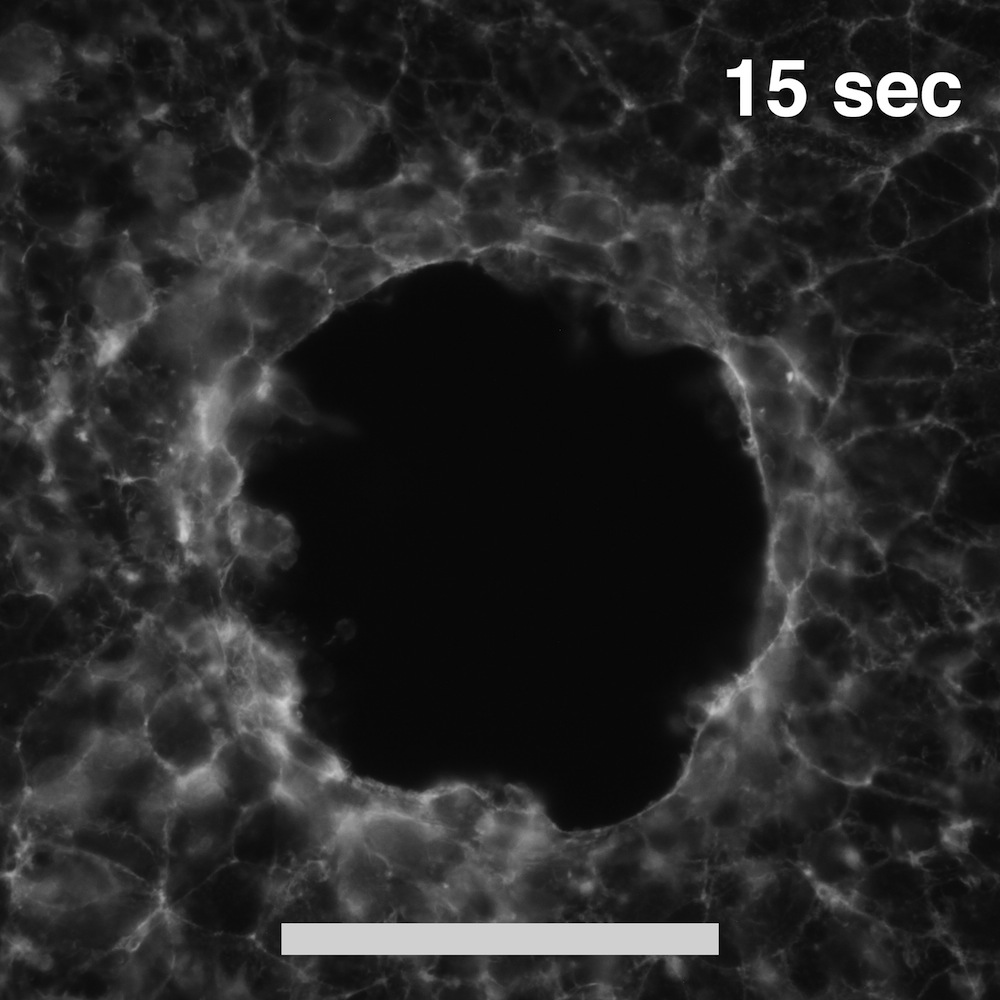}
        \put(1,3){\color{white}{\large (d)}} \end{overpic}}
  \subfloat{\label{fig:endo_1min}
    \begin{overpic}[width=0.30\textwidth]{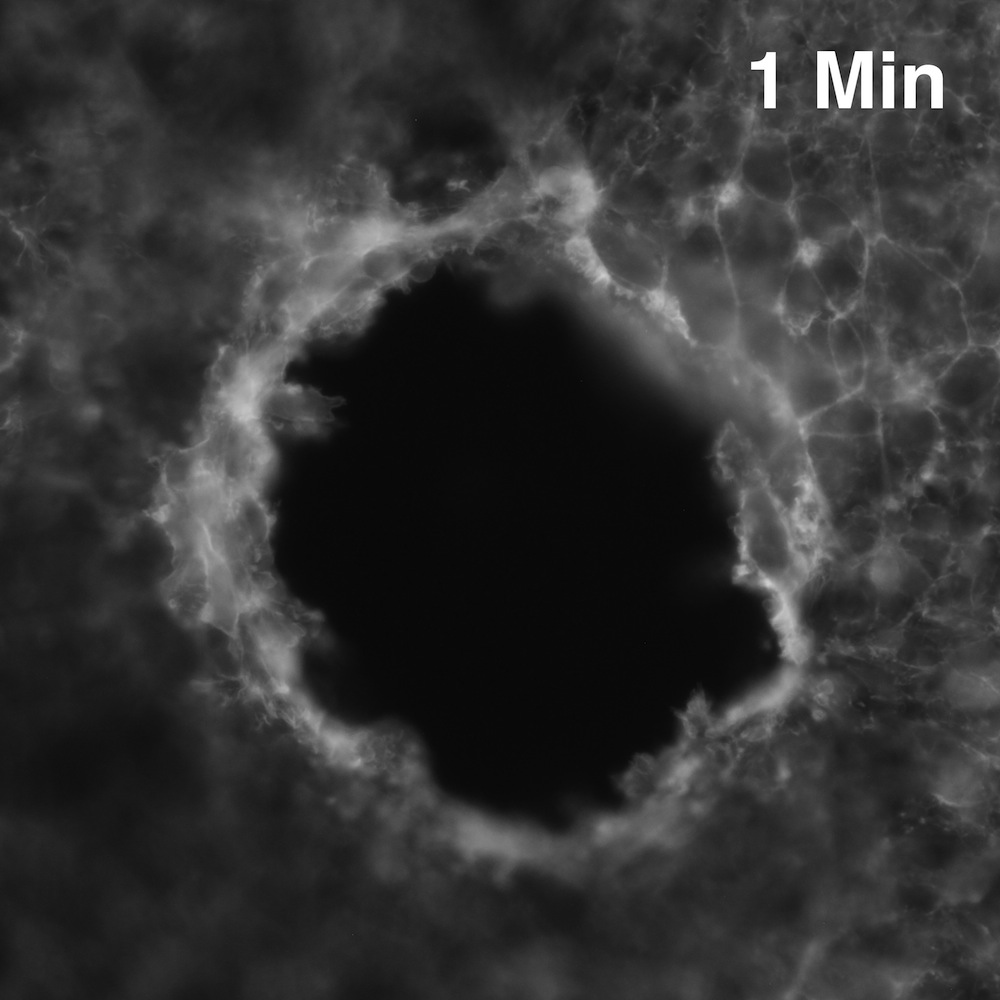}
        \put(1,3){\color{white}{\large (e)}} \end{overpic}}
  \subfloat{\label{fig:endo_11min}
    \begin{overpic}[width=0.30\textwidth]{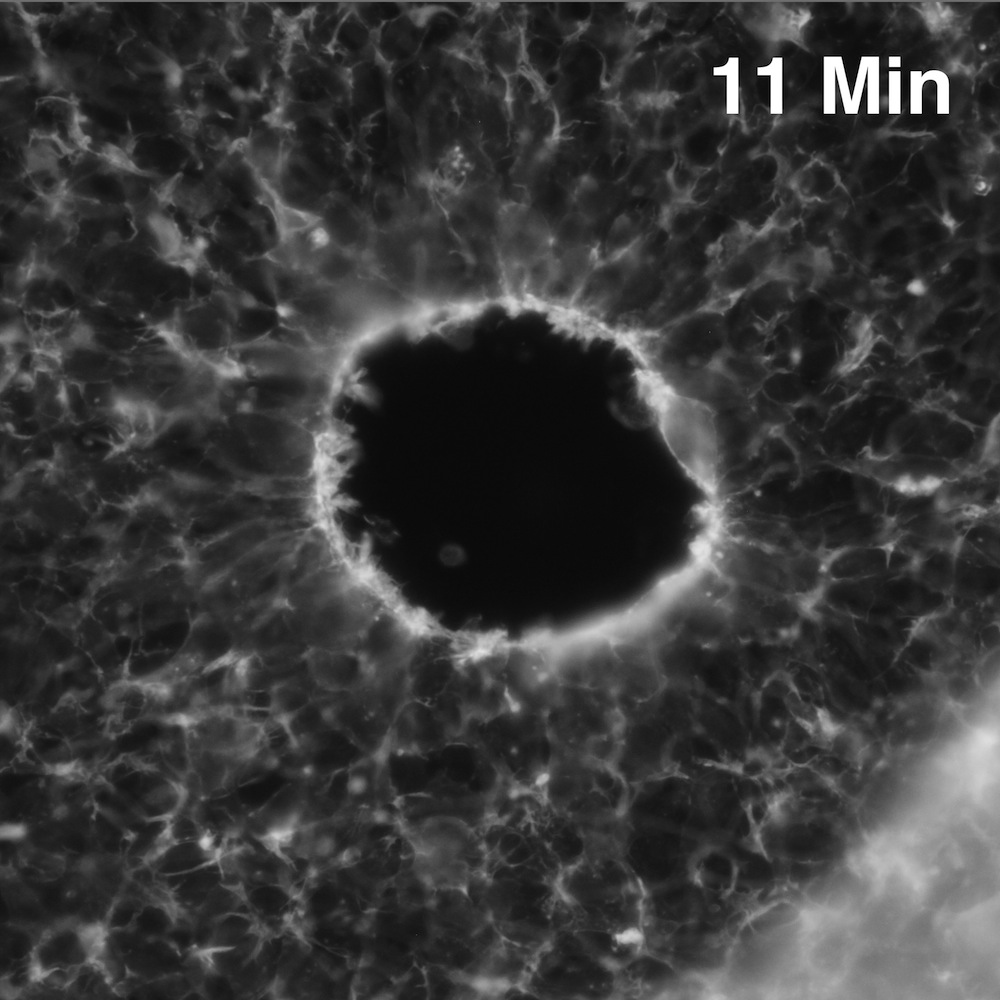}
        \put(1,3){\color{white}{\large (f)}} \end{overpic}}
  \end{minipage}
  \caption
{
Fluorescence images of circular wounds stained for actin at given times after wounding.
(\psubref*{fig:ecto_20sec}--\psubref*{fig:ecto_11min}) ectoderm;
(\psubref*{fig:endo_15sec}--\psubref*{fig:endo_11min}) endoderm.  Scale bars
100$\mu$m.  Panels \psubref*{fig:ecto_20sec} and \psubref*{fig:endo_15sec} are from different embryos,
while \psubref*{fig:ecto_1min} and \psubref*{fig:endo_1min} are same wound, and
\psubref*{fig:ecto_11min} and \psubref*{fig:endo_11min} are same wound.
} \label{fig:WoundActin}
\end{figure}
}

\newcommand{\figureGeometryTD}{  % Fig 5
\begin{figure}[htbp]
\begin{minipage}{\textwidth}
  \centering
  \subfloat{\label{fig:Geometry3Doverview}
      \begin{overpic}[width=0.30\textwidth]{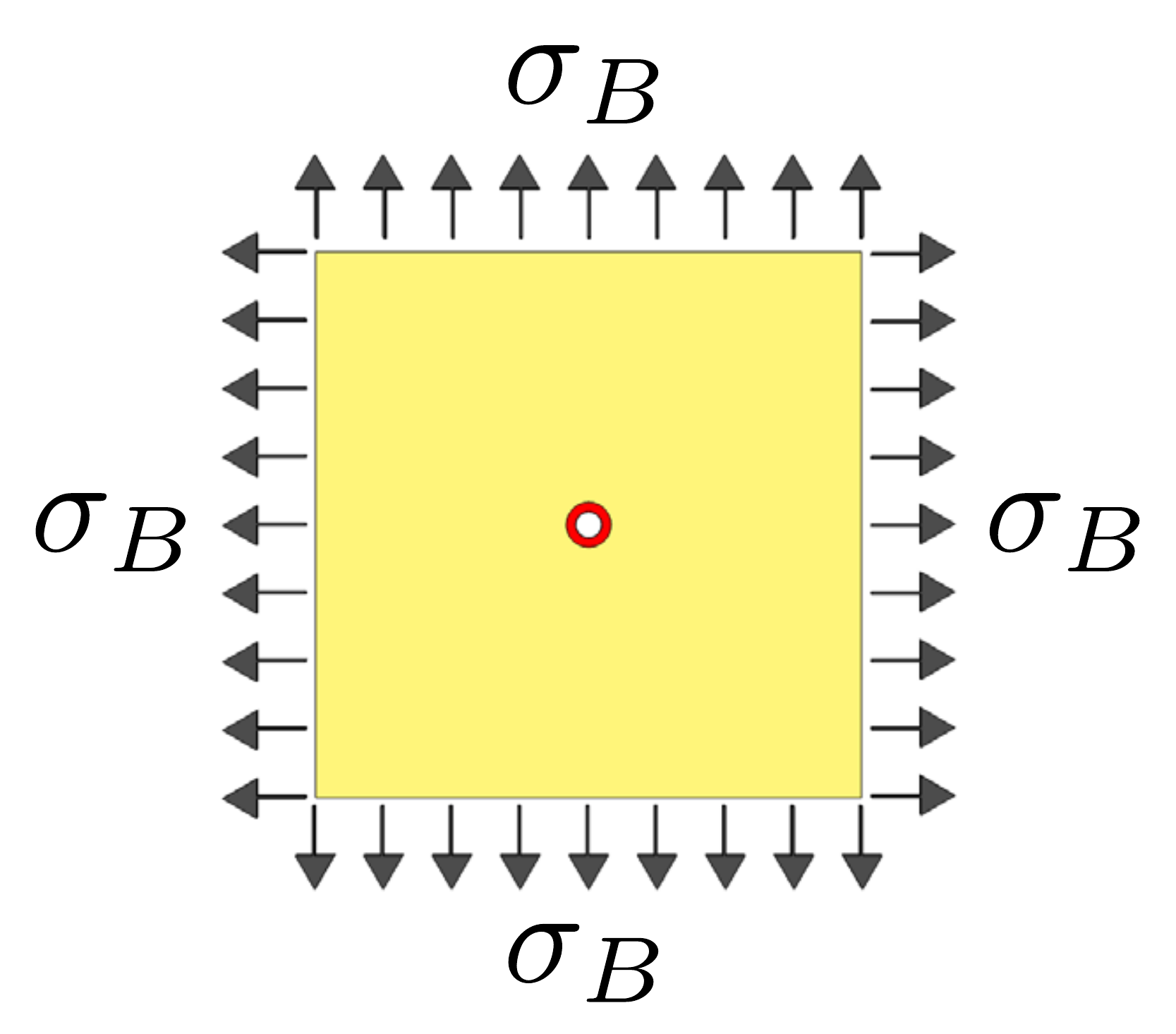}
        \put(1,3){\color{black}{\large (a)}} \end{overpic}}
  \subfloat{\label{fig:Geometry3Ddetail}
      \begin{overpic}[width=0.19\textwidth]{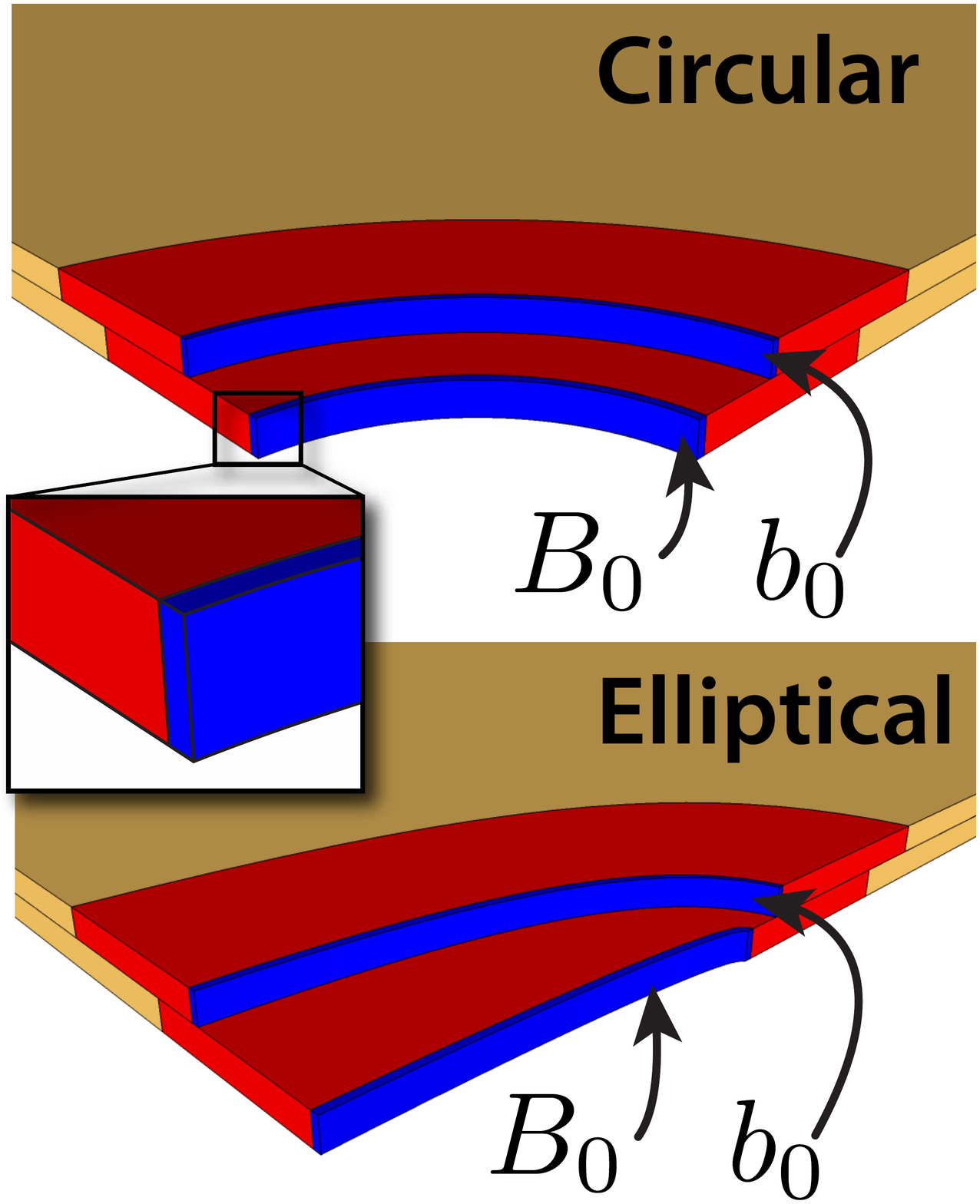}
        \put(1,3){\color{black}{\large (b)}} \end{overpic}}
  \subfloat{\label{fig:GeometryMesh}
      \begin{overpic}[width=0.27\textwidth]{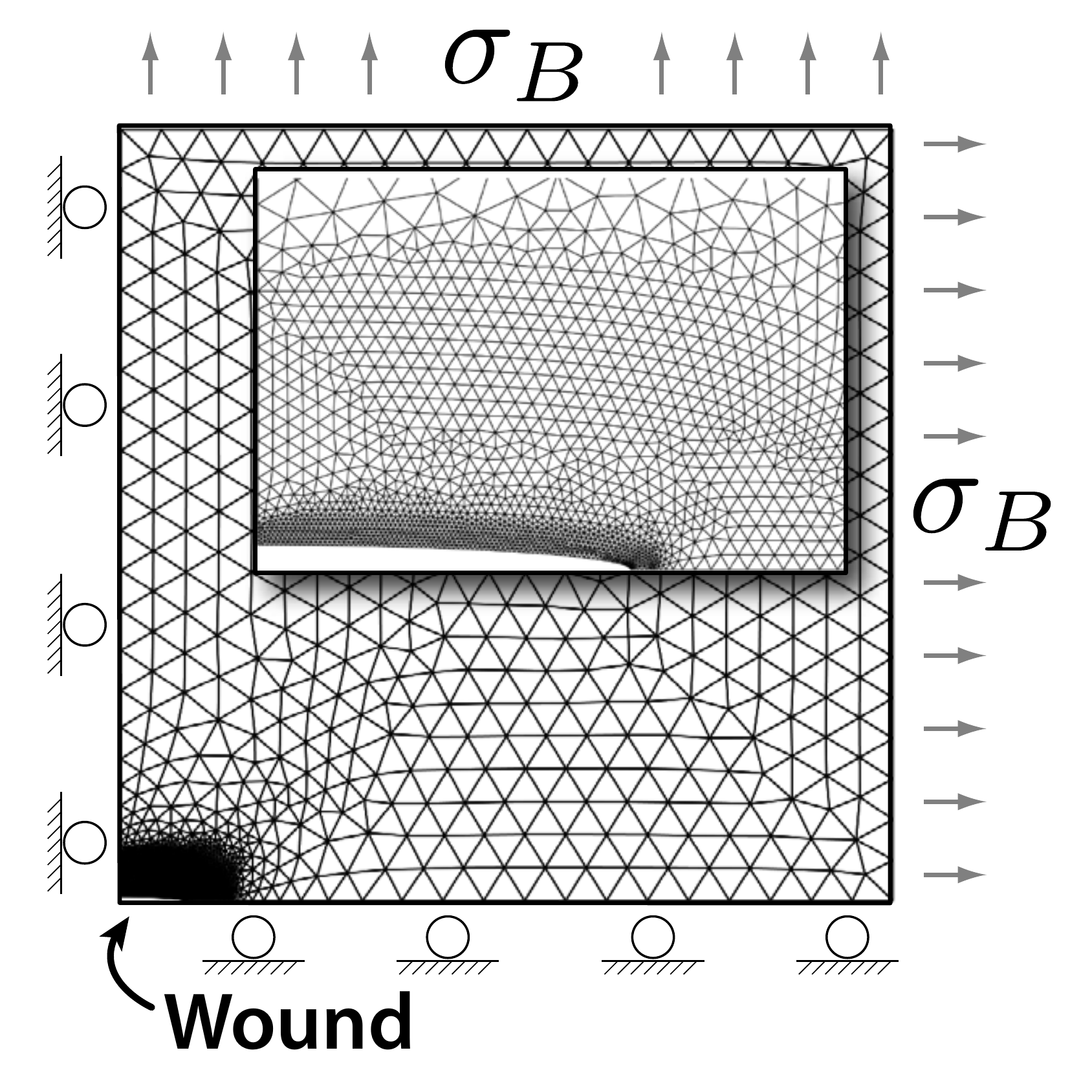}
        \put(1,3){\color{black}{\large (c)}} \end{overpic}}
\end{minipage}
  \caption
{
Schematic of model.
\psubref{fig:Geometry3Doverview} Geometry of circular wound model in reference
configuration $b_0$.  The blastoderm is modeled as a square single-layered
epithelial membrane under isotropic tension $\sigma_B$ with a centered wound.  
\psubref{fig:Geometry3Ddetail} Detail of wound region for circular
and elliptical wounds.  
The thick (cell) ring is shown in red, and the thin (fiber) ring in blue.  
The layers indicate the geometry in the
undeformed ($B_0$) and reference ($b_0$) configurations, illustrating
differences resulting from applied boundary stress: tension
$\sigma_B$ enlarges the wound and, in the elliptical case, decreases
the AR.  Inset illustrates relative width of thin ring (see also Fig.~\ref{fig:GeometryCoord})
\psubref{fig:GeometryMesh} 
Finite-element model detail, illustrating the mesh in the $B_0$ configuration
for the elliptical wound (circular wound is similar).  
Roller boundary
conditions on two internal edges enforce quarter-symmetry conditions, outer
edges have a specified stress $\sigma_B$, and wound edges are stress-free.
Inset details the wound region.  
} \label{fig:Geometry3D}
\end{figure}
}

\newcommand{\figureGrowthSchematicMultiFiber}{ % Fig 6
\begin{figure}[htbp]
\centering
\includegraphics[width=0.65\textwidth]{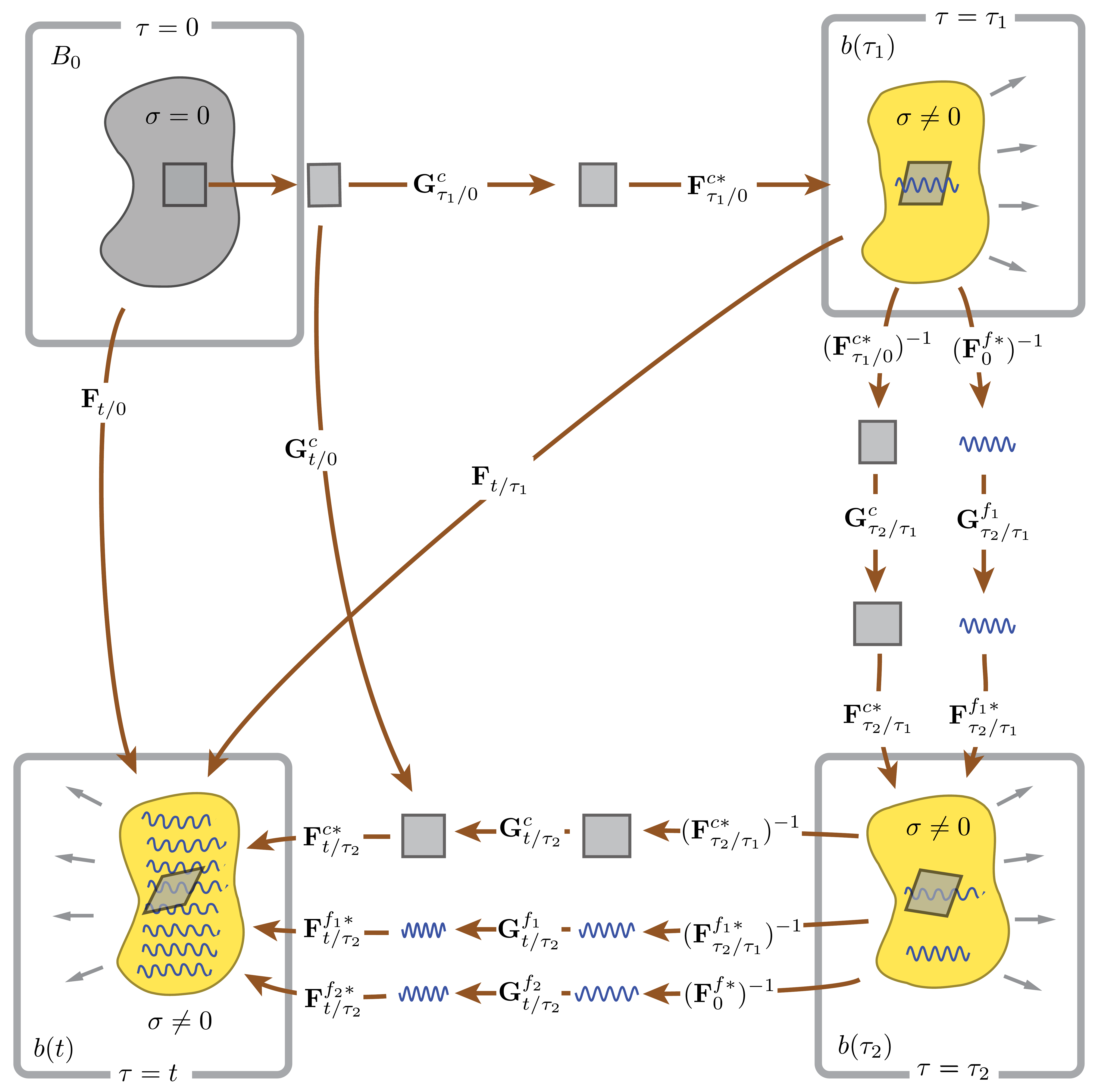}
\caption
{
Sequence of configurations which transform unloaded body $B_0$ into current
configuration $b(t)$, with fibers created at some intermediate time $\tau$.
See Table \ref{tab:Tensors} for definitions.  The mapping of the total
deformation between times $\tau_i$ and $\tau_j$ is given by
$\boldF_{\tau_j/\tau_i}$, while $\boldG_{\tau_j/\tau_i}$ and
$\boldF^*_{\tau_j/\tau_i}$ give the corresponding contraction and elastic
deformation, respectively.  The contraction and elastic deformation of cells
and each of the fibers is considered individually, but all components, once
created, undergo the same total deformation.  
The reference configuration    
$b(0)$ is a special case of $b(t)$, with no growth ($\bG^i=\boldI$) and only 
surface loads contributing to deformation.
}
\label{fig:GrowthSchematicMultiFiber}
\end{figure}
}

\newcommand{\figureGplots}{  % Fig 7
\begin{figure}[htbp]
  \centering
  \subfloat{\label{fig:Gc_Plot}
      \begin{overpic}[width=0.30\textwidth]{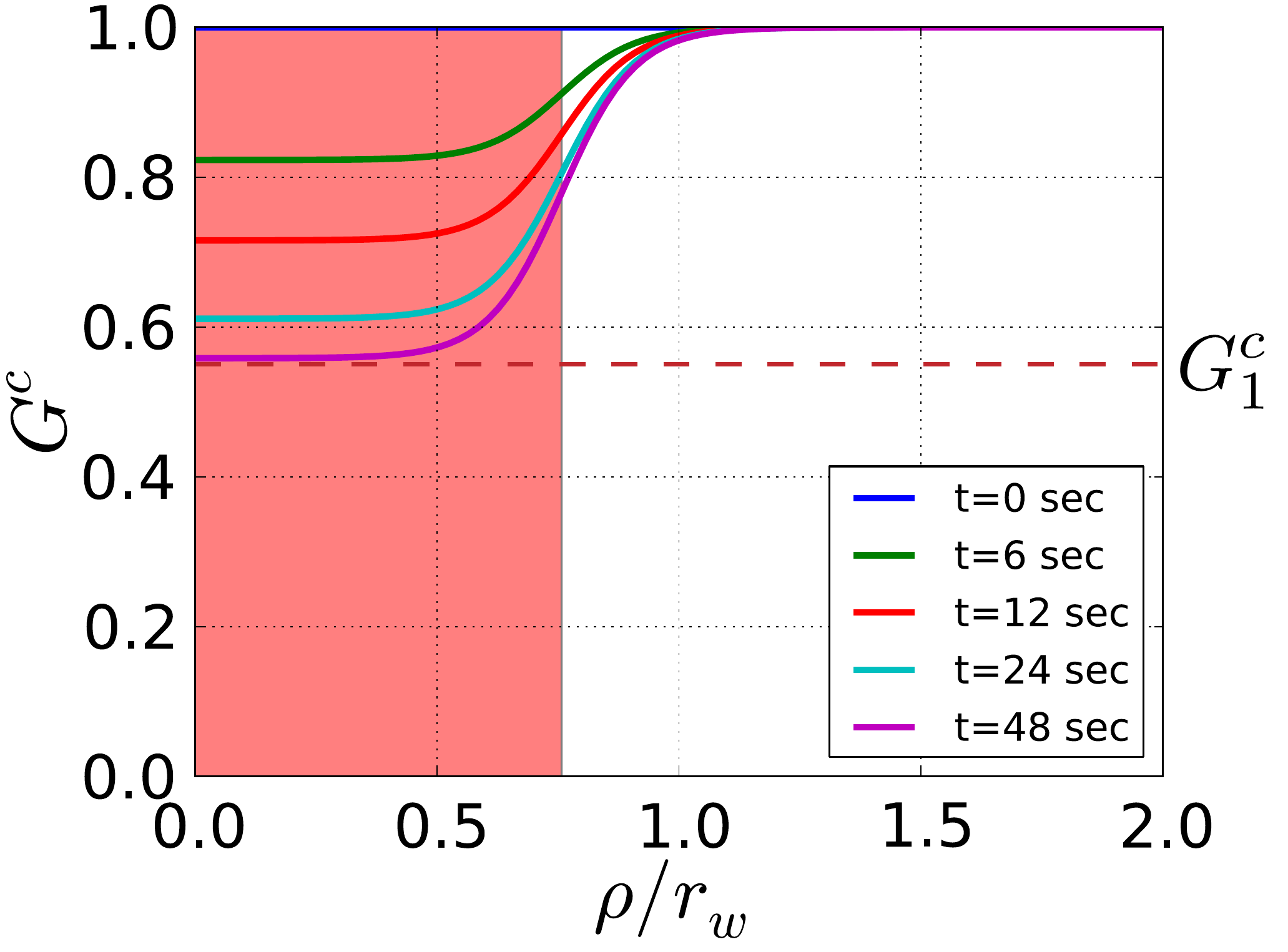}
        \put(0,0){\color{black}{\large (a)}} \end{overpic}}
  \subfloat{\label{fig:Gf_Plot}
      \begin{overpic}[width=0.30\textwidth]{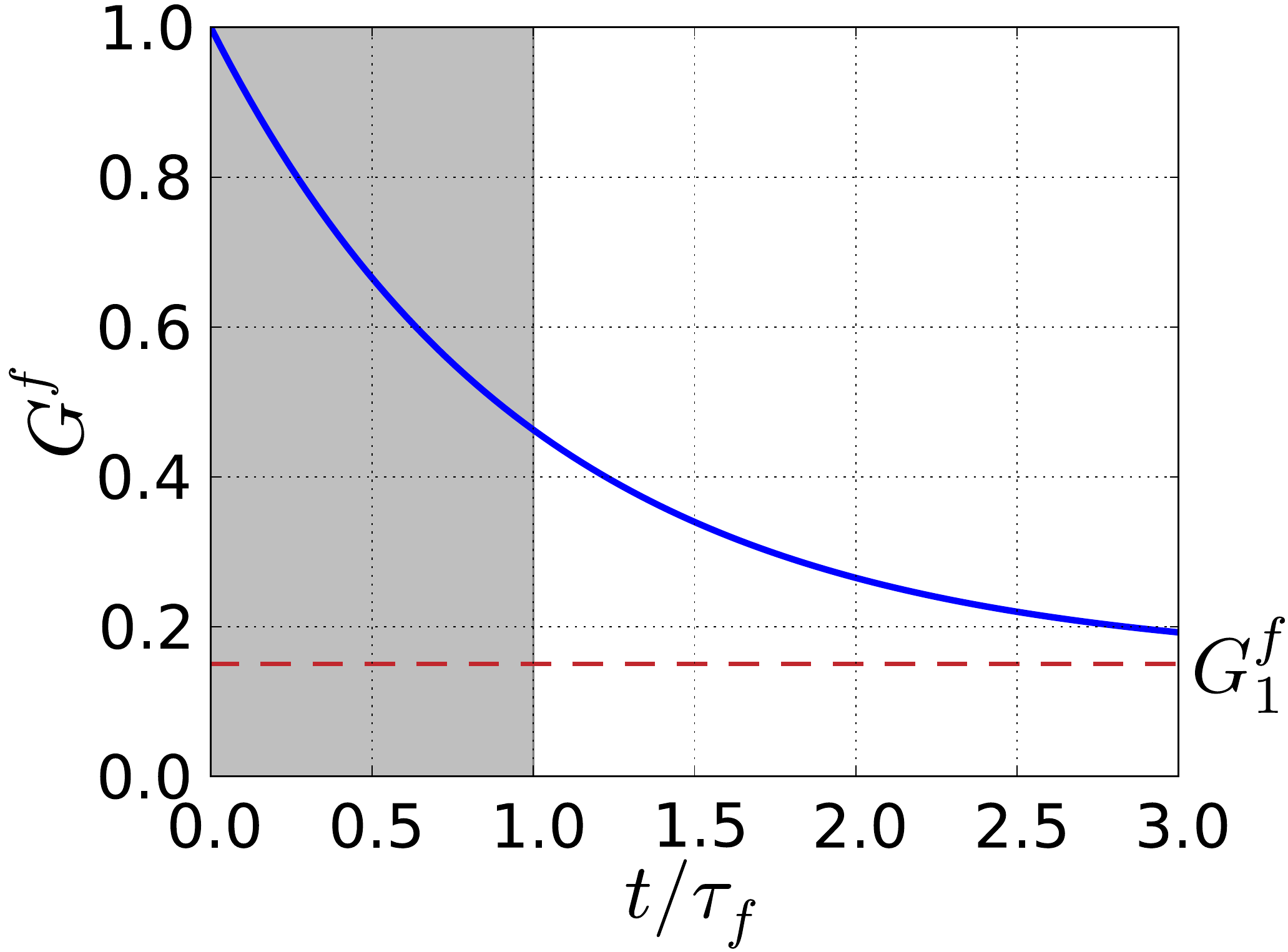}
        \put(0,0){\color{black}{\large (b)}} \end{overpic}}
  \subfloat{\label{fig:Phi_Plot}
      \begin{overpic}[width=0.30\textwidth]{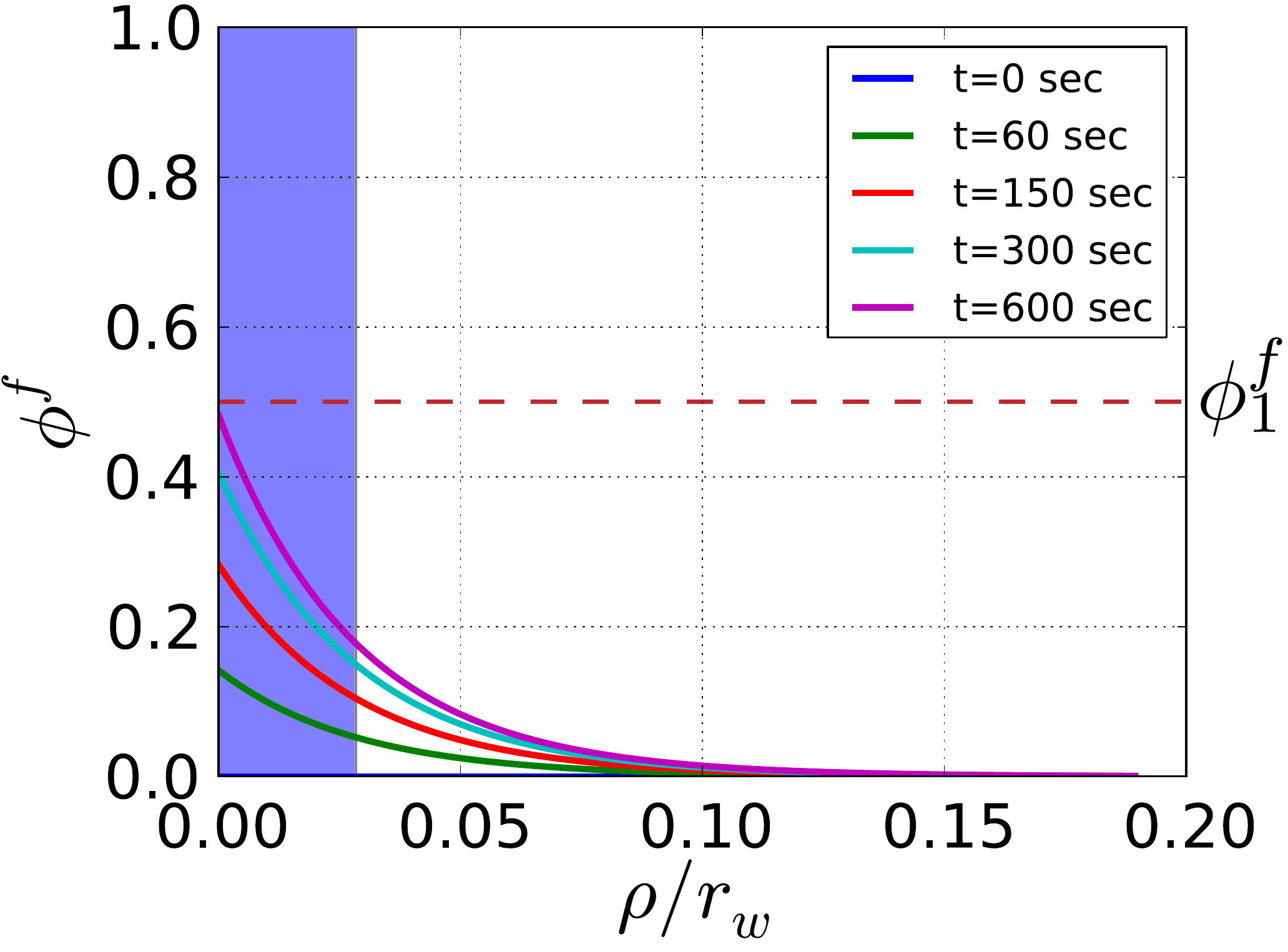}
        \put(0,0){\color{black}{\large (c)}} \end{overpic}}
  \caption
{
Cell and fiber contraction patterns, as well as fiber volume fraction, in the
vicinity of the wound in the undeformed configuration.  Parameters are given in
Tables~\ref{tab:Geometry} and \ref{tab:Parameters}.  \psubref{fig:Gc_Plot} Cell
contraction $G^c$ (Eq.~\ref{eqn:Gca}) as a function of distance from wound edge
$\rho$ for various times, with $\rho$ (see \ref{sec:coordinates}) normalized by
initial circular wound radius $r_w$.  Shaded region indicates domain of thick
cell ring.
\psubref{fig:Gf_Plot} Fiber contraction $G^f$ (Eq.~\ref{eqn:G_f}) as a function
of time, normalized by $\tau_f$.  At the end of the simulation (600 sec,
indicated by shaded region) $G^f$ has decreased from 1 to 0.46.  
\psubref{fig:Phi_Plot} Fiber volume fraction $\phi^f$ near the wound for
various times (Eq.~\ref{eqn:phi_f}).  Shaded region indicates nominal domain of
fiber ring.  Note that $G^f$ is spatially uniform, with the extent of fiber
ring given by $\phi^f$.
} \label{fig:Gplots}
\end{figure}
}

\newcommand{\figureModelCompare}{ % Fig 8
\begin{figure}[htbp]
  \begin{minipage}{\textwidth}
  \centering
  \subfloat{\label{fig:ModelCompare_schematic}
    \begin{overpic}[width=0.22\textwidth]{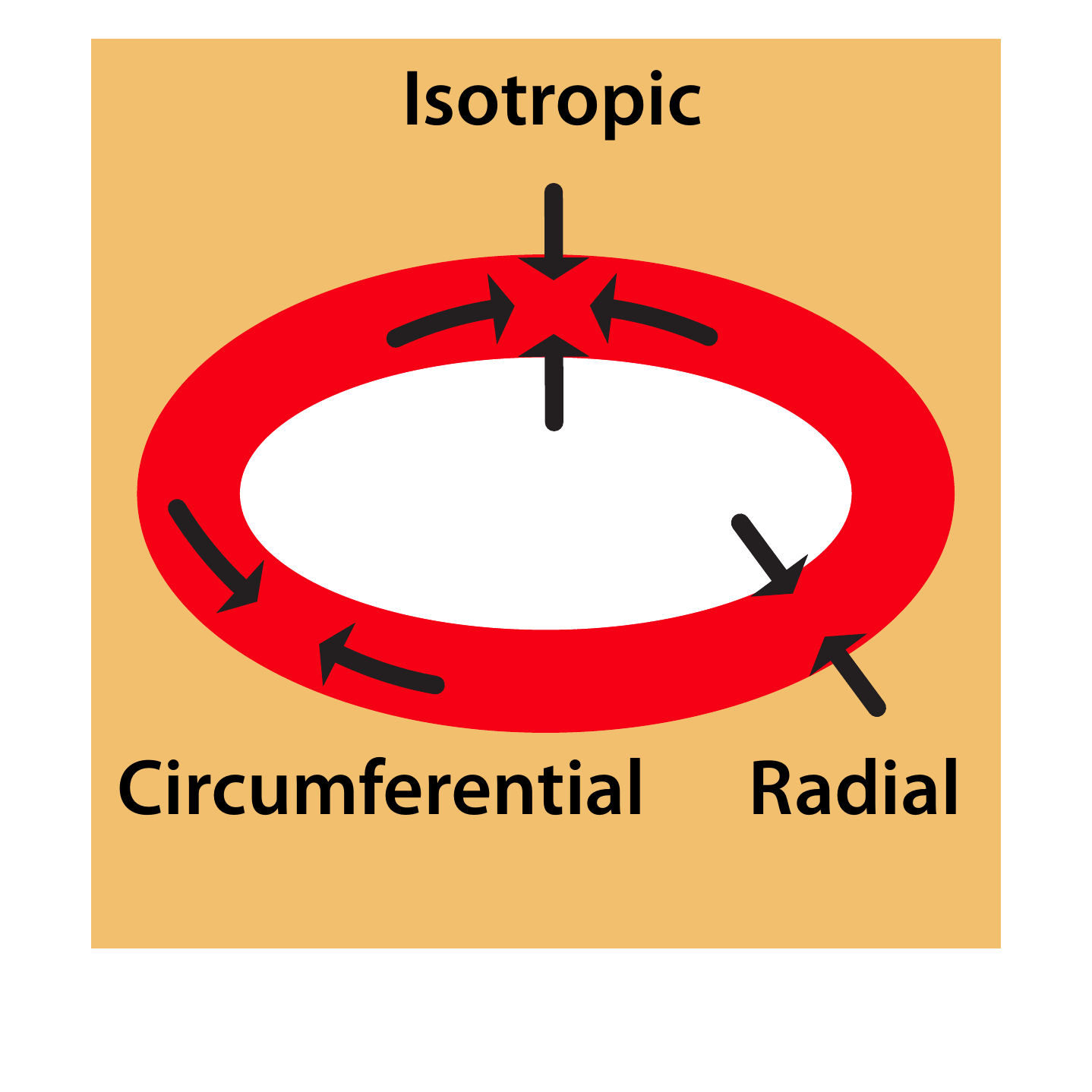}
        \put(0,0){\color{black}{\large (a)}} \end{overpic}}
  \subfloat{\label{fig:ModelCompare_area}
    \begin{overpic}[width=0.30\textwidth]{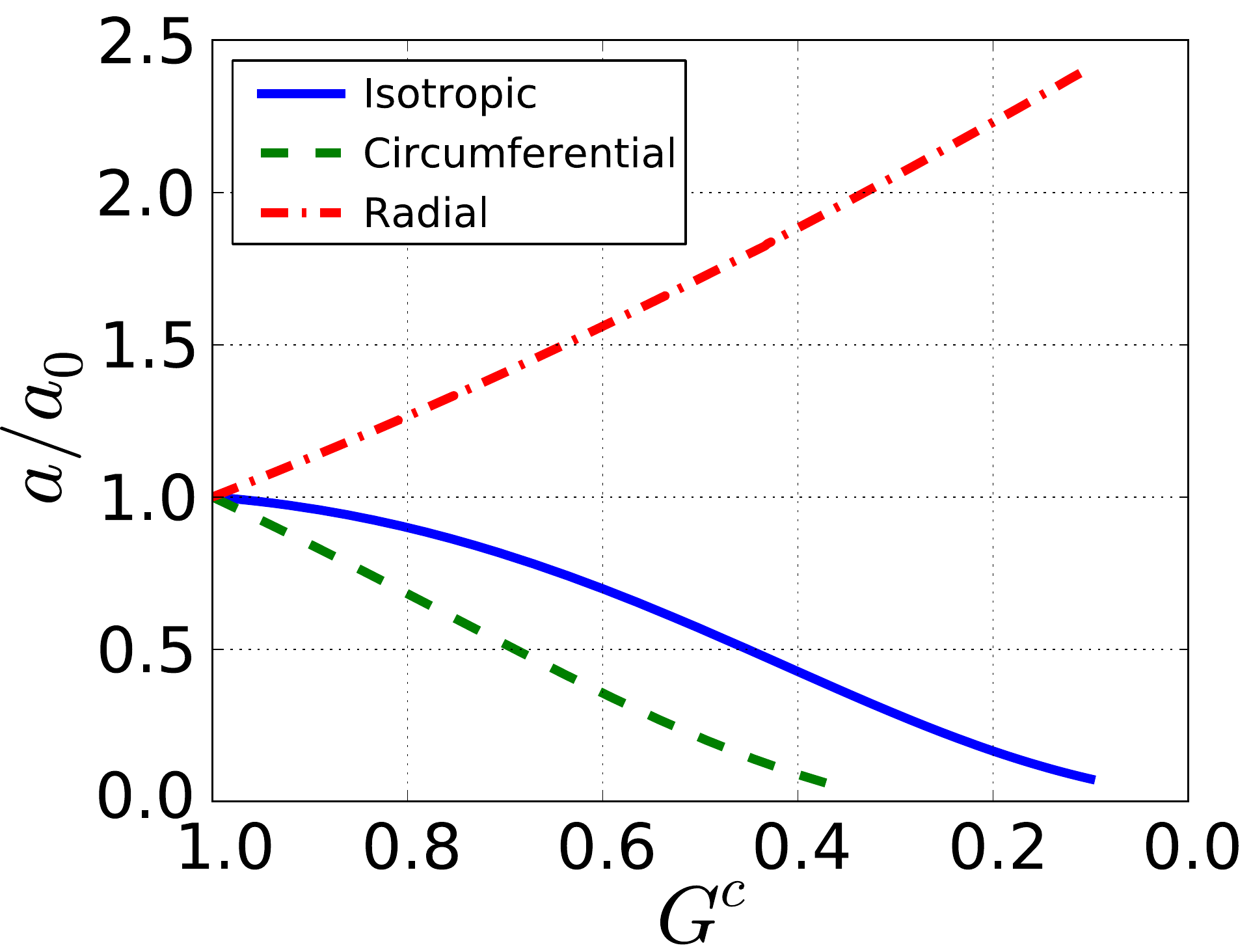}
        \put(0,0){\color{black}{\large (b)}} \end{overpic}}
  \subfloat{\label{fig:ModelCompare_AR}
    \begin{overpic}[width=0.30\textwidth]{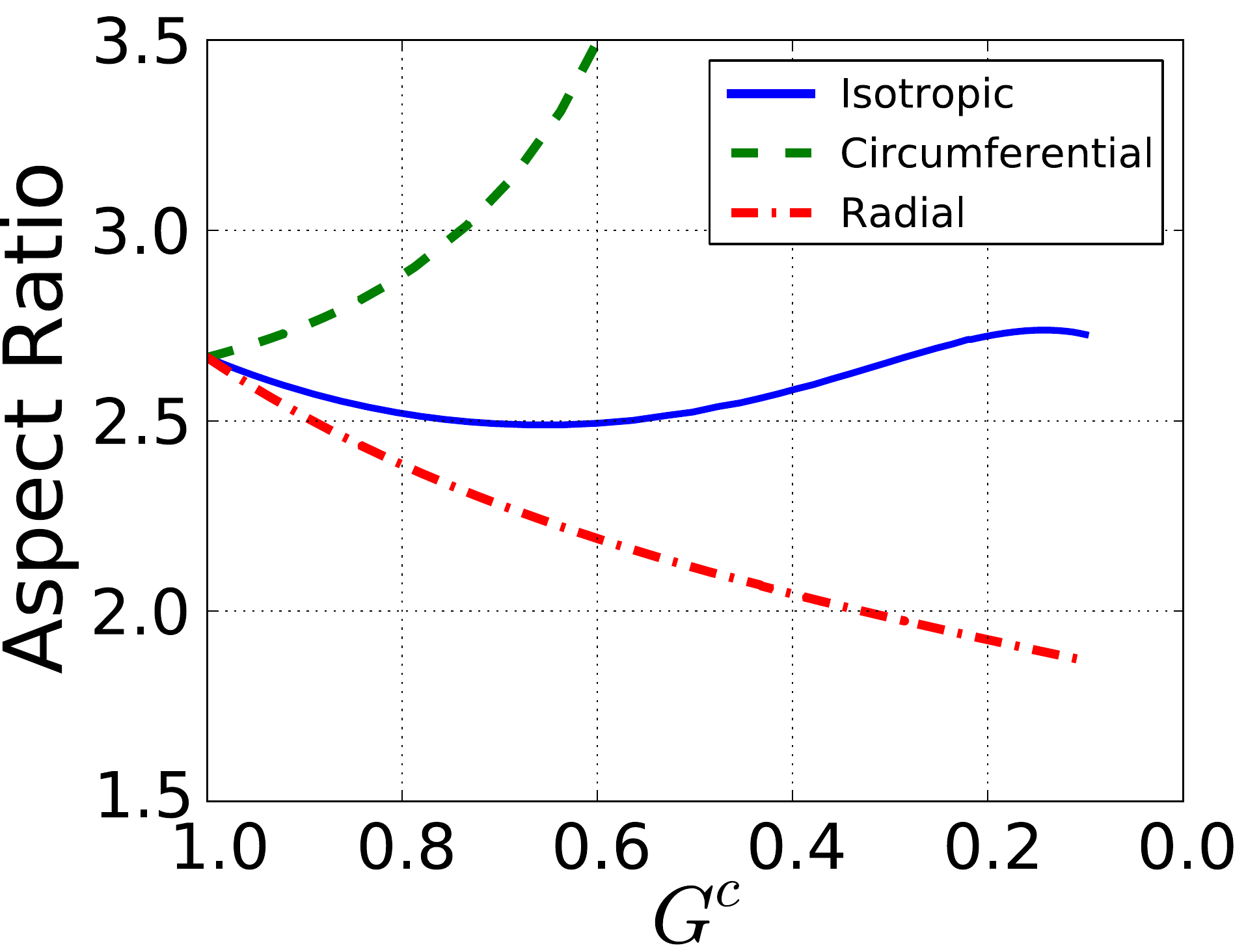}
        \put(0,0){\color{black}{\large (c)}} \end{overpic}}
  \end{minipage}
  \caption
{
Area and aspect ratio (AR) trends for various cell contraction schemes for
elliptical wound model.  No fibers are present in these simulations (fiber
fraction $\phi^f=0$) and cell contraction $G^c$ (Eq.~\ref{eqn:GcCases}) in the
ring is specified directly (note, in model $G^c < G^c_1=0.55$).  Wound area $a$
is normalized by area $a_0$ at $t=0$.
\psubref{fig:ModelCompare_schematic} The epithelial membrane in the cell ring
can contract in both directions (isotropic) or in one
direction only (radial or circumferential).  
(\psubref*{fig:ModelCompare_area}, \psubref*{fig:ModelCompare_AR}) While the wound
closes for both isotropic and circumferential schemes, only the isotropic case
yields AR trends consistent with experiment.  
These trends are independent of $\sigma_B$.
} \label{fig:ModelCompare}
\end{figure}
}

\newcommand{\figureModelFitWound}{  % Fig 9
\begin{figure}[htbp]
\centering
    \subfloat{\label{fig:ModelFitCircle} 
        \begin{overpic}[width=0.30\textwidth]{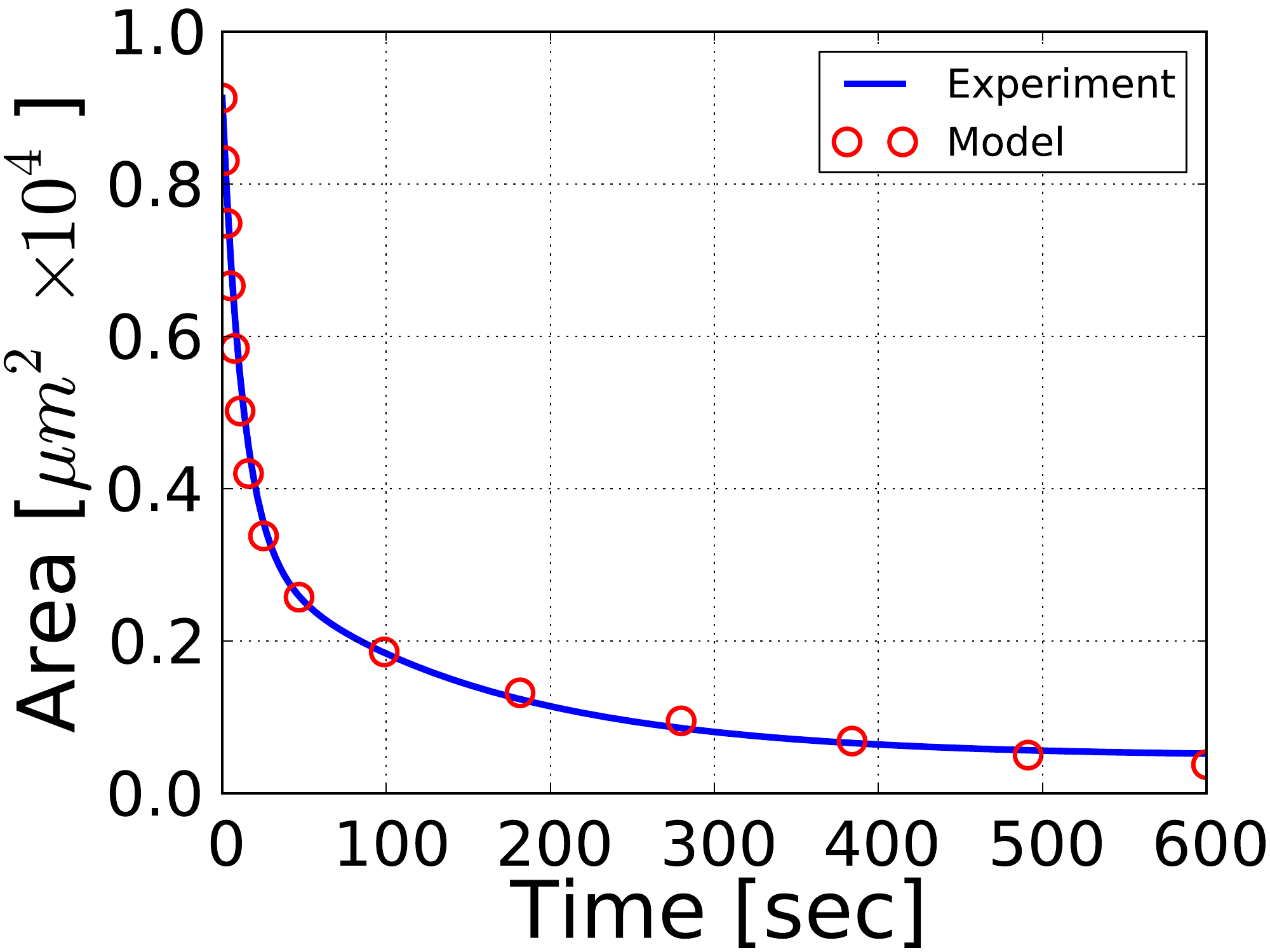}
        \put(0,0){\color{black}{\large (a)}} \end{overpic}}
    \subfloat{\label{fig:ModelFitEllipseArea}
        \begin{overpic}[width=0.30\textwidth]{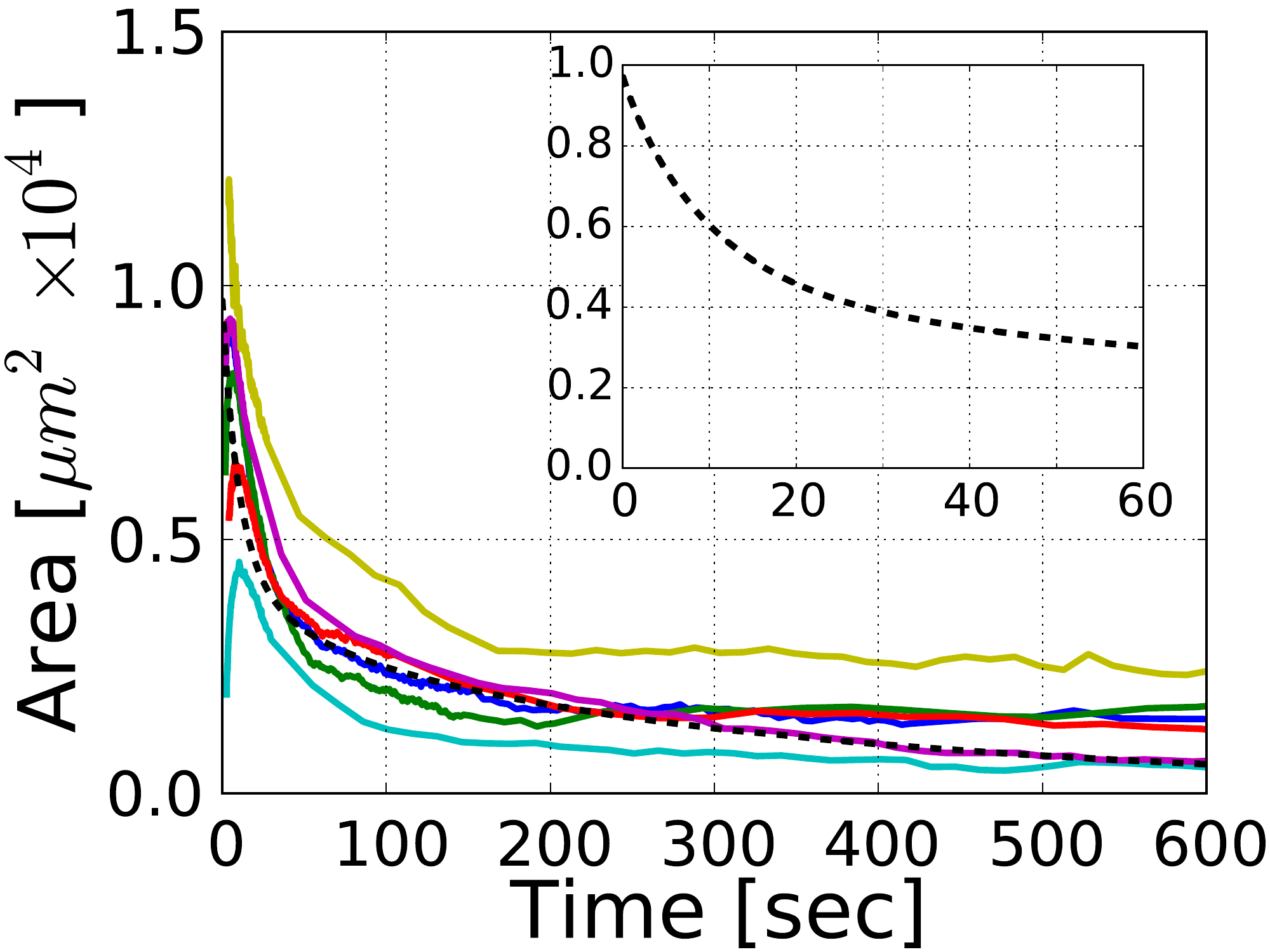}
        \put(0,0){\color{black}{\large (b)}} \end{overpic}}
    \subfloat{\label{fig:ModelFitEllipseAR}
        \begin{overpic}[width=0.30\textwidth]{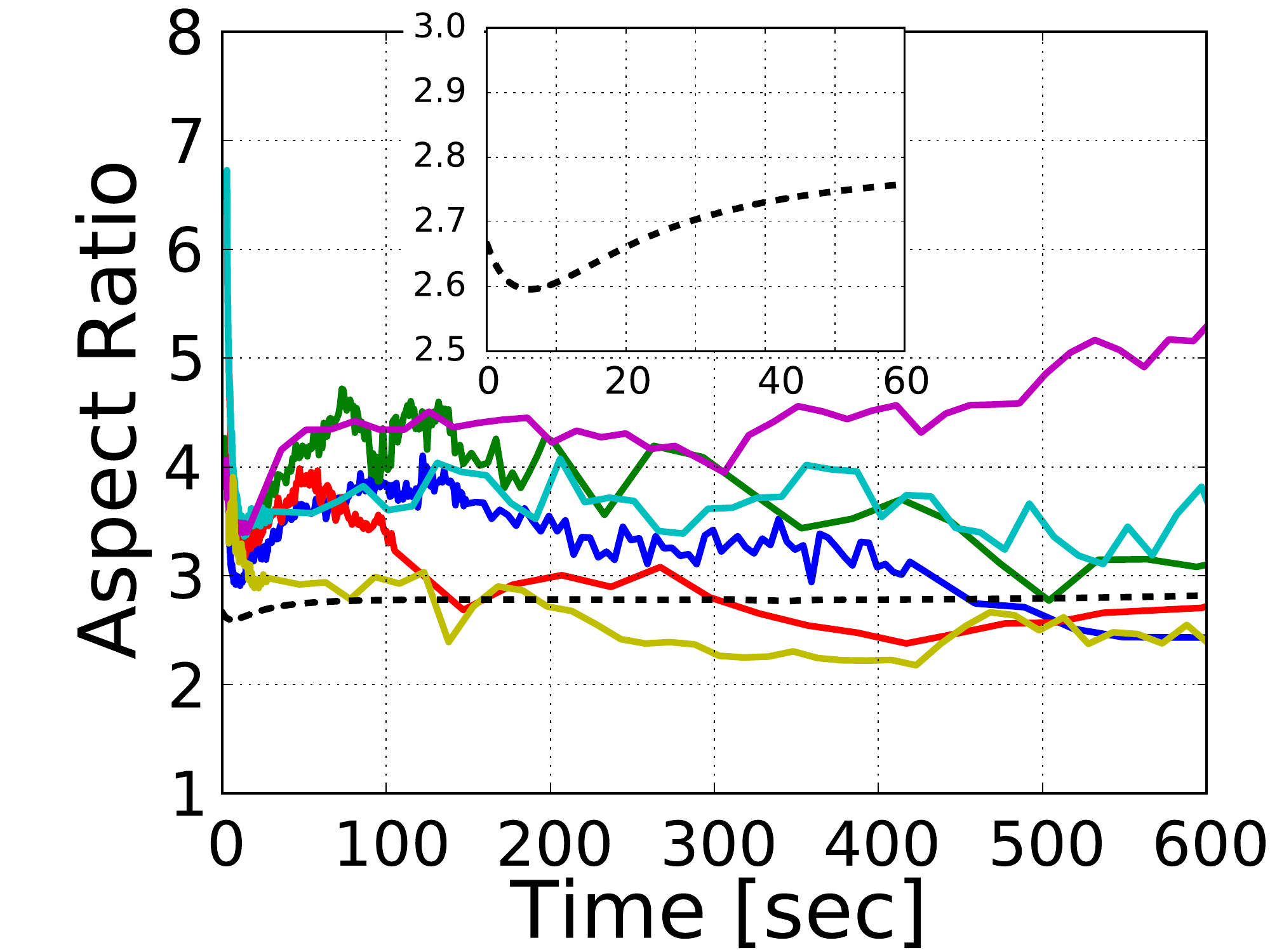}
        \put(0,0){\color{black}{\large (c)}} \end{overpic}}
\caption
{
Comparison of experimental and model wound healing behavior. (a) Circular wound.  Solid curve is
the experimental wound area given by Eq.~\eqref{eqn:fit} with mean parameters from
Table~\ref{tab:WoundFitAll}.  Circles indicate model results (parameters from
Table~\ref{tab:Parameters}). (b,c) 
Elliptical wound. Solid curves show experimental area (b) and AR (c).  Model
results (dashed curve) use same parameters as circular wound.  Insets detail model behavior
in first minute, with same axis labels as main plot.  
}
\label{fig:ModelFitWound}
\end{figure}
}

\newcommand{\figureModelStressCircular}{  % Fig 10
\begin{figure}[htbp]
\begin{minipage}{\textwidth}
\centering
    \subfloat{\label{fig:sig_rad_R}
        \begin{overpic}[width=0.40\textwidth]{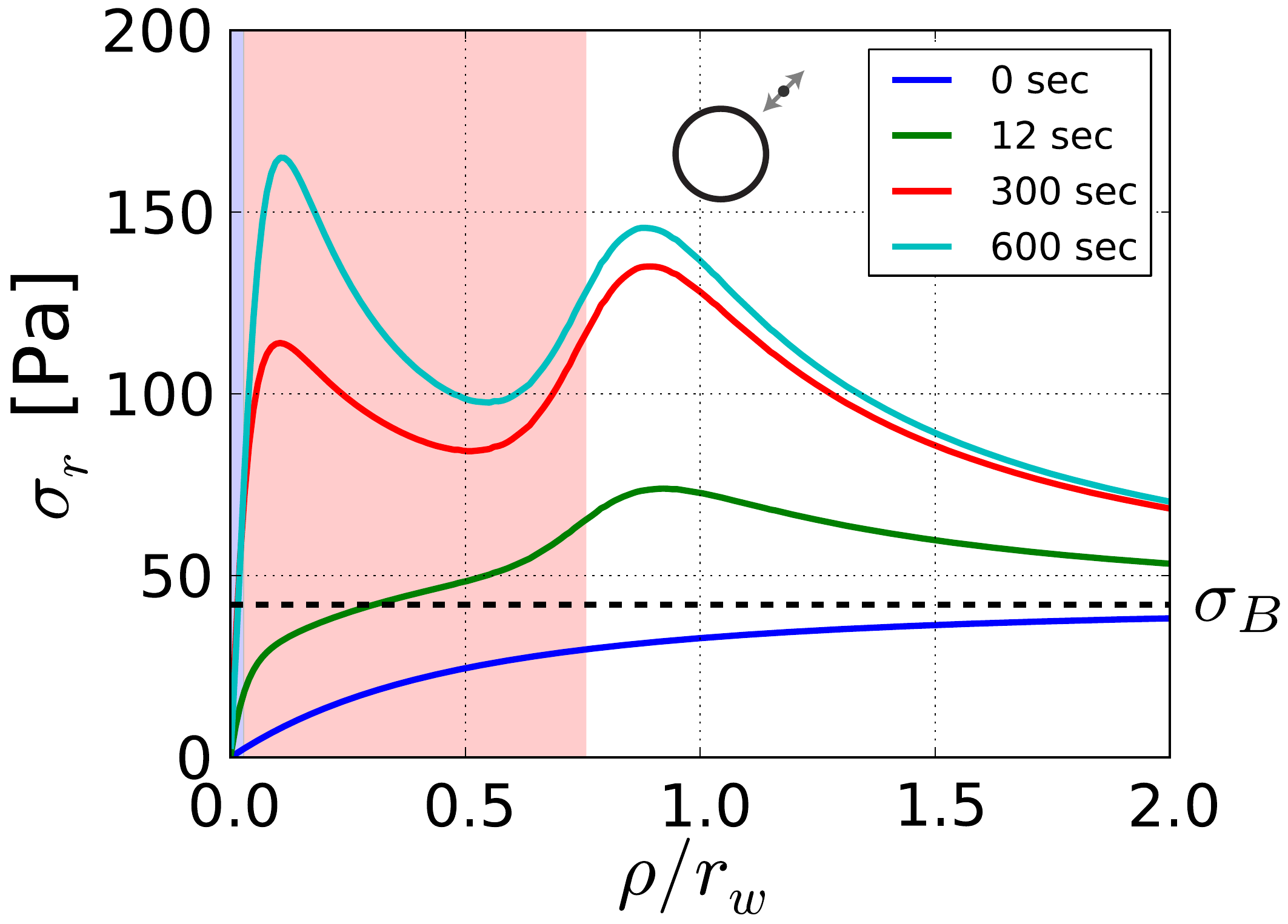}
        \put(1,3){\color{black}{\large (a)}} \end{overpic}}
    \subfloat{\label{fig:sig_circum_R}
        \begin{overpic}[width=0.40\textwidth]{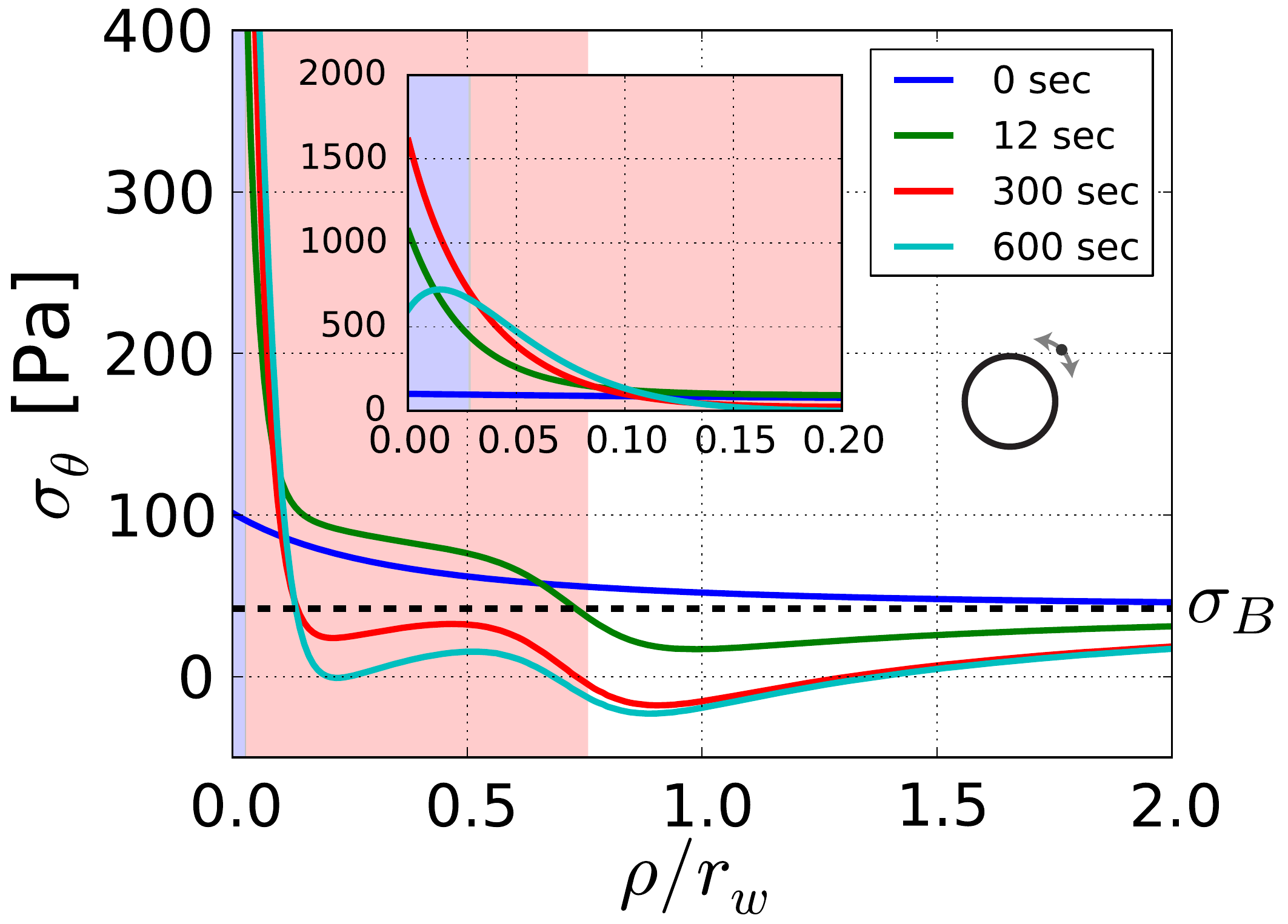}
        \put(1,3){\color{black}{\large (b)}} \end{overpic}}
\end{minipage}
\caption
{
Epithelial membrane stress in the vicinity of circular wound.  Icon indicates
stress component (radial $\sigma_r$, circumferential $\sigma_\theta$).  Distance from
wound edge in the reference configuration is given in multiples of the circular
wound radius $r_w$, and the fiber and cell regions are indicated with blue and
red shading, respectively (see Fig.~\ref{fig:Gplots}).  Inset in (b) details 
stress near wound boundary, with axes labels as in main figure.
Far from the wound,
both stress components approach the far-field value $\sigma_B = 42$~Pa.
}
\label{fig:ModelStressCircular}
\end{figure}
}

\newcommand{\figureModelFitSigB}{  % Fig 11
\begin{figure}[htbp]
\begin{minipage}{\textwidth}
\centering
  \subfloat{\label{fig:sigb_compare}
    \begin{overpic}[width=0.24\textwidth]{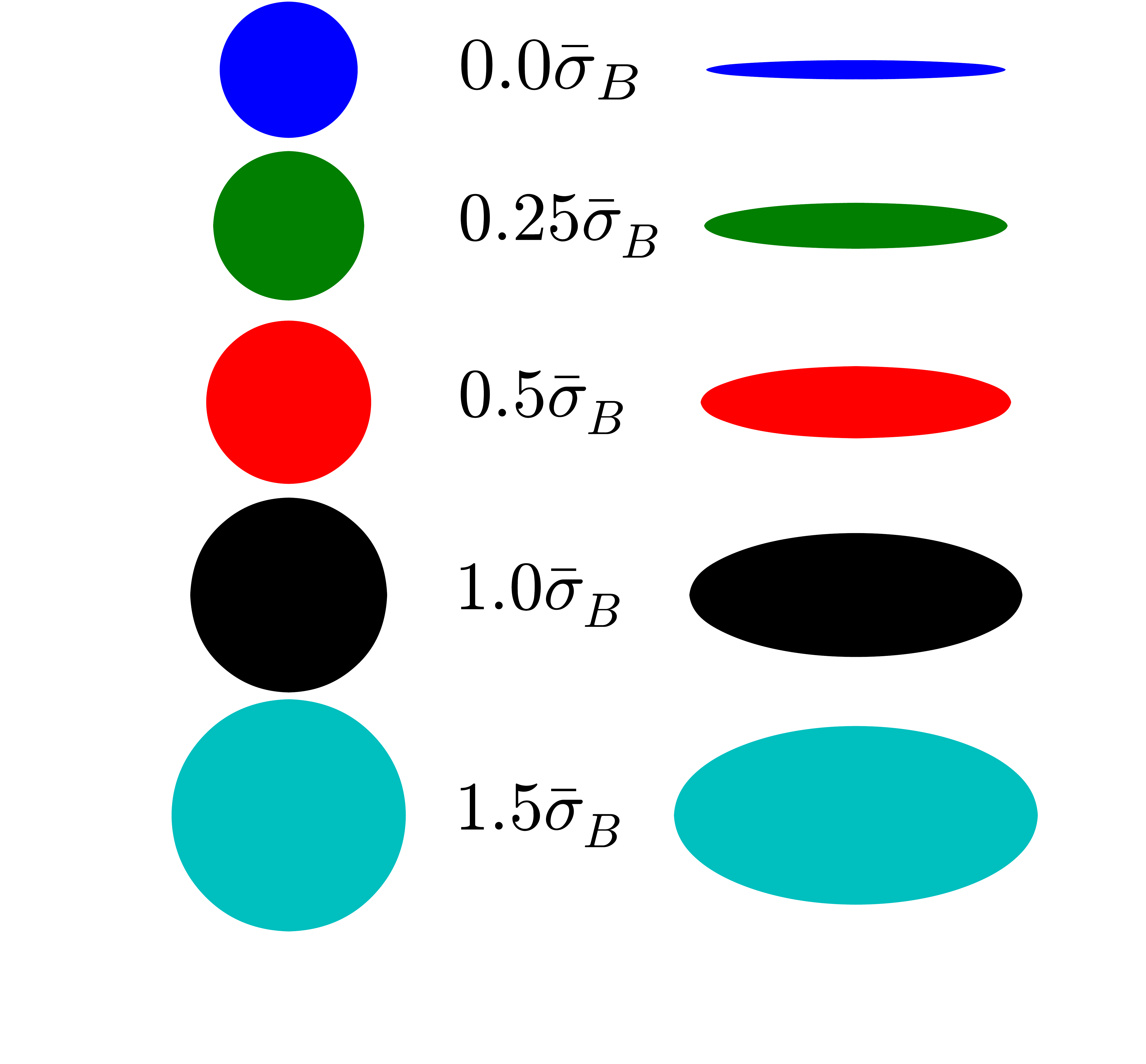}
        \put(0,0){\color{black}{\large (a)}} \end{overpic}}
  \subfloat{\label{fig:Fbnd_area_1min}
    \begin{overpic}[width=0.30\textwidth]{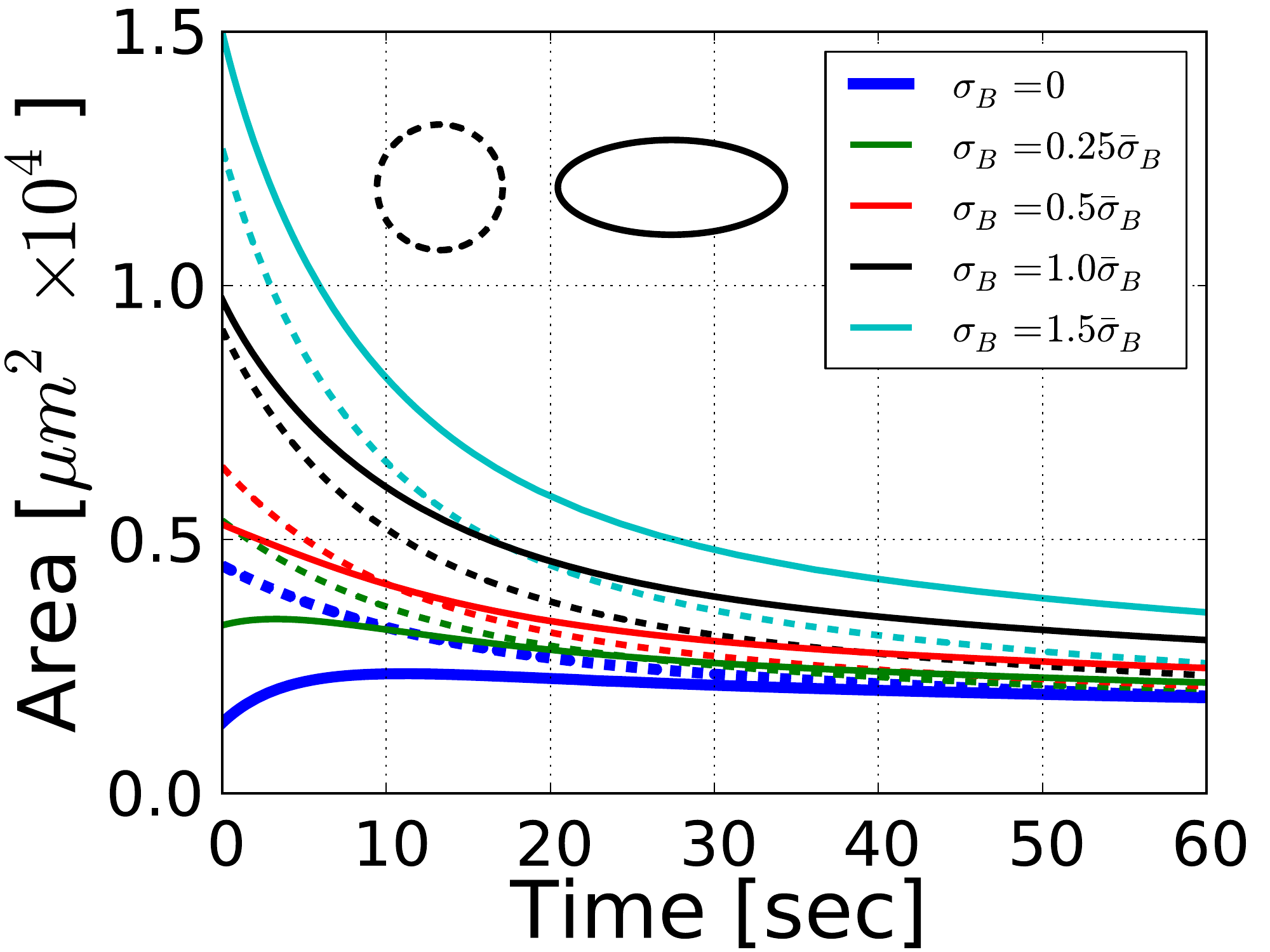}
        \put(0,0){\color{black}{\large (b)}} \end{overpic}}
  \subfloat{\label{fig:Fbnd_AR_1min}
    \begin{overpic}[width=0.30\textwidth]{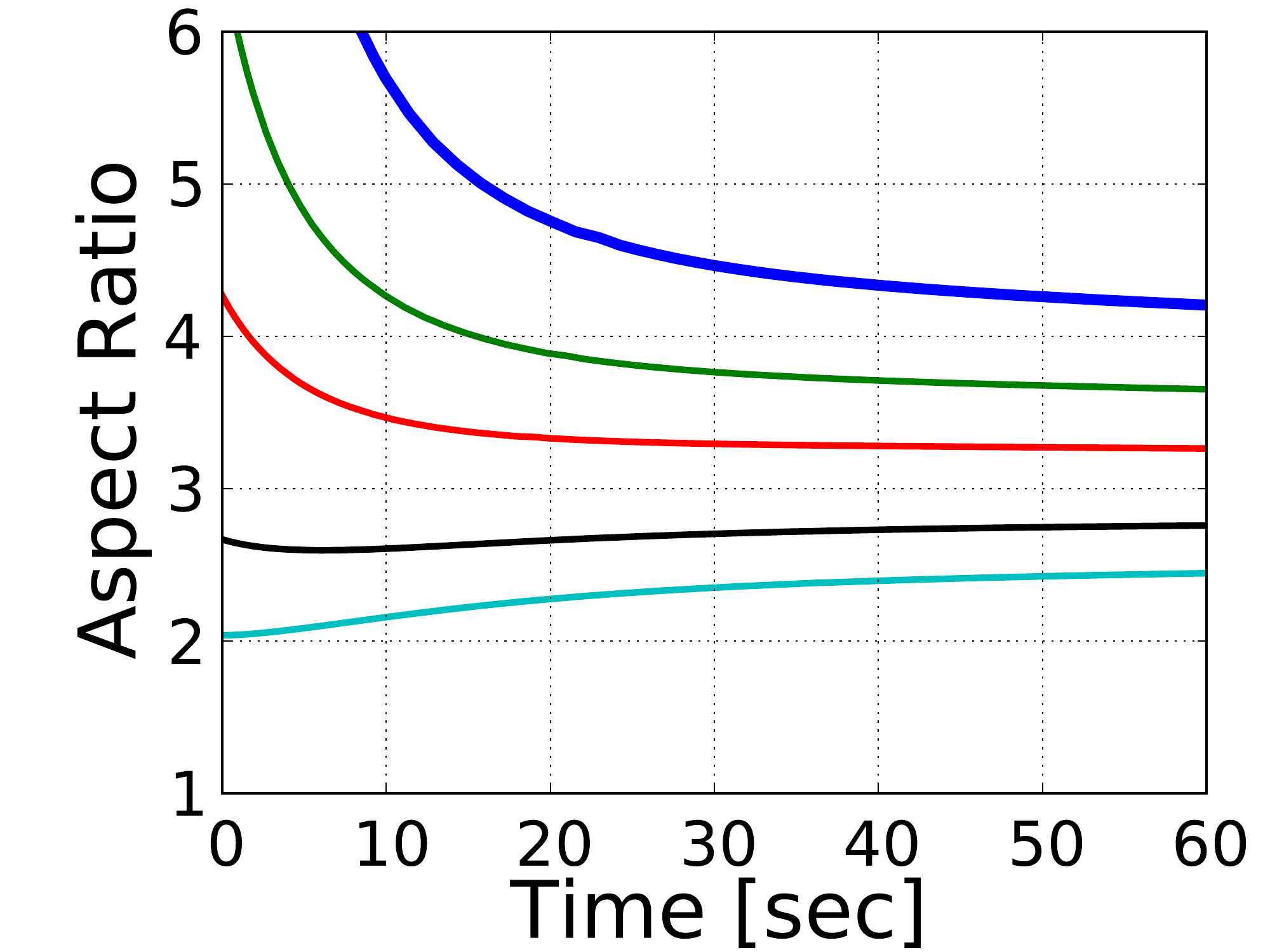}
        \put(0,0){\color{black}{\large (c)}} \end{overpic}}
\end{minipage}
\caption
{
Effect of tension $\sigma_B$ on circular and elliptical model wounds.
\psubref{fig:sigb_compare} Wound size and shape
at $t=0$ ($a_0$) for different $\sigma_B$.  
Tension is
relative to $\bar{\sigma}_B=42$ Pa (Table~\ref{tab:Parameters}).  
(b) Area for 
circular wounds (dashed curves) and elliptical wounds (solid curves) with different constant values of
$\sigma_B$.  
(c) Aspect ratio (AR) for elliptical wounds.
Low initial tension in
elliptical wounds results in a brief increase in wound area and a
sharp decrease in AR. 
}
\label{fig:ModelFitSigB}
\end{figure}
}

\newcommand{\figureGeometryCoord}{  % Fig 12
\begin{figure}[htbp]
  \centering
  \subfloat{\label{fig:GeometryCylindrical}
      \begin{overpic}[width=0.40\textwidth]{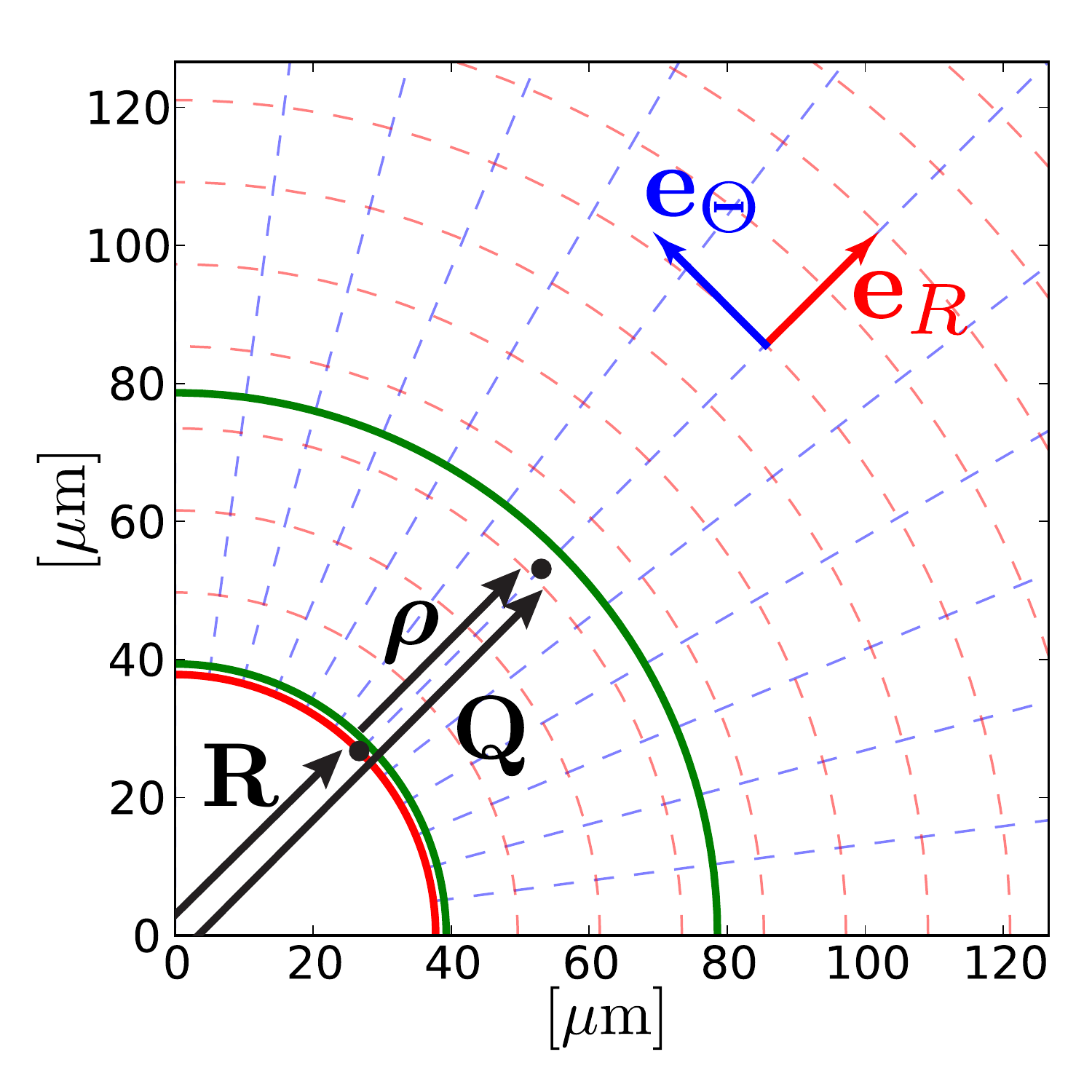}
        \put(1,3){\color{black}{\large (a)}} \end{overpic}}
  \subfloat{\label{fig:GeometryElliptical}
      \begin{overpic}[width=0.40\textwidth]{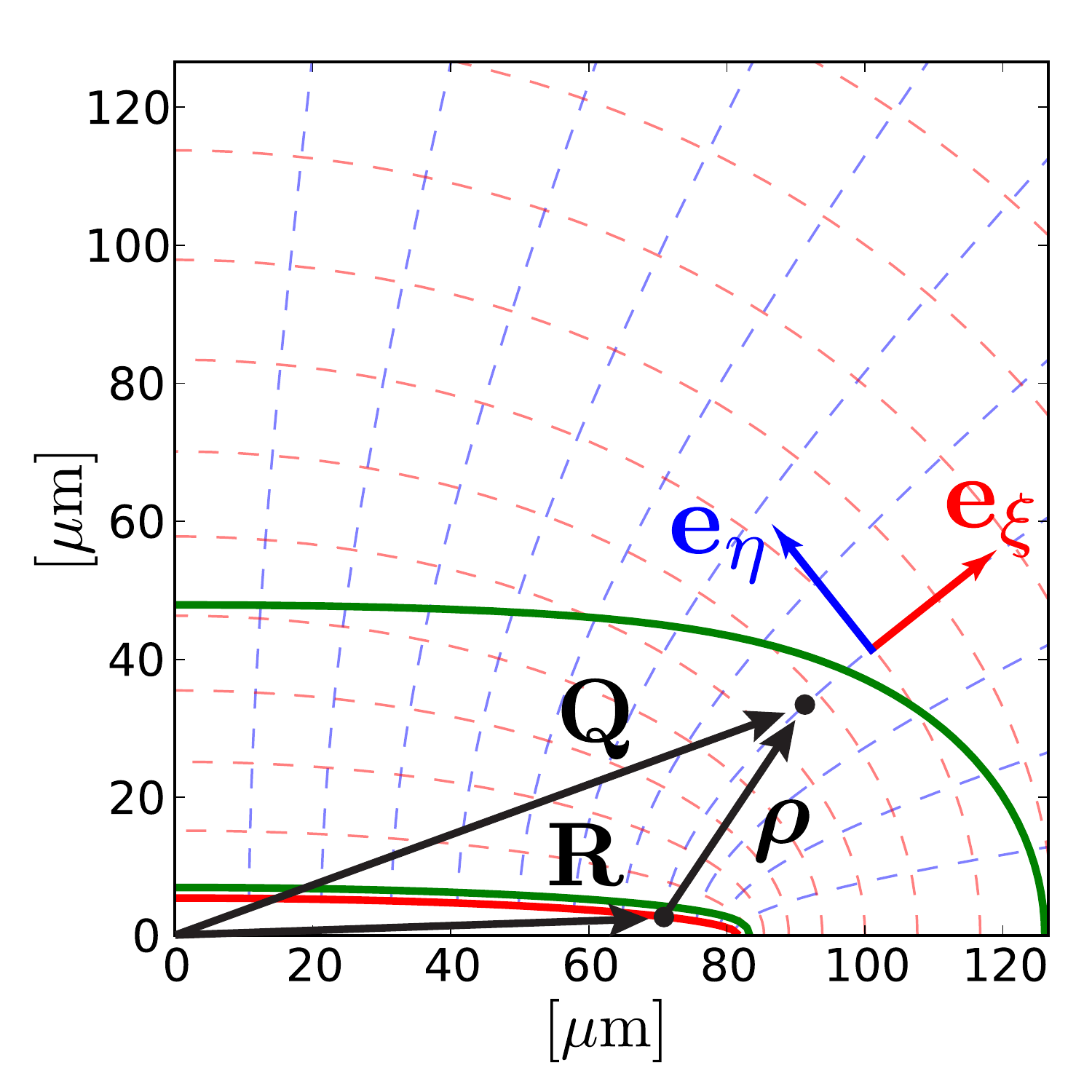}
        \put(1,3){\color{black}{\large (b)}} \end{overpic}}
  \caption
{
Detail of model geometry in the immediate vicinity of the  wound in the
undeformed ($B_0$) configuration for circular \psubref{fig:GeometryCylindrical}
and elliptical \psubref{fig:GeometryElliptical} wounds.  Because of symmetry,
only one quadrant of the membrane is considered, and the wound lies
in the lower-left corner.  The solid red curve indicates the wound boundaries,
and the green curves mark the boundaries of the thin and thick rings.
Dashed blue and red curves indicate
direction along the circumferential ($\bolde_\Theta$ and $\bolde_\eta$) and radial ($\bolde_R$ and $\bolde_\xi$)
directions, respectively.  The vector $\boldrho$ marks the distance between an
arbitrary vector $\boldQ$ and a corresponding vector $\boldR$ on the wound edge.
In panel \psubref{fig:GeometryCylindrical} these colinear vectors have been separated for clarity.
} \label{fig:GeometryCoord}
\end{figure}
}

\newcommand{\figureFiberStress}{ % Fig 13
\begin{figure}[htbp]
\centering
    \includegraphics[width=0.35\textwidth]{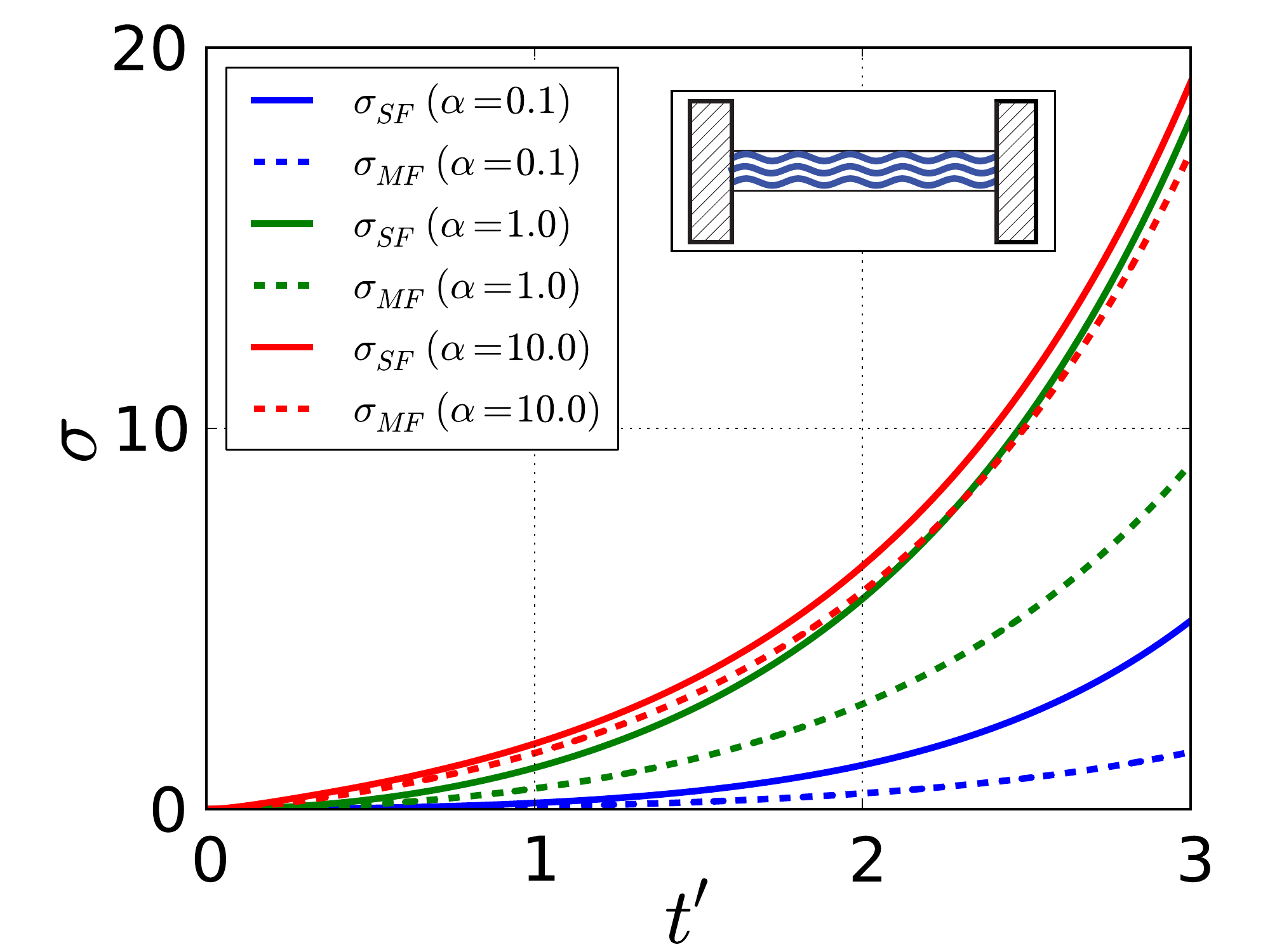}
\caption
{
Fiber stress in a homogeneous bar with fixed ends (inset) as a function of
non-dimensional time $t'$ for different fiber models and values of $ \alpha$.
In all cases $\sigma_{SF} \ge \sigma_{MF}$, but this difference decreases with
large $\alpha$.
}
\label{fig:FiberStress}
\end{figure}
}

\newcommand{\figureWoundColor}{  % Fig S1
\begin{figure}[htbp]
  \centering
    \includegraphics[width=0.40\textwidth]{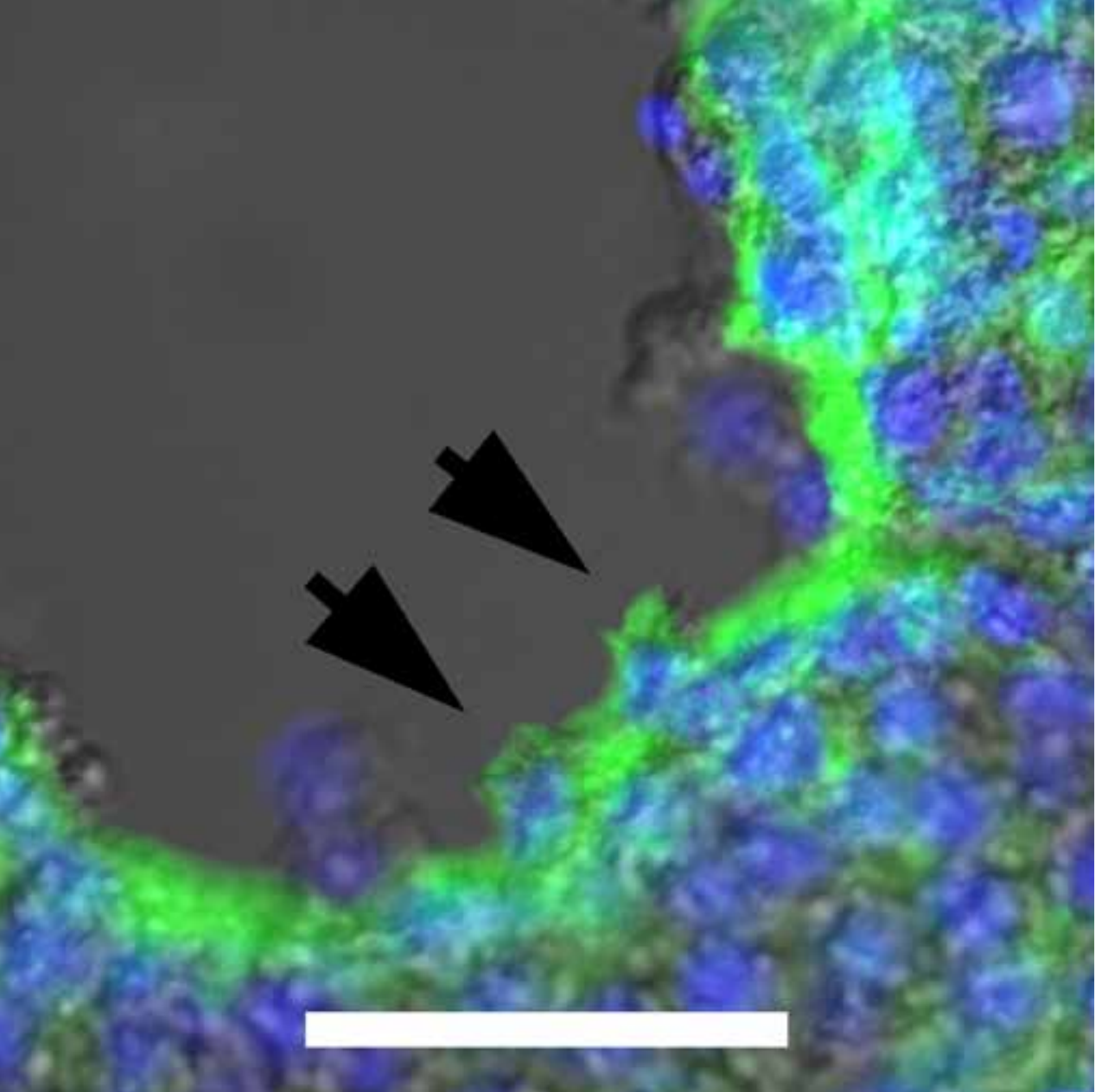}
  \caption
{
Detail of actin staining in Fig.~\ref{fig:endo_1min} (endoderm, fixed 1 min after wounding) with cells (gray), nuclei
(blue), and actin (green).  Arrows indicate lamellipodia; scale bar 50 $\mu$m.
} \label{fig:endo_1min_s}
\end{figure}
}

\newcommand{\figureWoundpMLC}{  % Fig S2
\begin{figure}[htbp]
  \begin{minipage}{\textwidth}
  \centering
  \subfloat{\label{fig:pmlc_10sec}
    \begin{overpic}[width=0.40\textwidth]{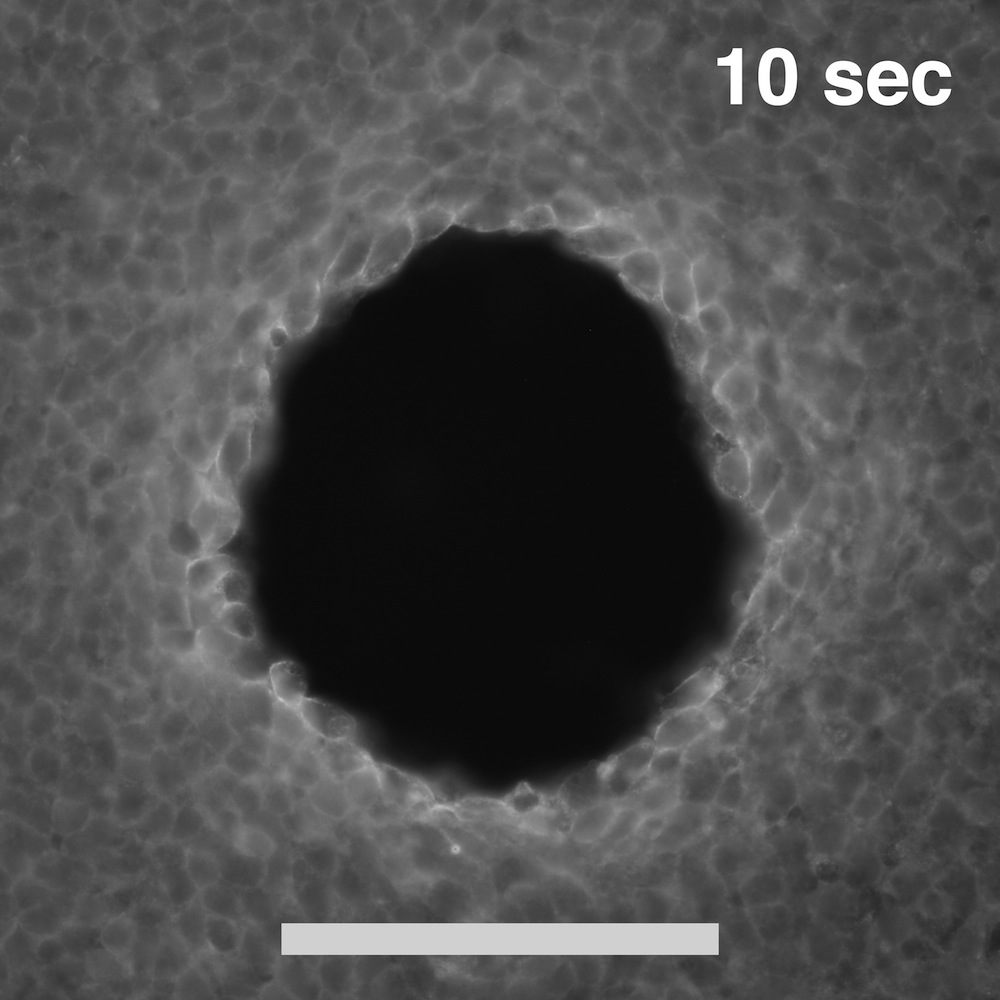}
        \put(1,3){\color{white}{\large (a)}} \end{overpic}}
  \subfloat{\label{fig:pmlc_10min}  % see notebook p. 150 for discussion of ID
    \begin{overpic}[width=0.40\textwidth]{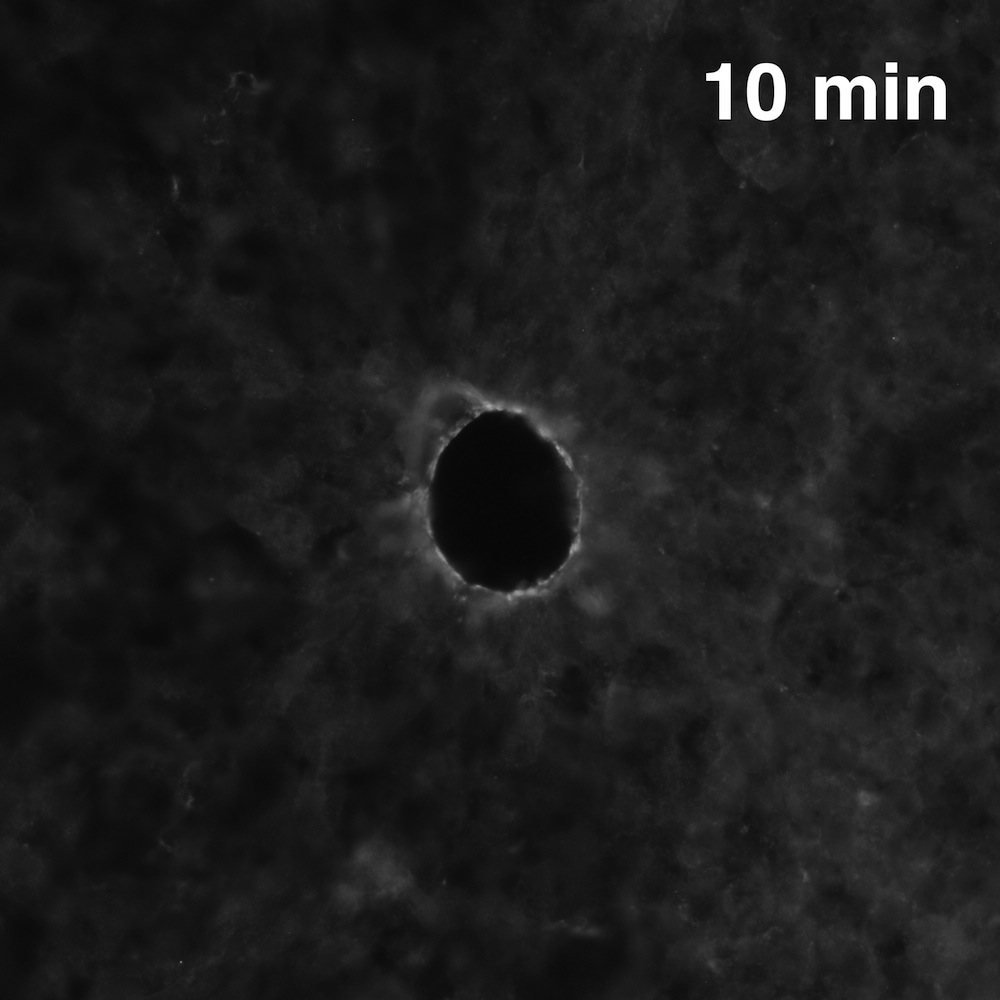}
        \put(1,3){\color{white}{\large (b)}} \end{overpic}}
  \end{minipage}
  \caption
{
Fluorescence images of ectoderm  near circular wound stained for phosphorylated
myosin light chain (pMLC) at given times post wounding. Scale bar 100 $\mu m$.
} \label{fig:WoundPMLC}
\end{figure}
}

\newcommand{\figureWoundActinElliptical}{  % Fig S3
\begin{figure}[htbp]
  \begin{minipage}{\textwidth}
  \centering
  \subfloat{\label{fig:ecto_3min}
    \begin{overpic}[width=0.30\textwidth]{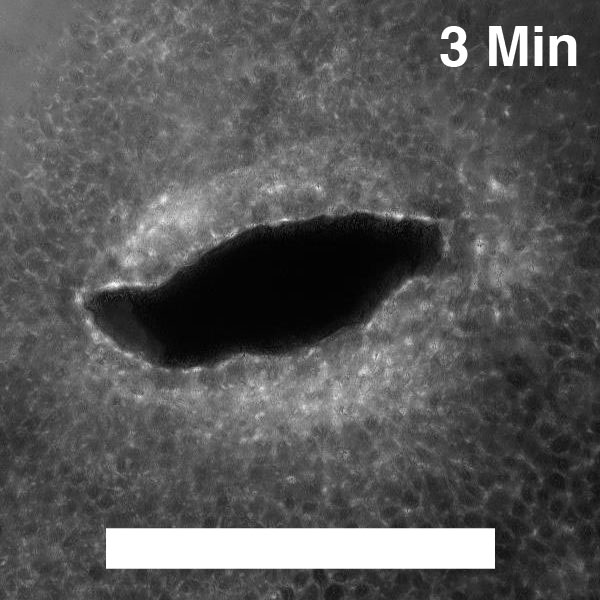}
        \put(1,3){\color{white}{\large (a)}} \end{overpic}}
  \subfloat{\label{fig:ecto_15min}
    \begin{overpic}[width=0.30\textwidth]{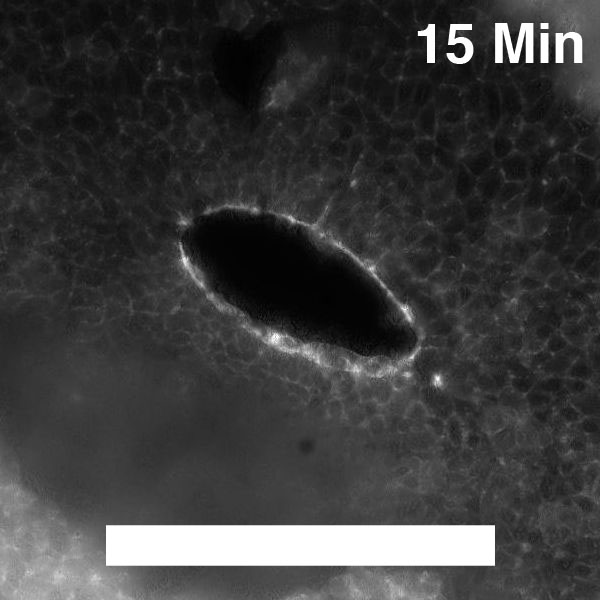}
        \put(1,3){\color{white}{\large (b)}} \end{overpic}}
  \subfloat{\label{fig:endo_15min}
    \begin{overpic}[width=0.30\textwidth]{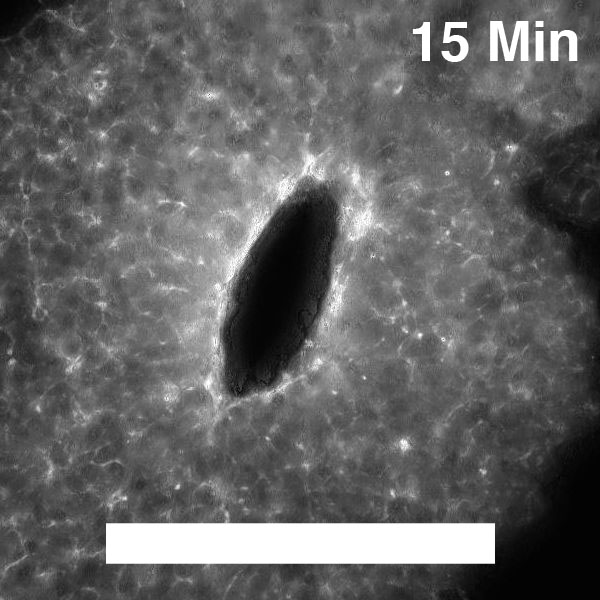}
        \put(1,3){\color{white}{\large (c)}} \end{overpic}}
  \end{minipage}
  \caption
{
Fluorescence images of elliptical wounds stained for actin at given times after wounding.
(\psubref*{fig:ecto_3min},\psubref*{fig:ecto_15min}) ectoderm;
(\psubref*{fig:endo_15min}) endoderm.  Scale bars
200$\mu$m.  Panels \psubref*{fig:ecto_15min} and \psubref*{fig:endo_15min}
are same wound.  Images have been processed with Extended Depth of Field ImageJ plugin~\citep{Forster:2004}.
Both the cell and fiber rings are visible at 3 minutes, and by 15 minutes a well developed fiber ring
is present in both the endoderm and ectoderm.
} \label{fig:WoundActinElliptical}
\end{figure}
}

\newcommand{\figureModelStressElliptical}{ % Fig S4
\begin{figure}[htbp]
\begin{minipage}{\textwidth}
\centering
    \subfloat{\label{fig:sig_rad_X}
        \begin{overpic}[width=0.40\textwidth]{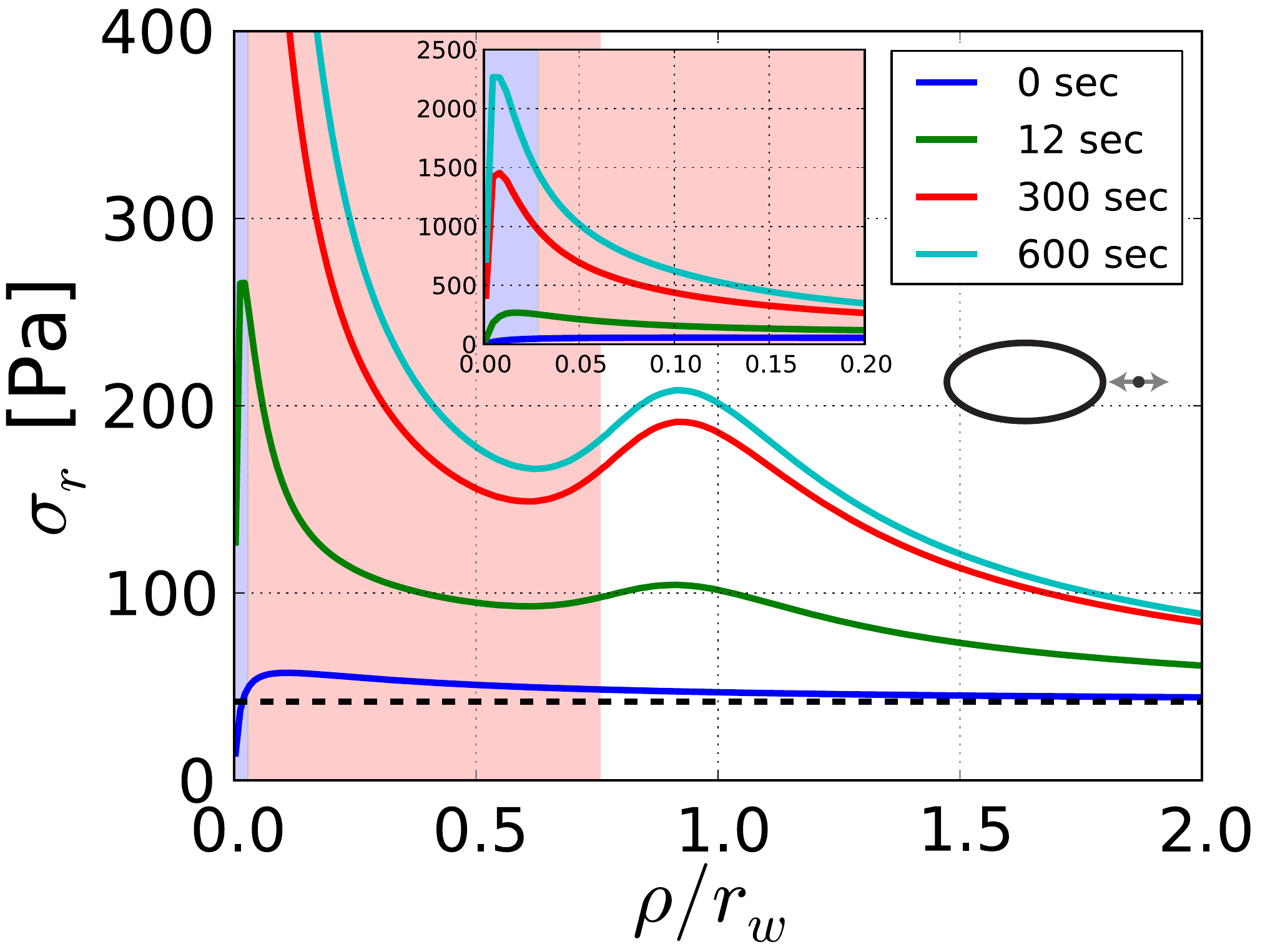}
        \put(1,3){\color{black}{\large (a)}} \end{overpic}}
    \subfloat{\label{fig:sig_circum_X}
        \begin{overpic}[width=0.40\textwidth]{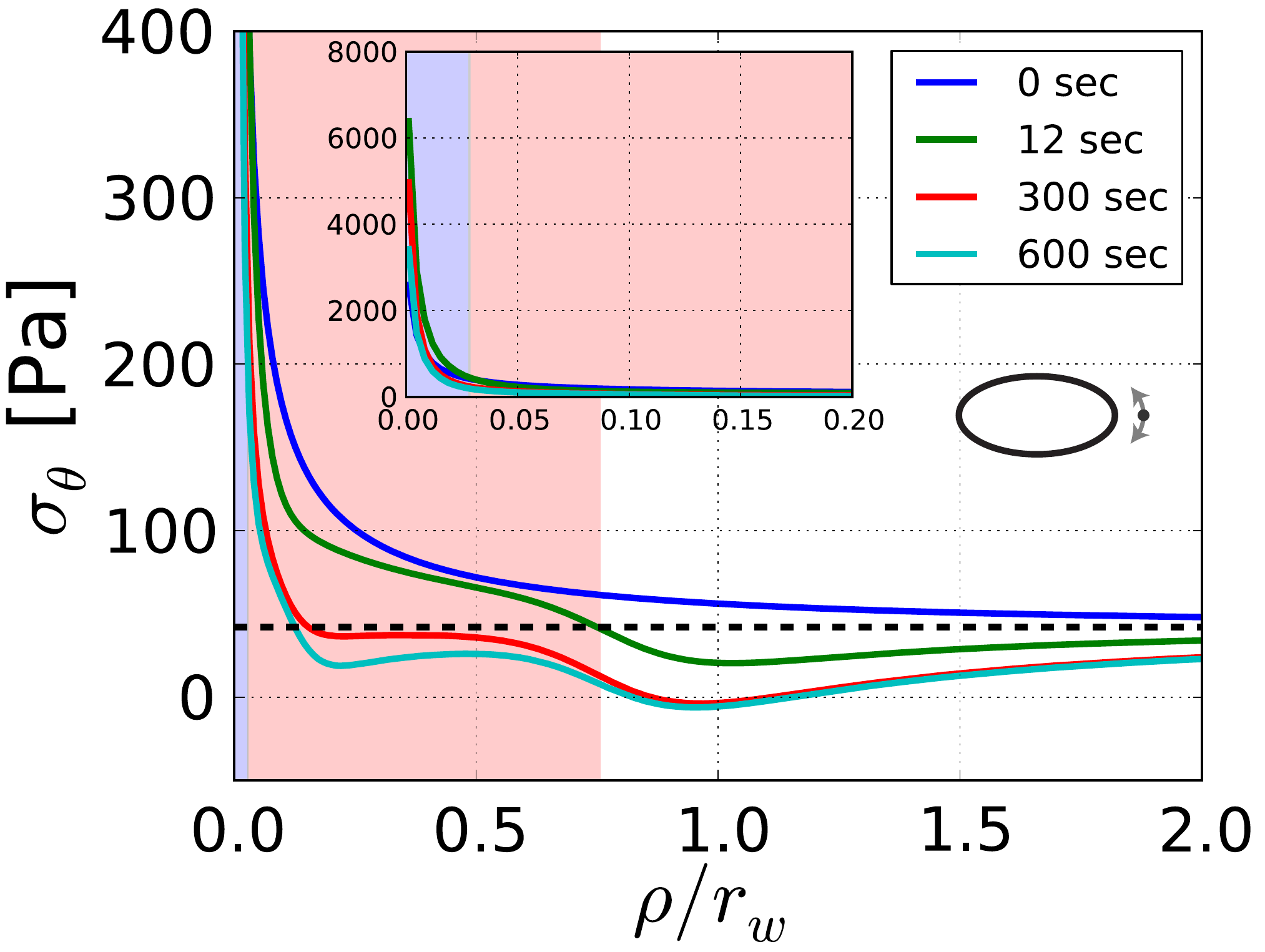}
        \put(1,3){\color{black}{\large (b)}} \end{overpic}}
\end{minipage}
\begin{minipage}{\textwidth}
\centering
    \subfloat{\label{fig:sig_rad_Y}
        \begin{overpic}[width=0.40\textwidth]{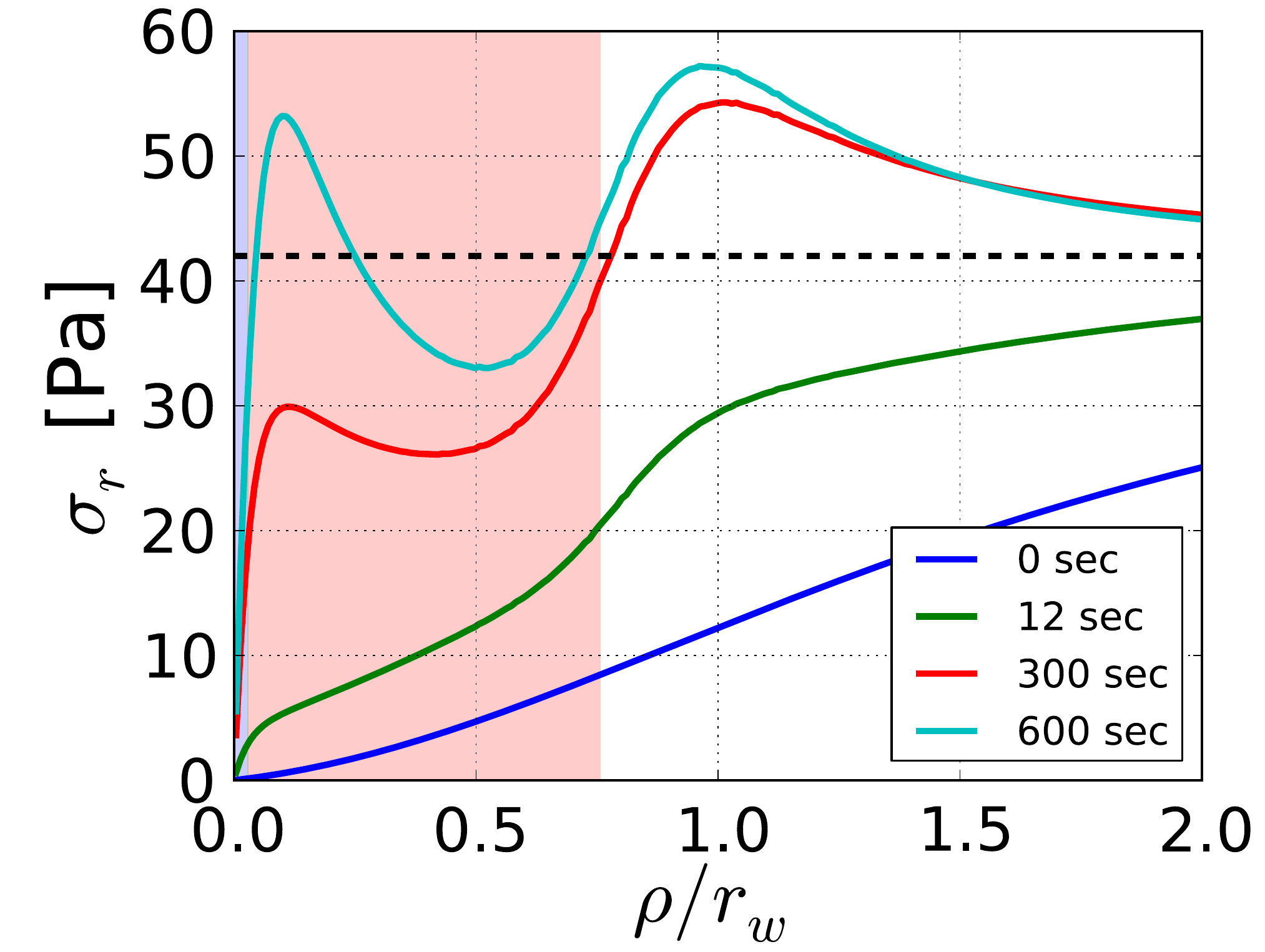}
        \put(1,3){\color{black}{\large (c)}} \end{overpic}}
    \subfloat{\label{fig:sig_circum_Y}
        \begin{overpic}[width=0.40\textwidth]{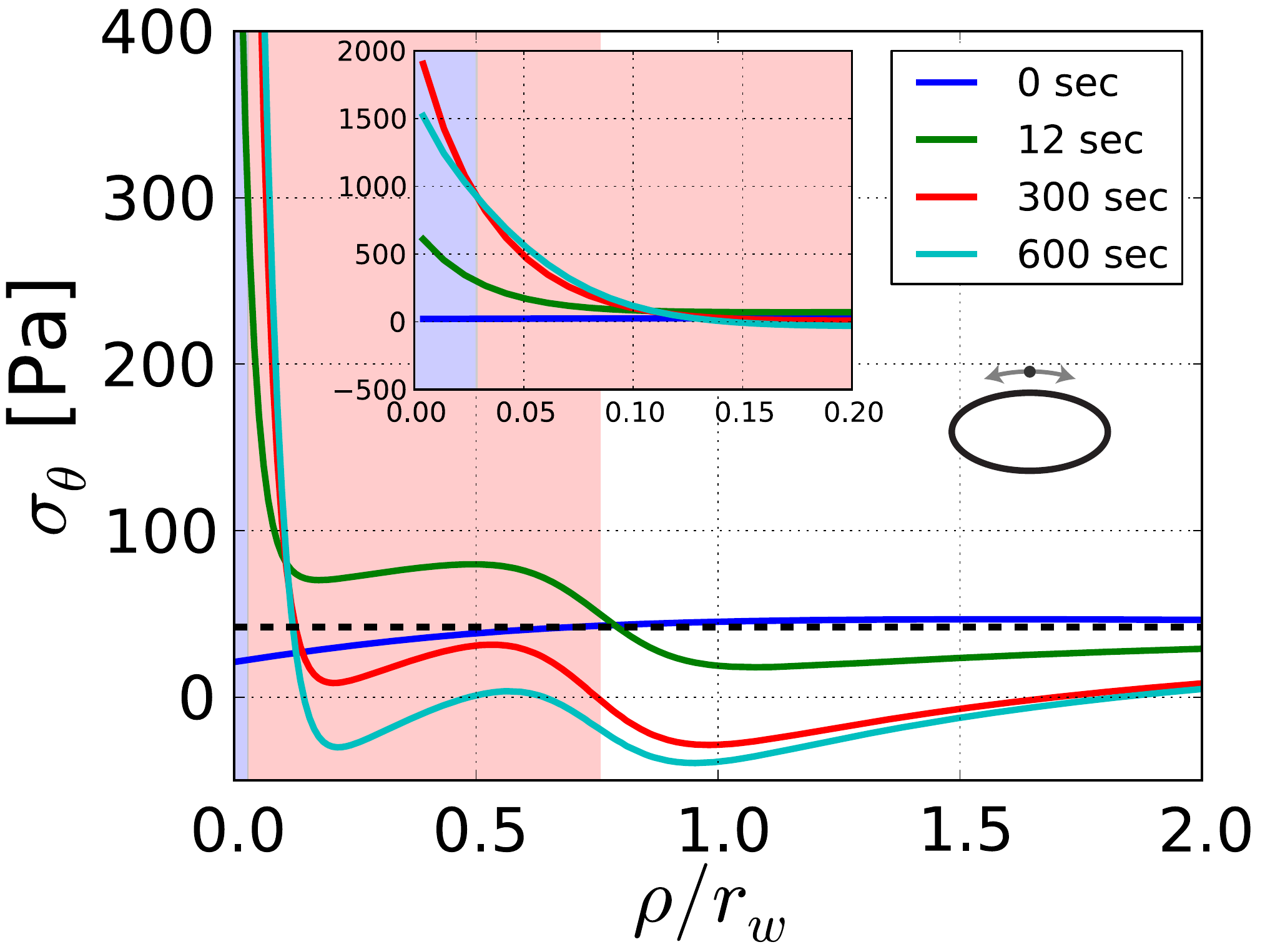}
        \put(1,3){\color{black}{\large (d)}} \end{overpic}}
\end{minipage}
\caption
{
Epithelial membrane stress in the vicinity of elliptical wound.  Icons indicate
position (along minor or major axis for elliptical wound), and stress component
(radial $\sigma_r$, circumferential $\sigma_\theta$).  Distance from wound edge
in the reference configuration is given in multiples of the circular wound
radius $r_w$, and the fiber and cell regions are indicated with blue and red
shading, respectively (see Fig.~\ref{fig:Gplots}).  Far from the wound both
stress components approach the far-field value $\sigma_B = 42$~Pa, indicated
by black dashed line.  
}
\label{fig:ModelStressElliptical}
\end{figure}
}

\newcommand{\figureLogMisesCs}{ % Fig S5
\begin{figure}[htbp]
\begin{minipage}{\textwidth}
\centering
    \subfloat{\label{fig:logmisesCs_0.1}
        \begin{overpic}[width=0.40\textwidth]{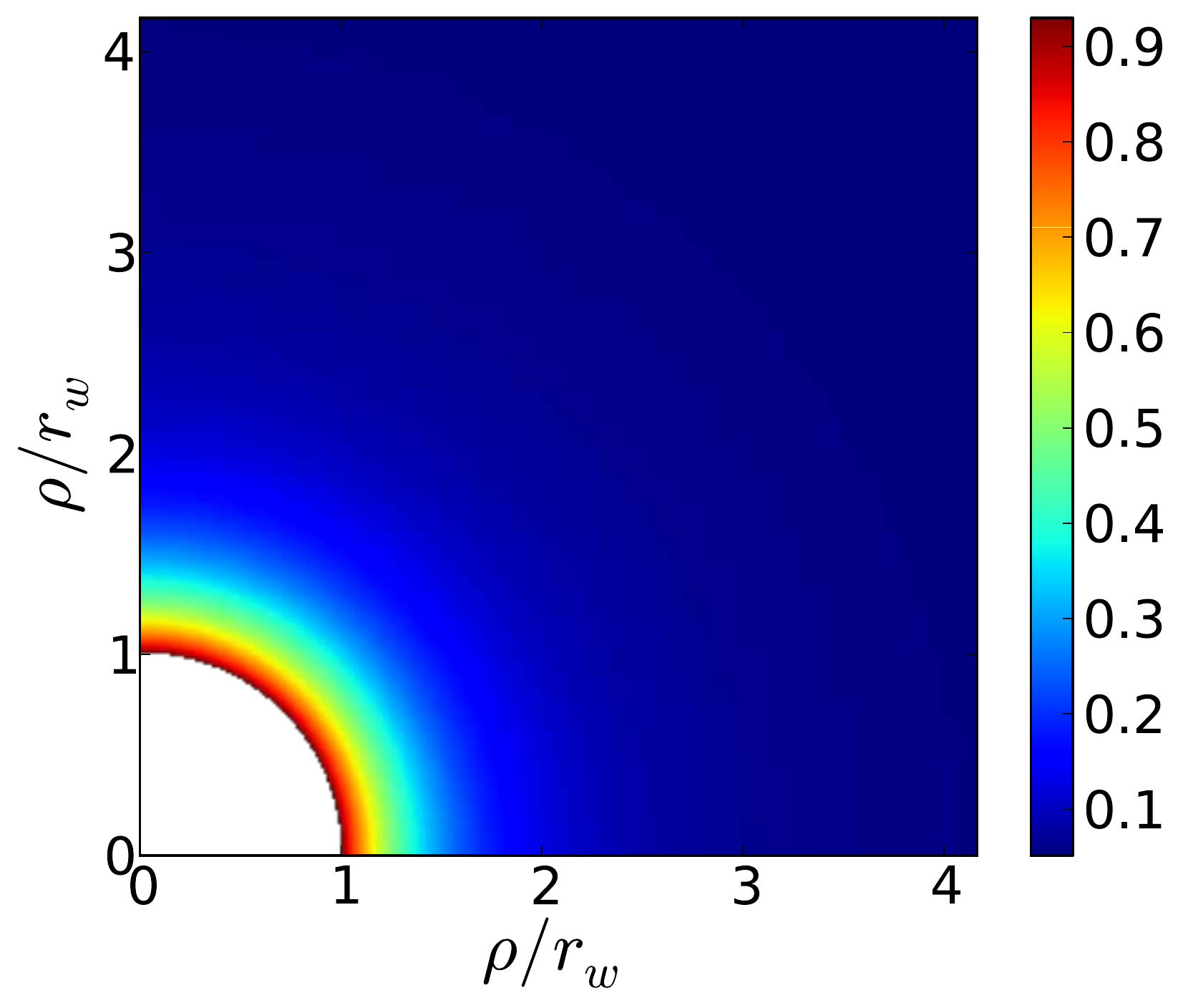}
        \put(1,3){\color{black}{\large (a)}} \end{overpic}}
    \subfloat{\label{fig:logmisesCs0.12}
        \begin{overpic}[width=0.40\textwidth]{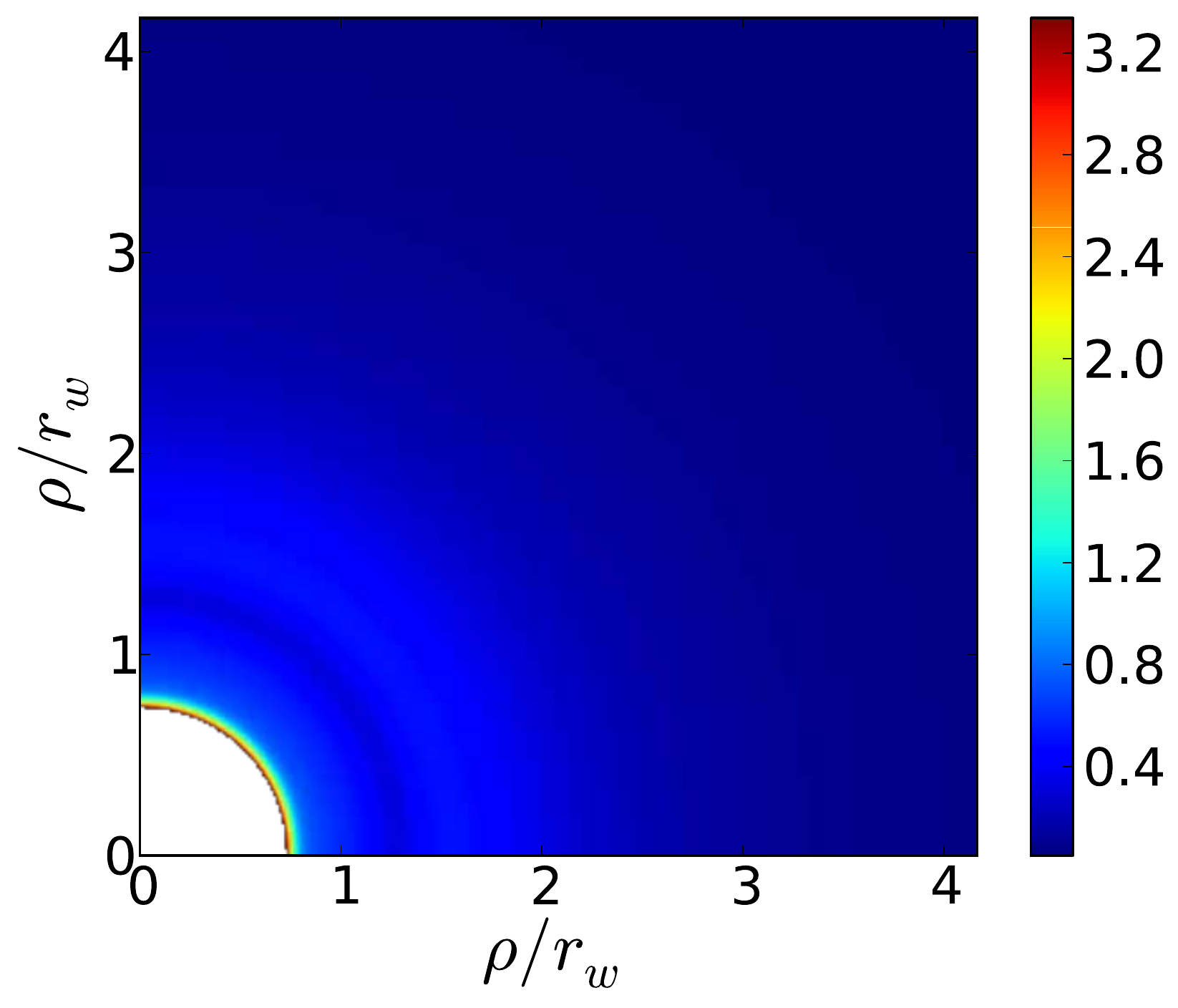}
        \put(1,3){\color{black}{\large (b)}} \end{overpic}}
\end{minipage}
\begin{minipage}{\textwidth}
\centering
    \subfloat{\label{fig:logmisesCs_0.6}
        \begin{overpic}[width=0.40\textwidth]{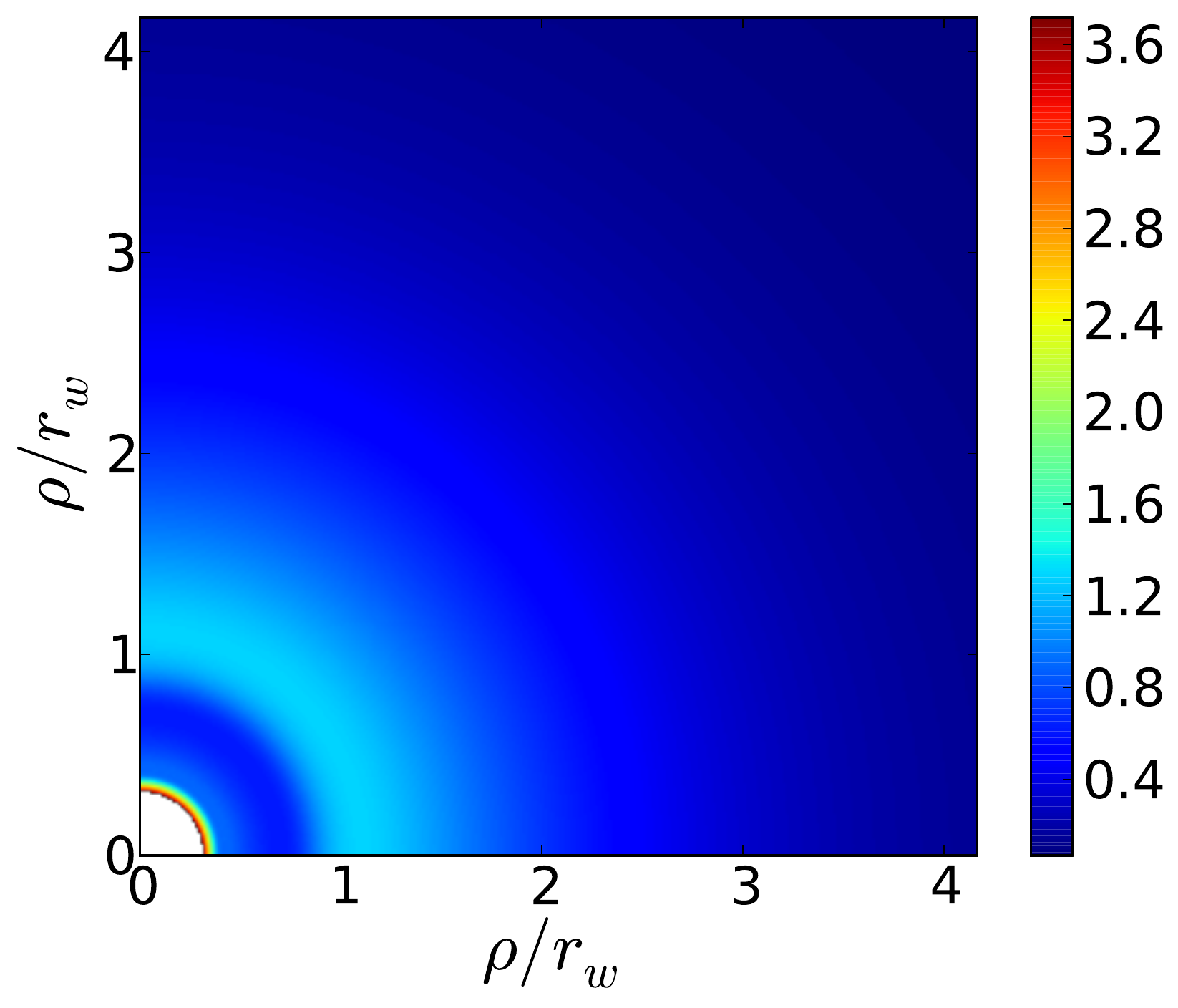}
        \put(1,3){\color{black}{\large (c)}} \end{overpic}}
    \subfloat{\label{fig:logmisesCs_1.1}
        \begin{overpic}[width=0.40\textwidth]{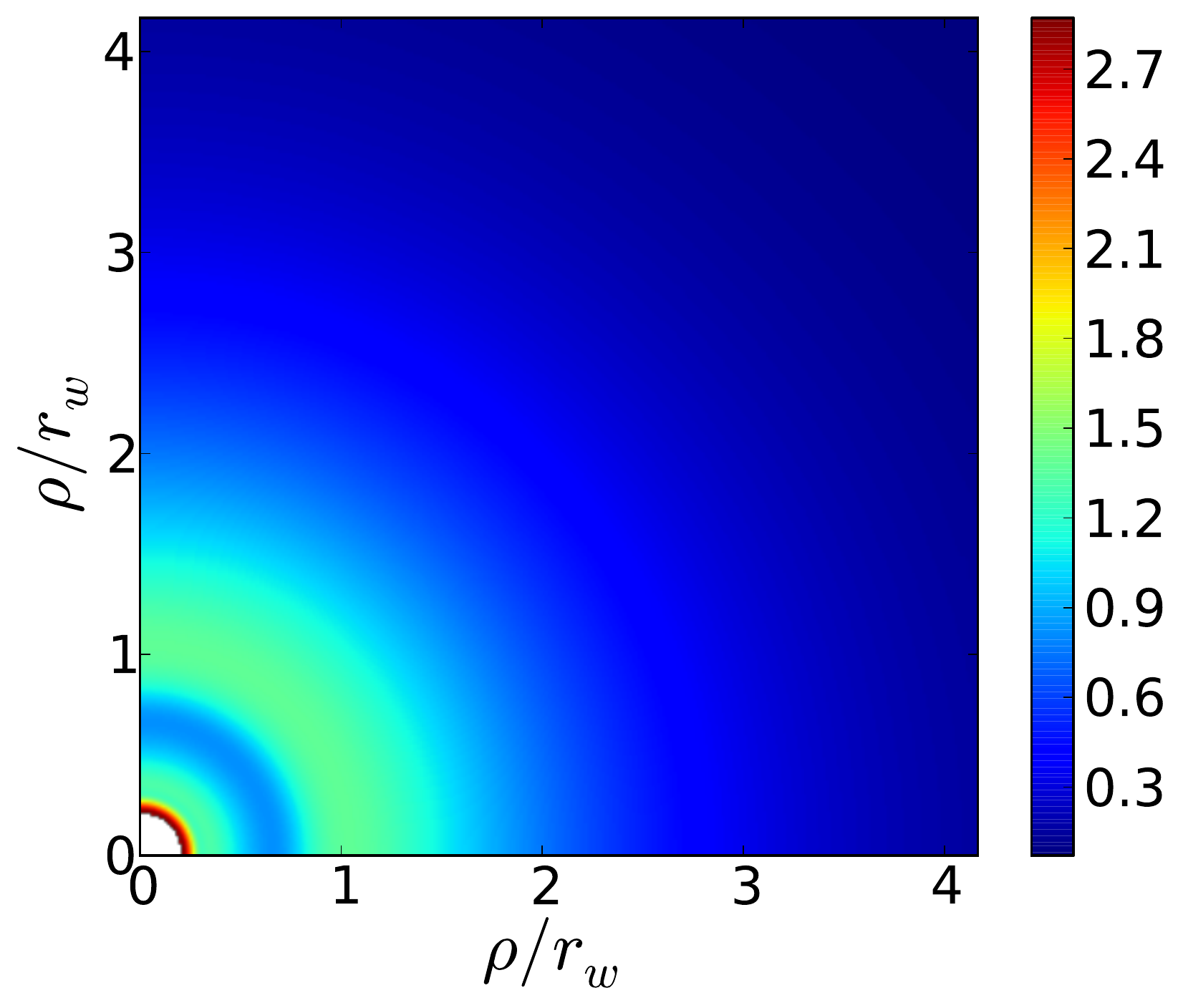}
        \put(1,3){\color{black}{\large (d)}} \end{overpic}}
\end{minipage}
\caption
{
Stress around closing circular wound.  Logarithm of Von Mises stress plotted for t=0, 12, 300, and 600 seconds
in panels a-d, respectively.
}
\label{fig:logmisesCs}
\end{figure}
}

\newcommand{\figureDispCs}{ % Fig S6
\begin{figure}[htbp]
\begin{minipage}{\textwidth}
\centering
    \subfloat{\label{fig:dispCs0.12}
        \begin{overpic}[width=0.30\textwidth]{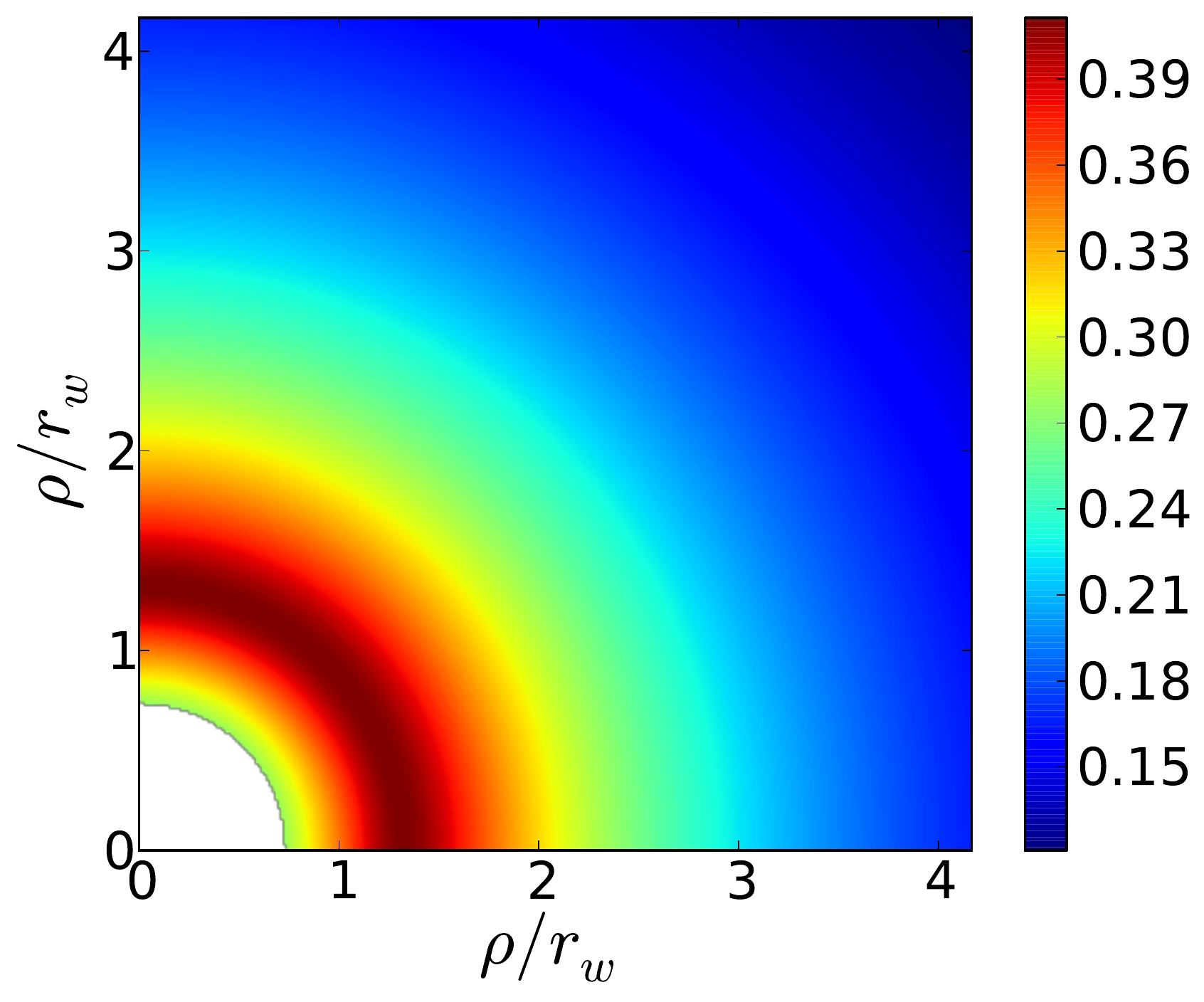}
        \put(1,3){\color{black}{\large (a)}} \end{overpic}}
    \subfloat{\label{fig:dispCs_0.6}
        \begin{overpic}[width=0.30\textwidth]{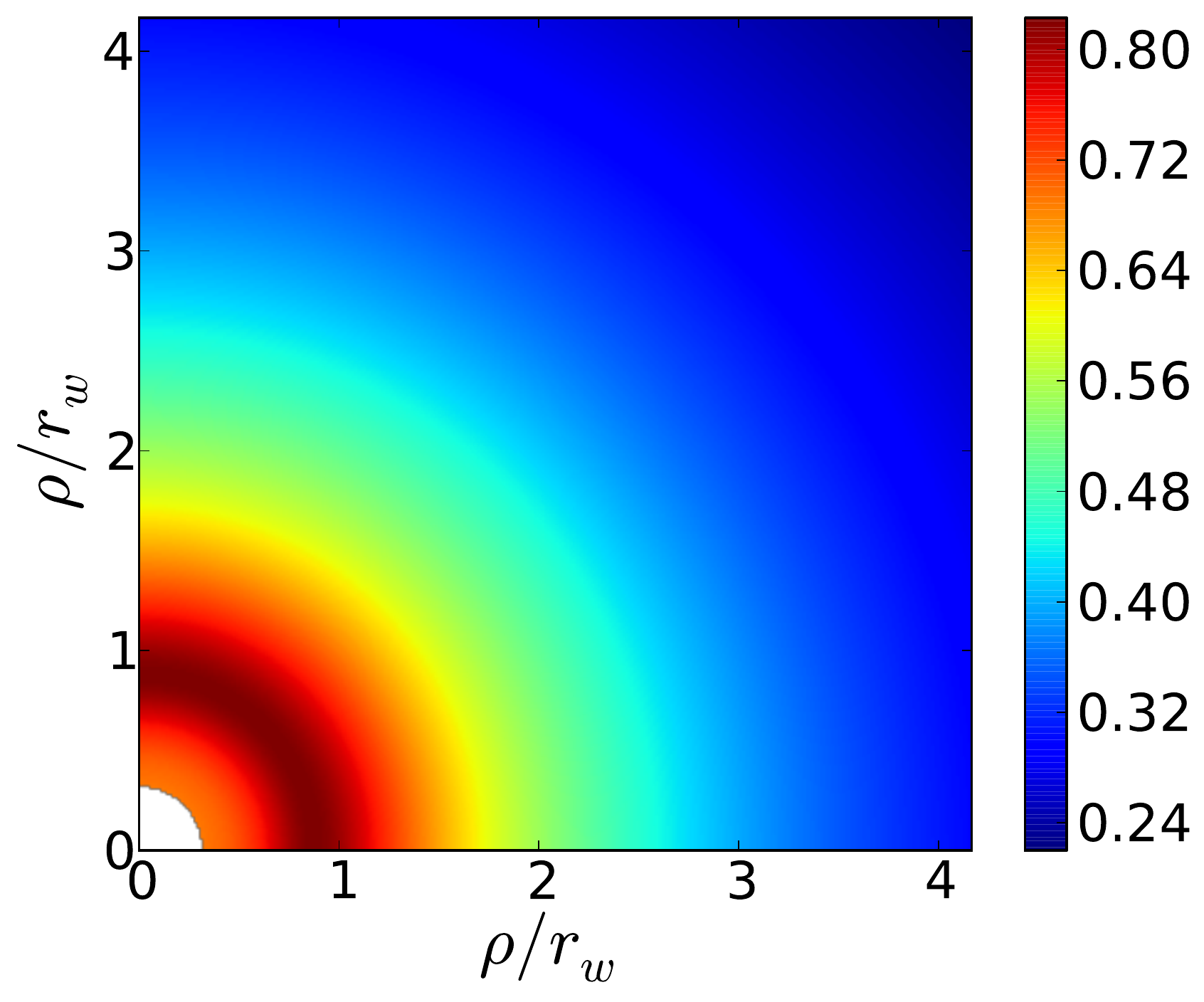}
        \put(1,3){\color{black}{\large (b)}} \end{overpic}}
    \subfloat{\label{fig:dispCs_1.1}
        \begin{overpic}[width=0.30\textwidth]{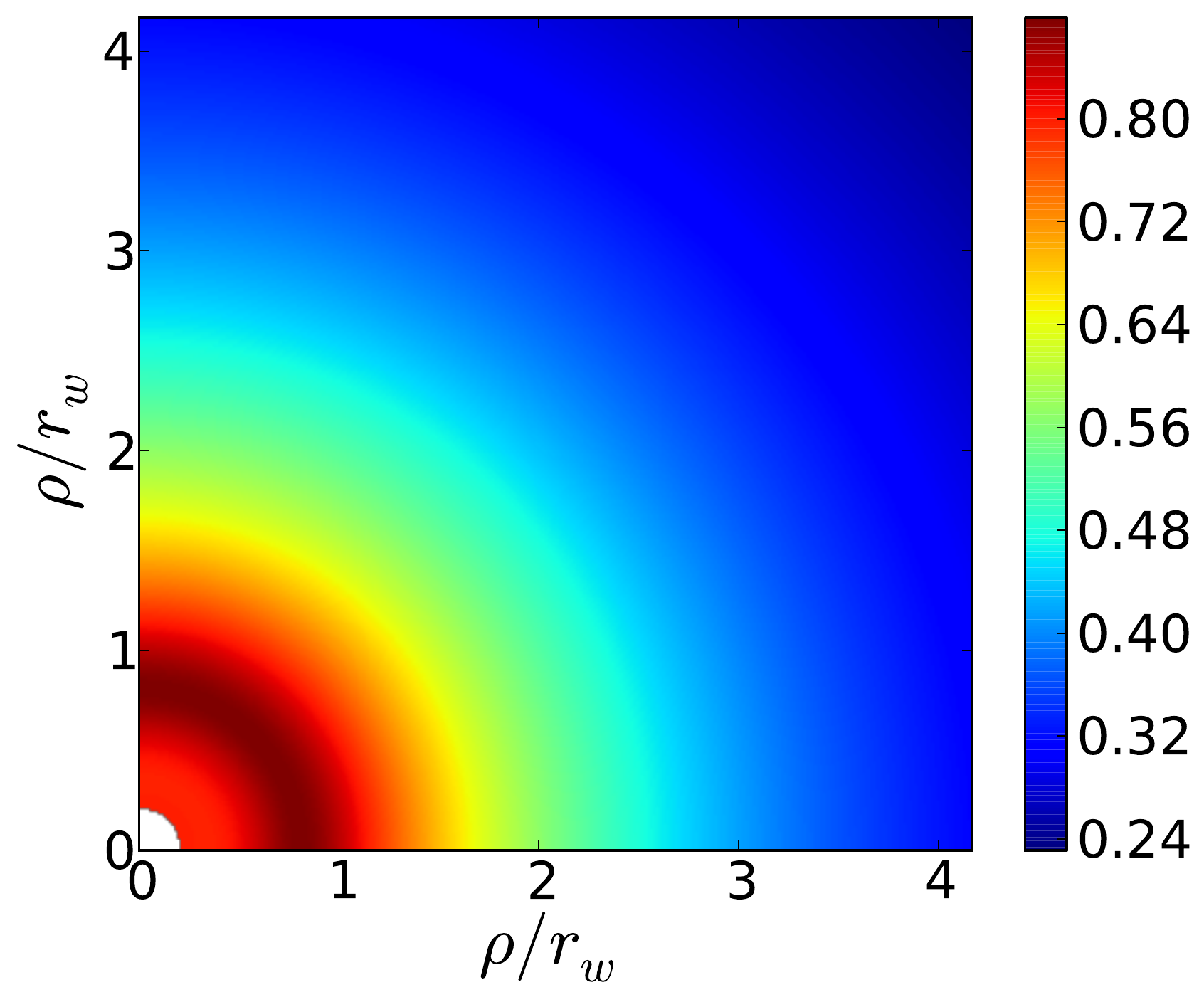}
        \put(1,3){\color{black}{\large (c)}} \end{overpic}}
\end{minipage}
\caption
{
Displacement around closing circular wound relative to the 
reference (t=0) configuration
for t=12, 300, and 600 seconds in panels a-c, respectively.
Displacement and distance are relative to circular wound radius $r_w$.
}
\label{fig:dispCs}
\end{figure}
}

\newcommand{\figureLogMisesEs}{ % Fig S7
\begin{figure}[htbp]
\begin{minipage}{\textwidth}
\centering
    \subfloat{\label{fig:logmisesEs_0.1}
        \begin{overpic}[width=0.40\textwidth]{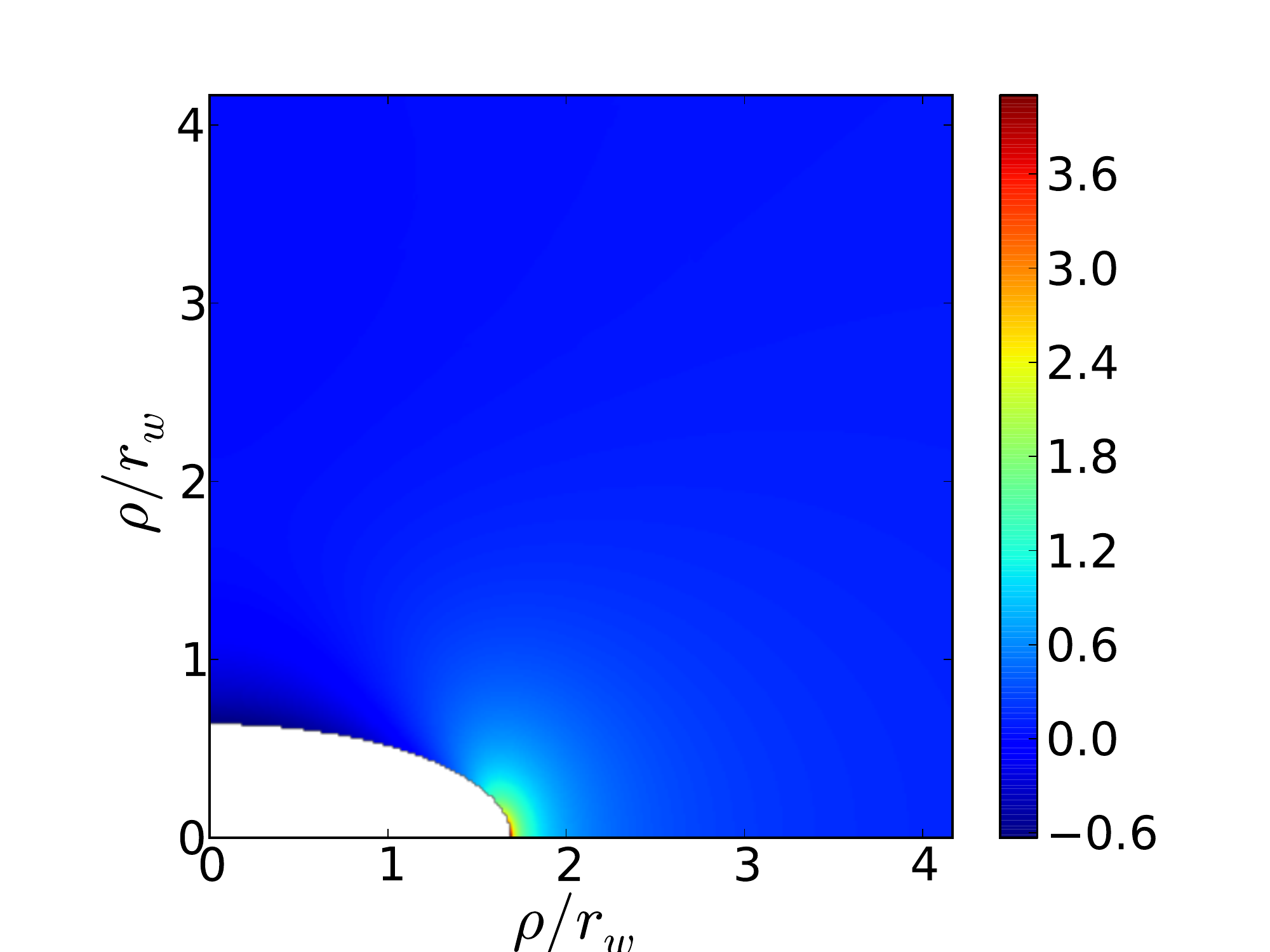}
        \put(1,3){\color{black}{\large (a)}} \end{overpic}}
    \subfloat{\label{fig:logmisesEs0.12}
        \begin{overpic}[width=0.40\textwidth]{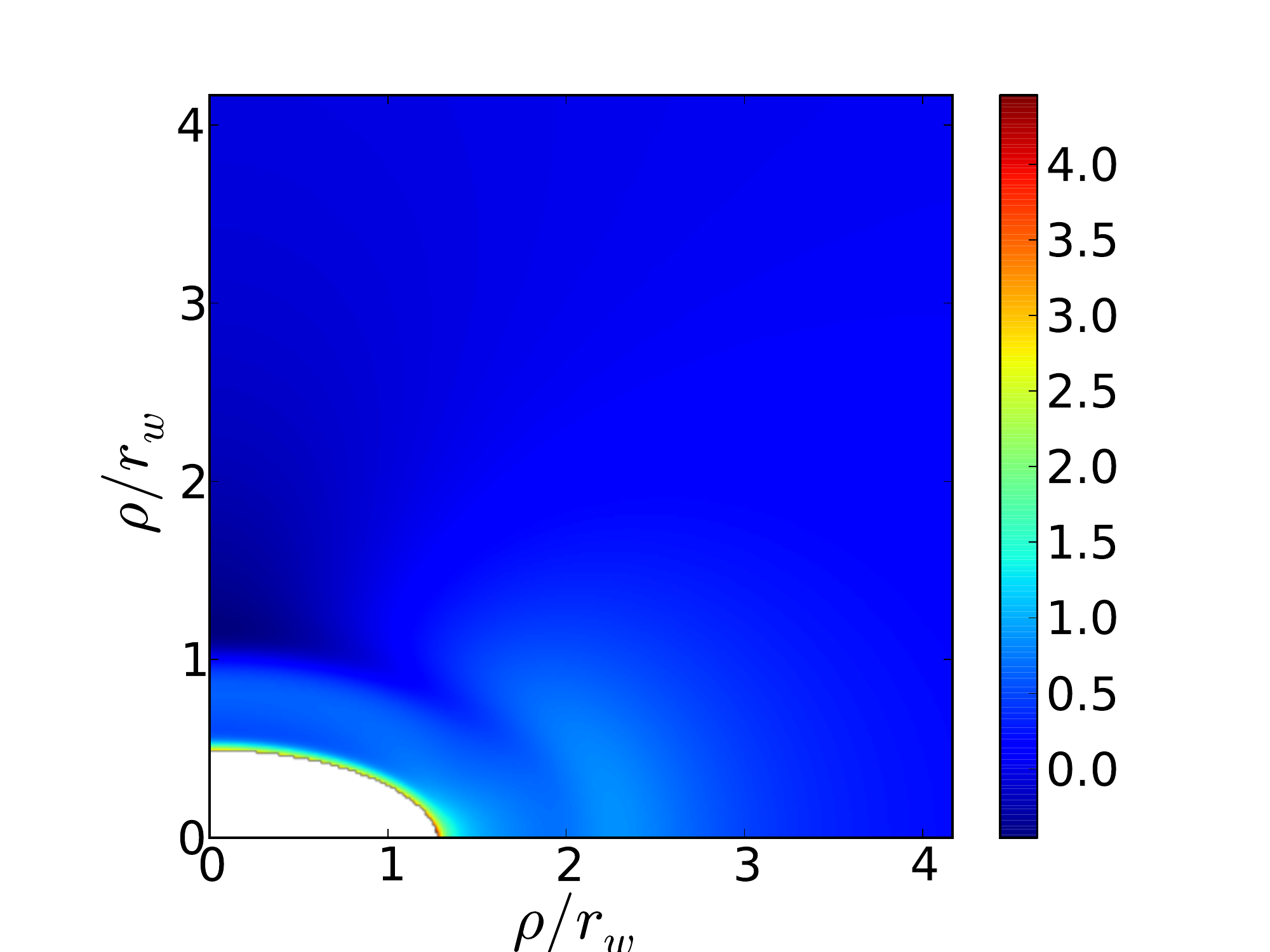}
        \put(1,3){\color{black}{\large (b)}} \end{overpic}}
\end{minipage}
\begin{minipage}{\textwidth}
\centering
    \subfloat{\label{fig:logmisesEs_0.6}
        \begin{overpic}[width=0.40\textwidth]{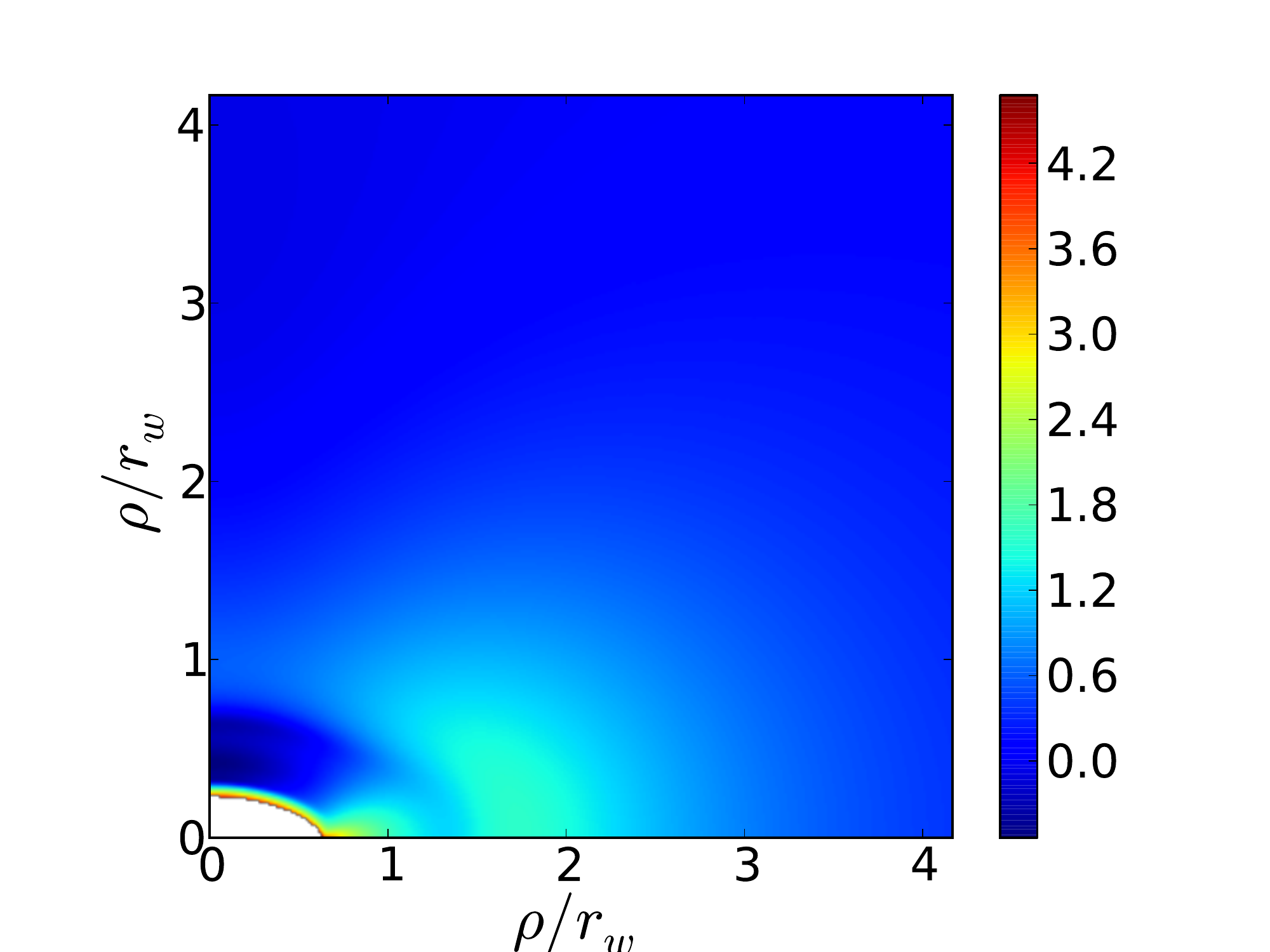}
        \put(1,3){\color{black}{\large (c)}} \end{overpic}}
    \subfloat{\label{fig:logmisesEs_1.1}
        \begin{overpic}[width=0.40\textwidth]{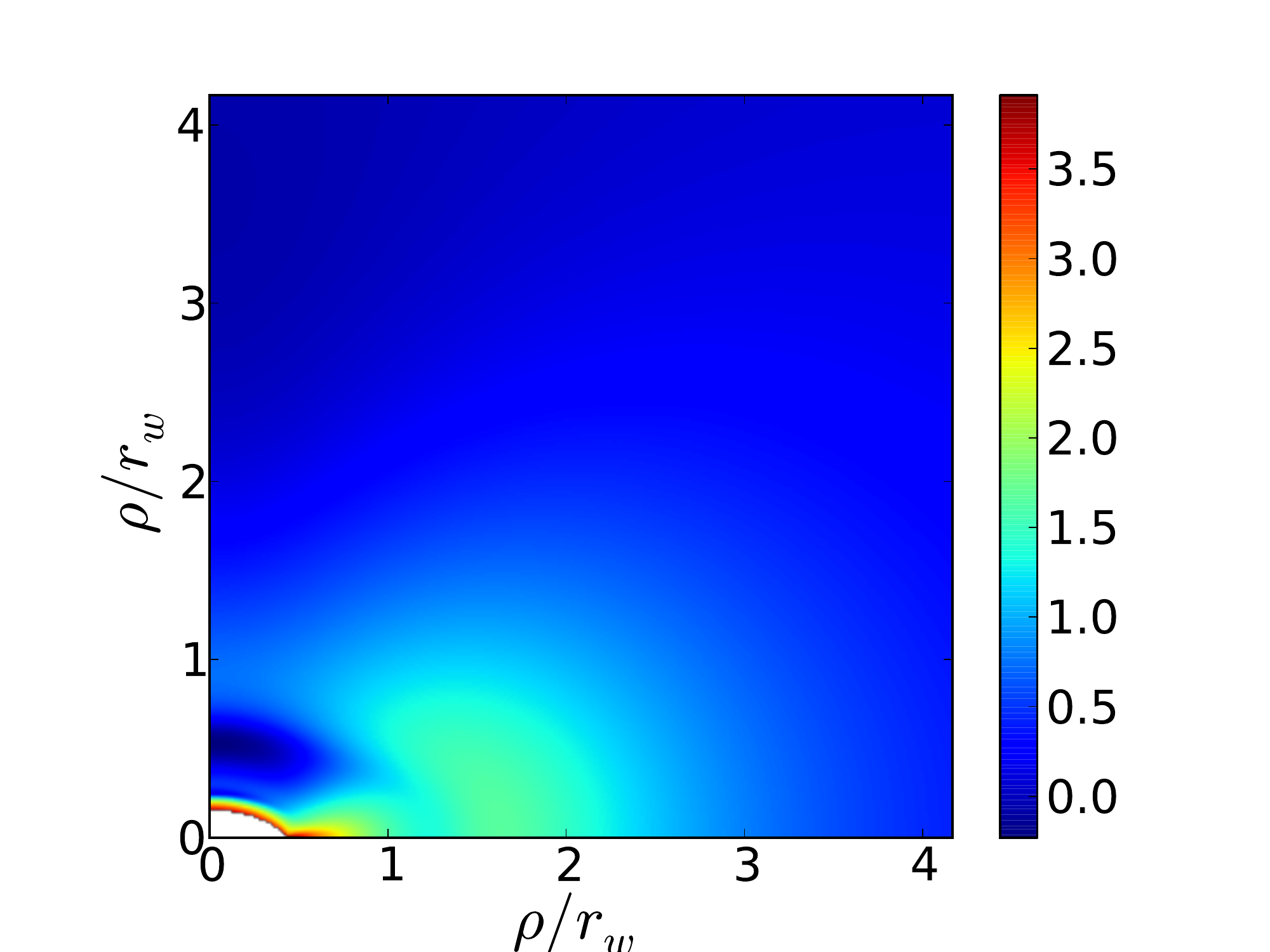}
        \put(1,3){\color{black}{\large (d)}} \end{overpic}}
\end{minipage}
\caption
{
Stress around closing elliptical wound.  Logarithm of Von Mises stress plotted for t=0, 12, 300, and 600 seconds
in panels a-d, respectively.
}
\label{fig:logmisesEs}
\end{figure}
}

\newcommand{\figureDispEs}{ % Fig S8
\begin{figure}[htbp]
\begin{minipage}{\textwidth}
\centering
    \subfloat{\label{fig:dispEs0.12}
        \begin{overpic}[width=0.30\textwidth]{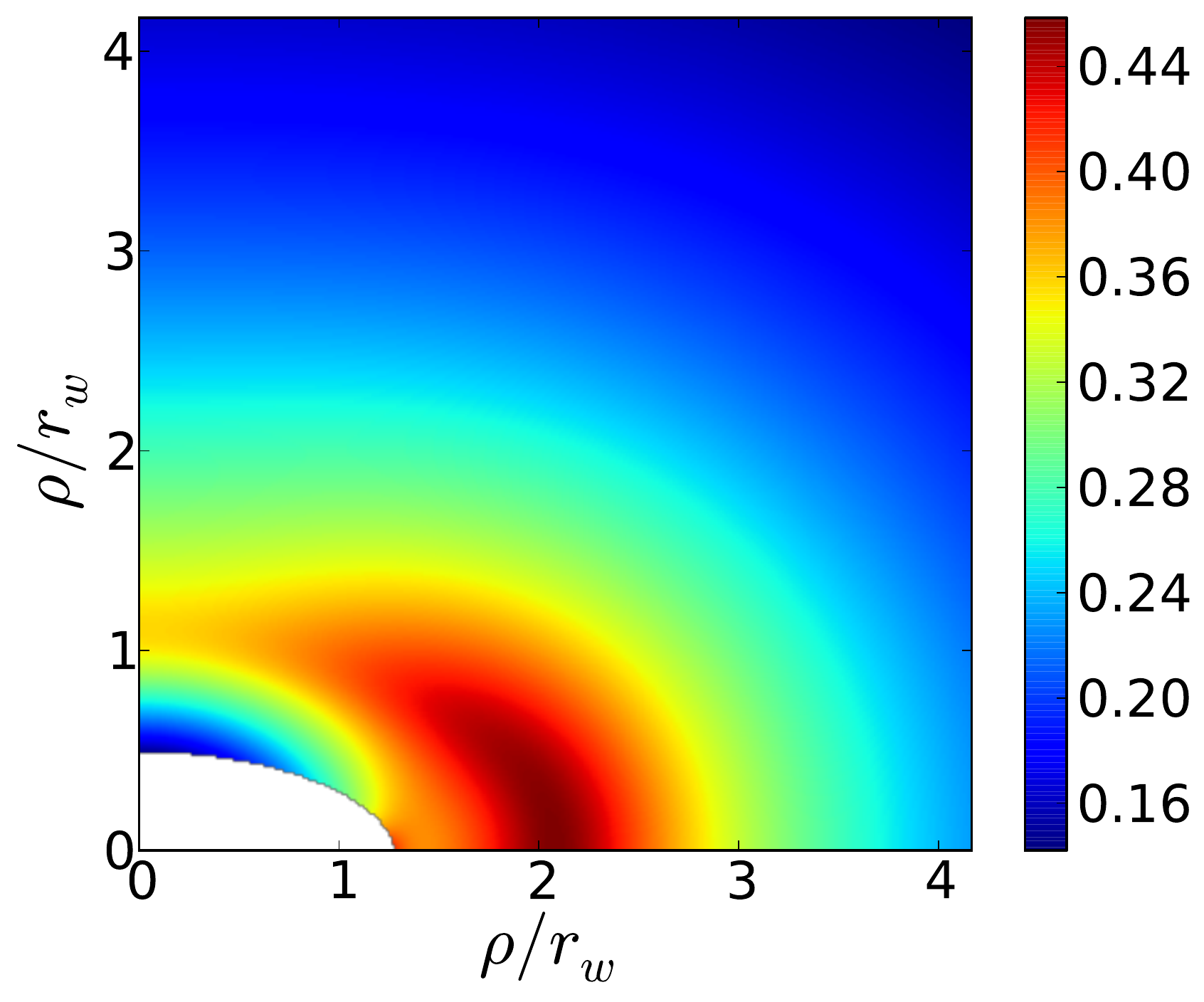}
        \put(1,3){\color{black}{\large (a)}} \end{overpic}}
    \subfloat{\label{fig:dispEs_0.6}
        \begin{overpic}[width=0.30\textwidth]{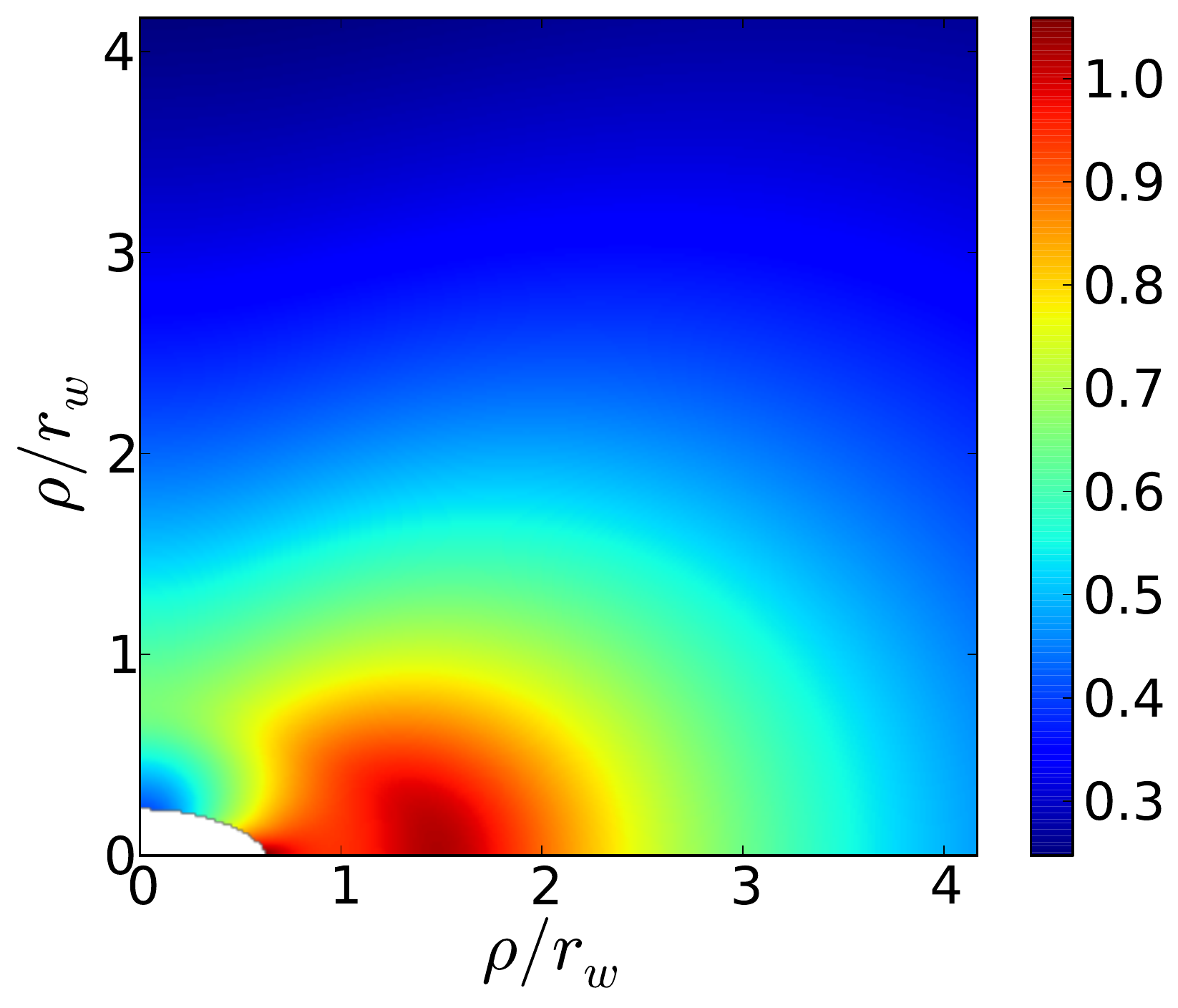}
        \put(1,3){\color{black}{\large (b)}} \end{overpic}}
    \subfloat{\label{fig:dispEs_1.1}
        \begin{overpic}[width=0.30\textwidth]{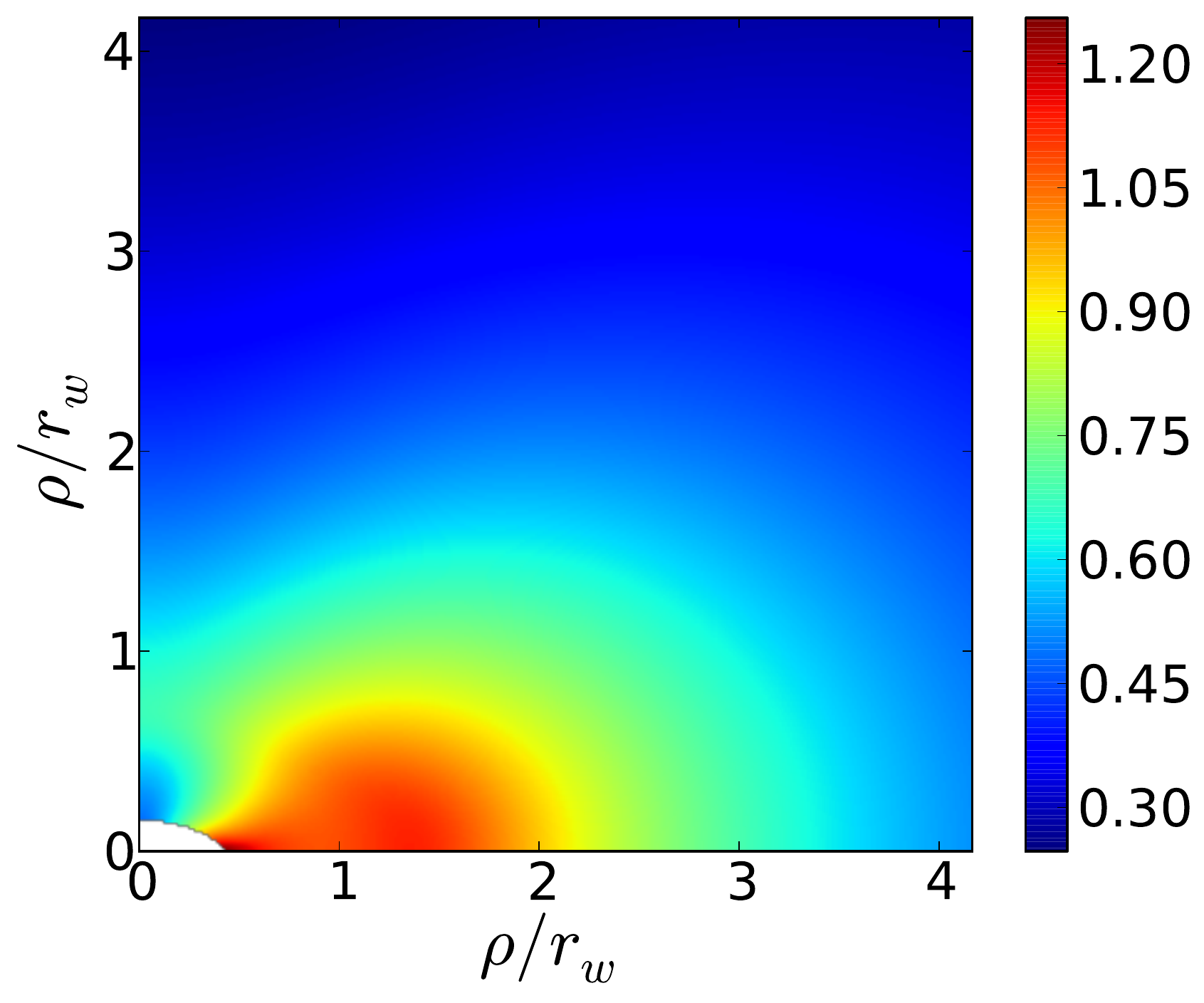}
        \put(1,3){\color{black}{\large (c)}} \end{overpic}}
\end{minipage}
\caption
{
Displacement around closing elliptical wound relative to the 
reference (t=0) configuration
for t=12, 300, and 600 seconds in panels a-c, respectively.
Displacement and distance are relative to circular wound radius $r_w$.
}
\label{fig:dispEs}
\end{figure}
}

\newcommand{\figureRadiusCompare}{  % Fig S9
\begin{figure}[htbp]
\centering
    \subfloat{\label{fig:ExptRadiusCompare}
        \begin{overpic}[width=0.40\textwidth]{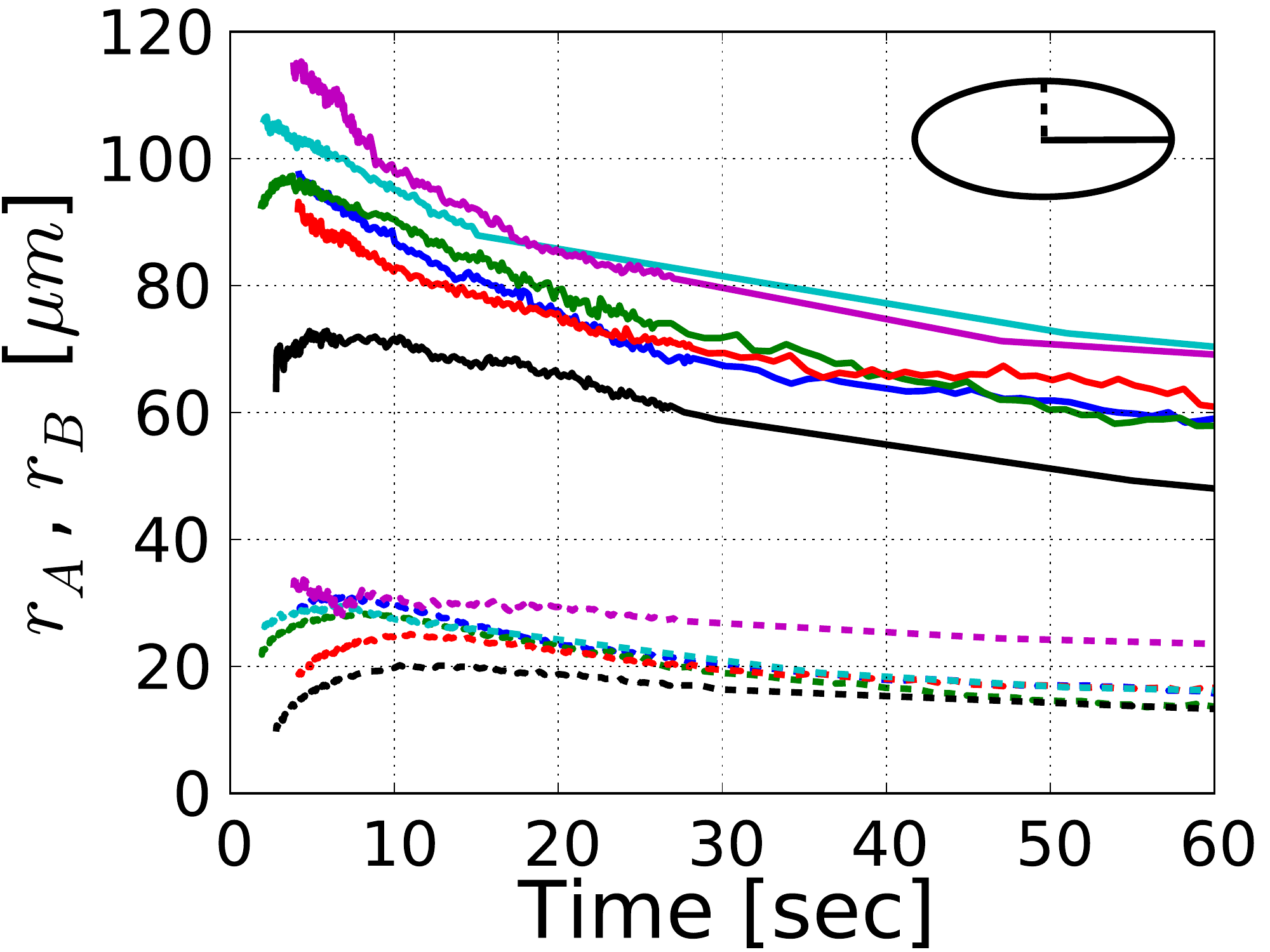}
        \put(0,0){\color{black}{\large (a)}} \end{overpic}}
    \subfloat{\label{fig:ModelRadiusCompare}
        \begin{overpic}[width=0.40\textwidth]{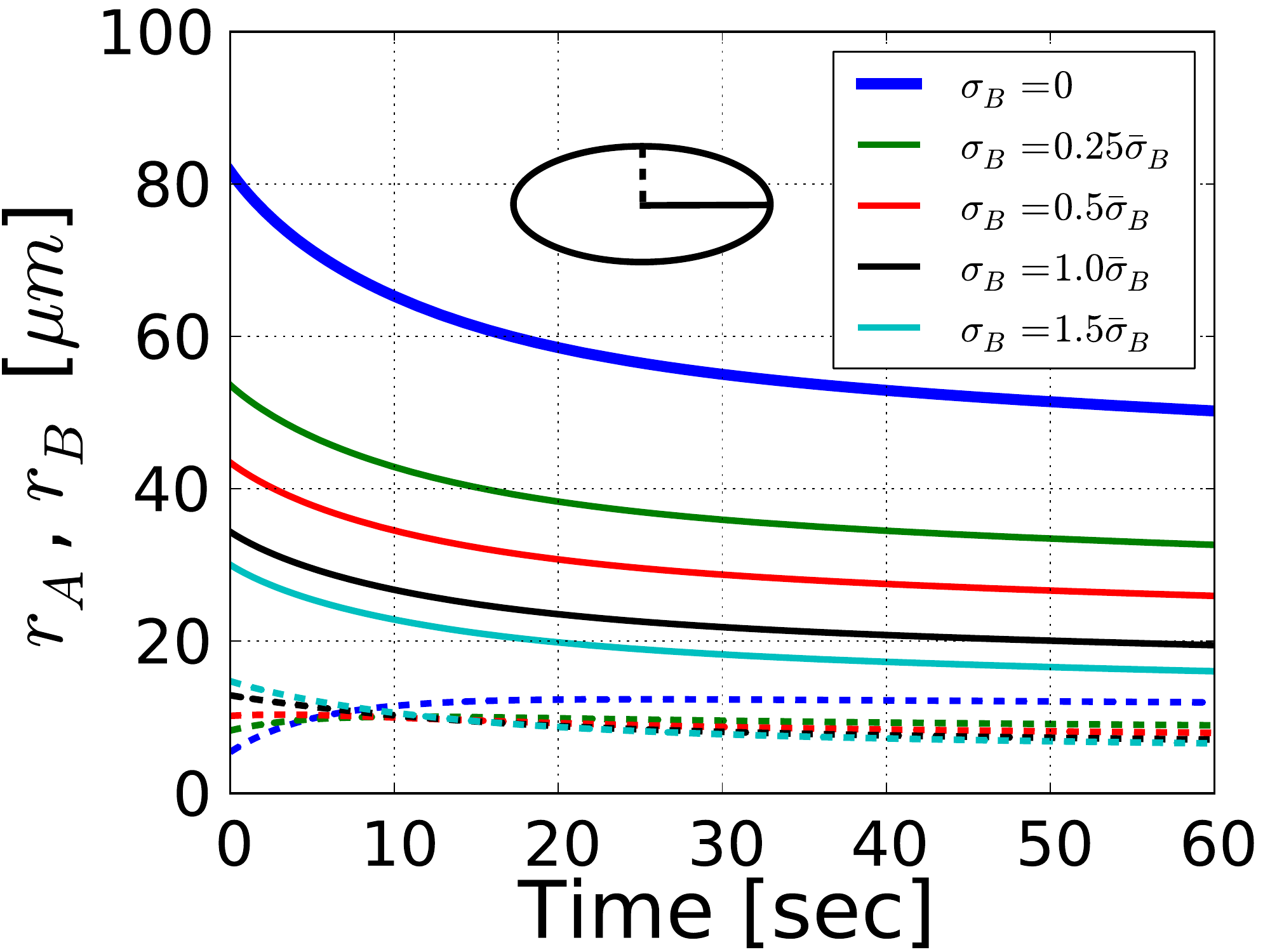}
        \put(0,0){\color{black}{\large (b)}} \end{overpic}}
\caption
{
Major and minor radii of elliptical wounds in initial stages of closing.  Solid
and dashed curves represent major and minor radii, respectively, as indicated
by ellipse icons.
\psubref{fig:ExptRadiusCompare} Radii of experimental wounds.
Colors represent individual experiments (data from Fig.~\ref{fig:EllipseAreaAR}).  
\psubref{fig:ModelRadiusCompare} Radii of model wounds, with colors indicating
boundary tension $\sigma_B$, data from Fig.~\ref{fig:ModelFitSigB}.
}
\label{fig:RadiusCompare}
\end{figure}
}

\newcommand{\figureModelFitR}{ % Fig S10
\begin{figure}[htbp]
\begin{minipage}{\textwidth}
\centering
  \subfloat{\label{fig:rab_area}
    \begin{overpic}[width=0.40\textwidth]{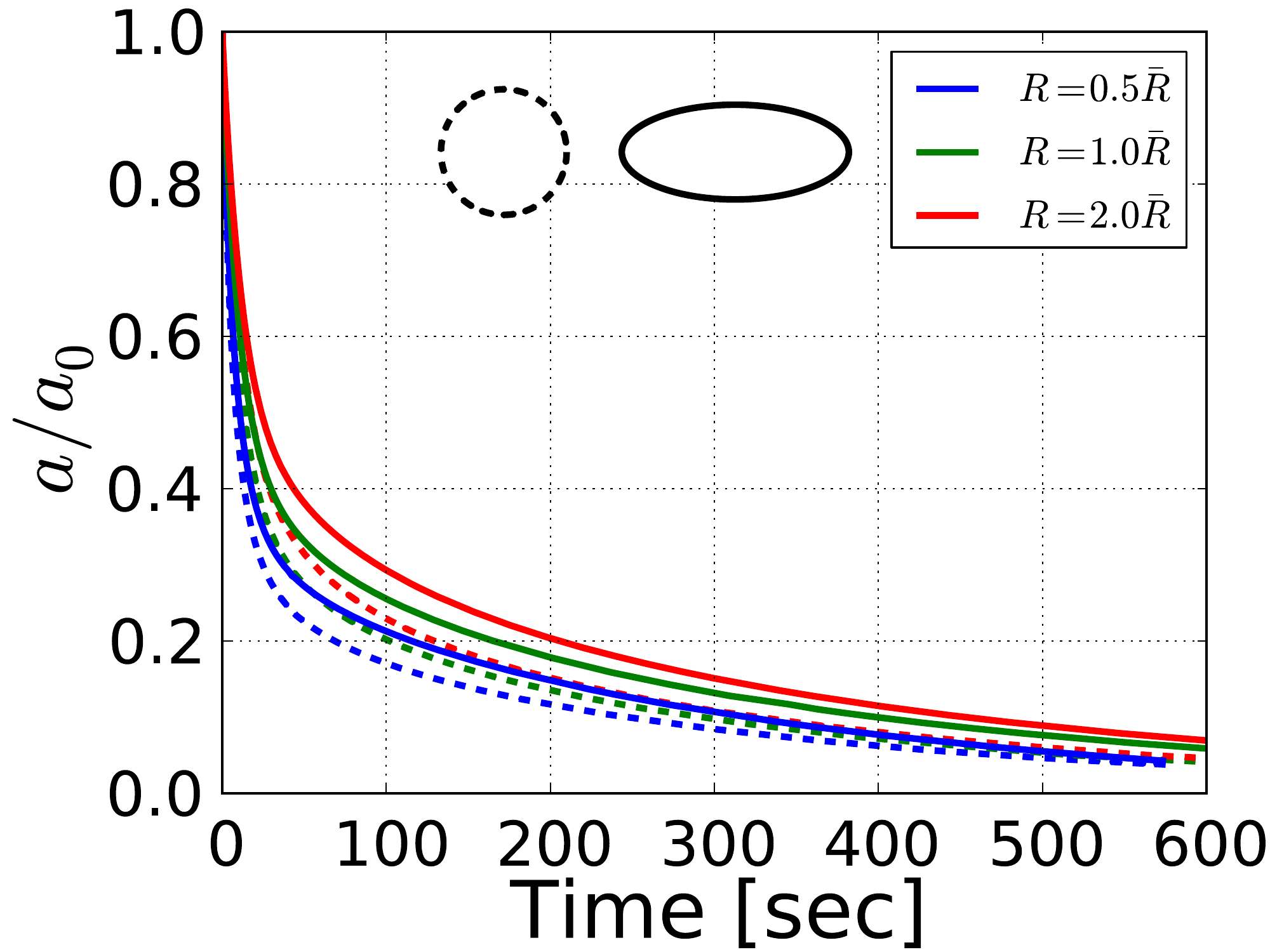}
        \put(0,0){\color{black}{\large (a)}} \end{overpic}}
  \subfloat{\label{fig:rab_AR}
    \begin{overpic}[width=0.40\textwidth]{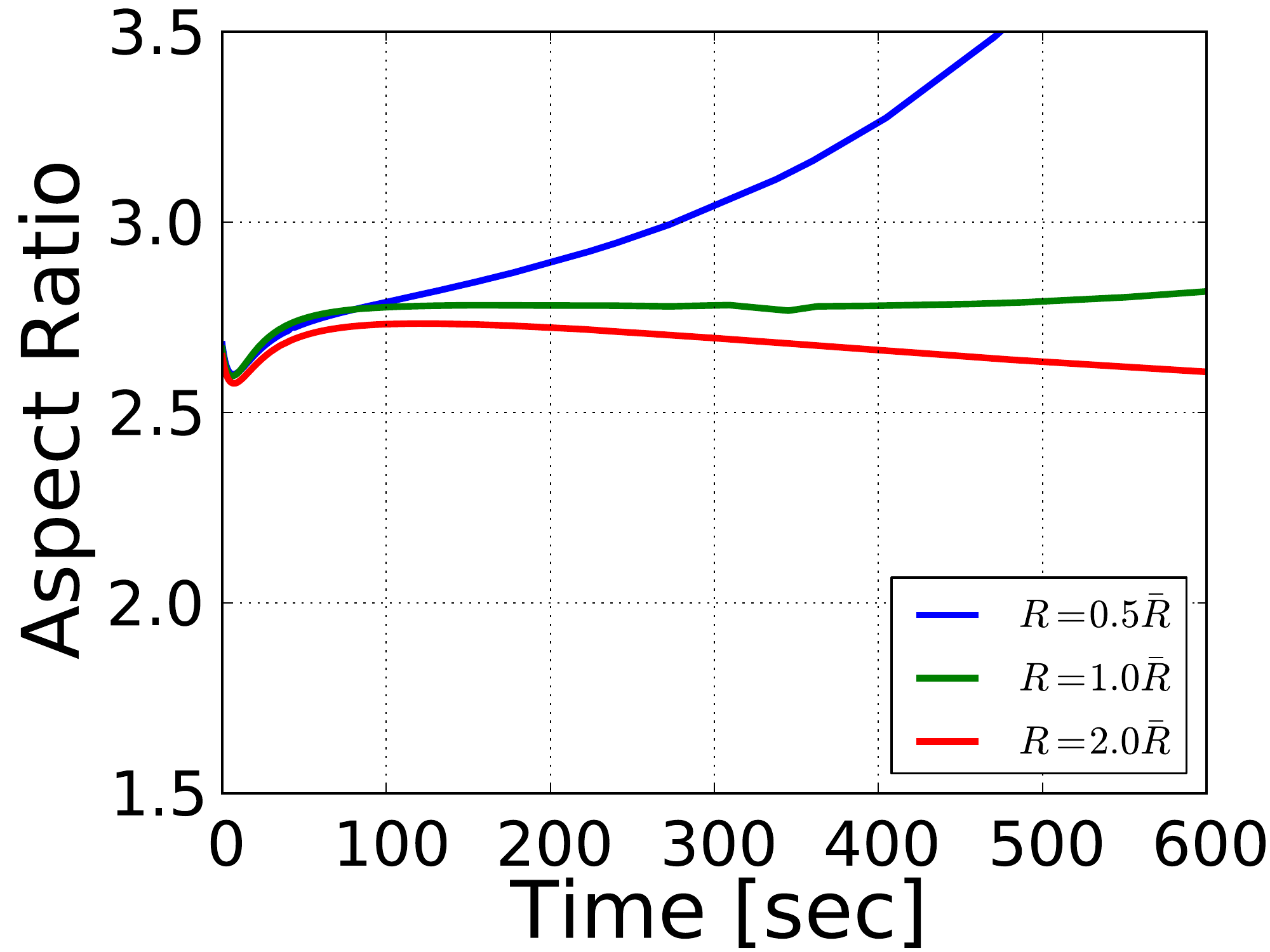}
        \put(0,0){\color{black}{\large (b)}} \end{overpic}}
\end{minipage}
\caption
{
Wound area and aspect ratio (AR) with varying initial area (same $\sigma_B$). Dashed
and solid curves, correspond to circular and elliptical wounds, respectively, as
indicated by icons.  Undeformed wound radii ($R_w$ for circular wound, $R_A$
and $R_B$ for elliptical) are varied relative to nominal values
(Table~\ref{tab:Geometry}).  Decreasing wound size increases effective width of
fiber and cell rings, and results in slightly faster wound closing.
}
\label{fig:ModelFitR}
\end{figure}
}

\newcommand{\figureModelFitAR}{  % Fig S11
\begin{figure}[htbp]
\begin{minipage}{\textwidth}
\centering
  \subfloat{\label{fig:rb_area}
    \begin{overpic}[width=0.40\textwidth]{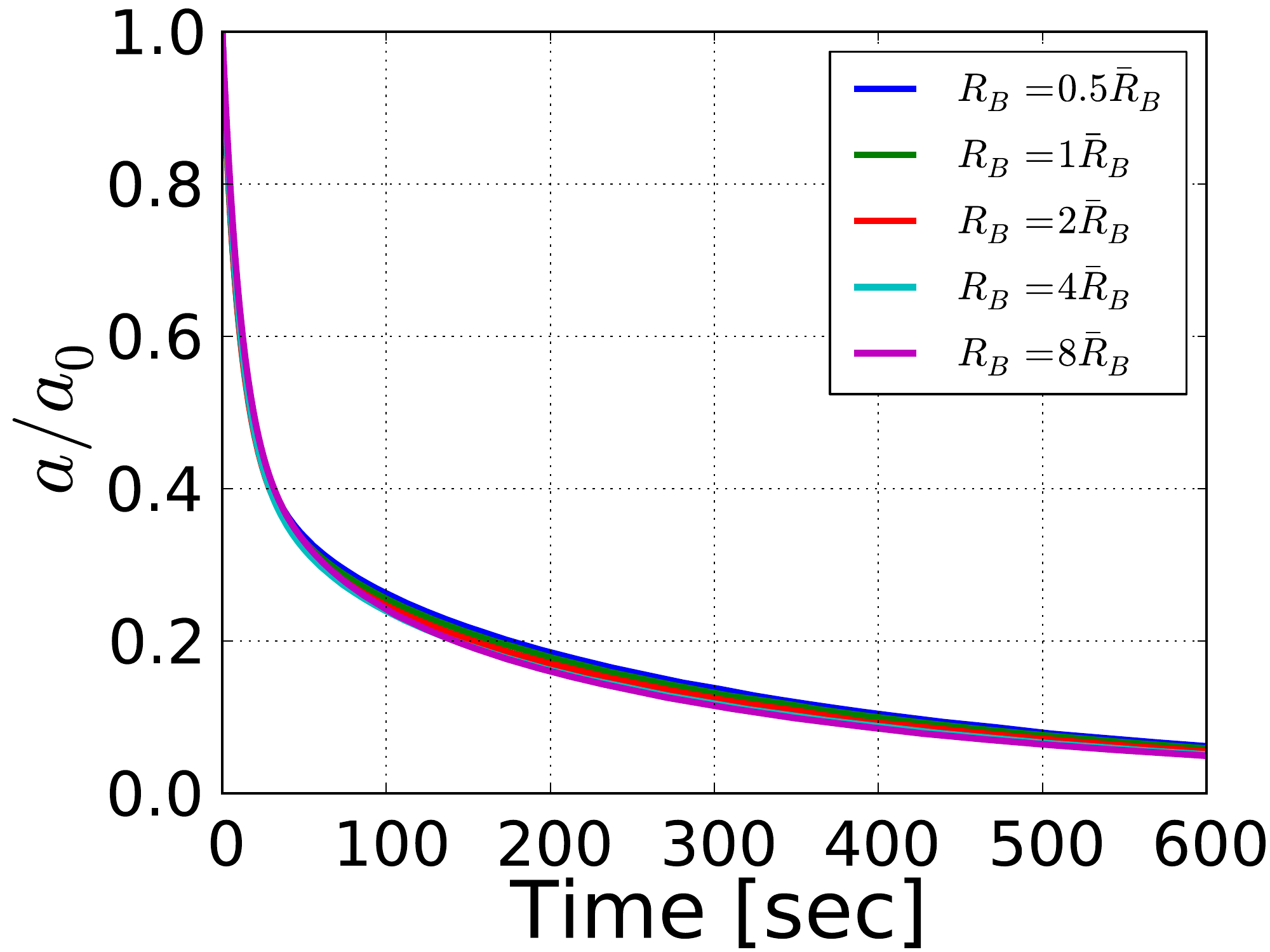}
        \put(0,0){\color{black}{\large (a)}} \end{overpic}}
  \subfloat{\label{fig:rb_AR}
    \begin{overpic}[width=0.40\textwidth]{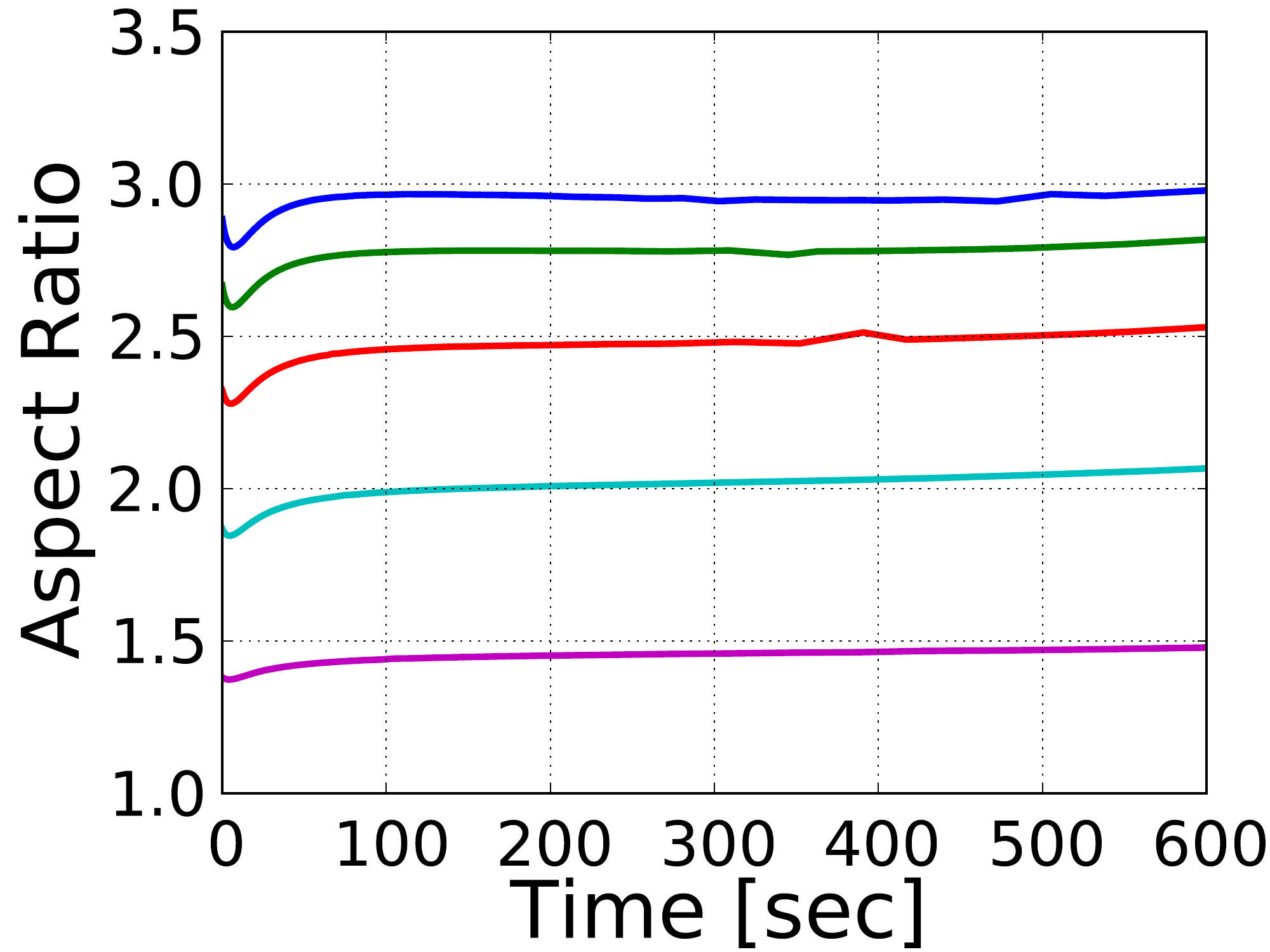}
        \put(0,0){\color{black}{\large (b)}} \end{overpic}}
\end{minipage}
\caption
{
Area and aspect ratio (AR) of model elliptical wounds with varying initial AR.
The minor radius in undeformed configuration $R_B$ is varied relative to
nominal value (Table~\ref{tab:Geometry}), i.e.,   $R_B = 0.5 \bar{R}_B$
specifies a wound with increased AR, and a circular wound is equivalent to $R_B
\simeq 15 \bar{R}_B$.  Wound geometry variation is reflected mainly in initial
AR and has little effect on relative area or AR trends.
}
\label{fig:ModelFitAR}
\end{figure}
}

\newcommand{\figureAltParams}{ % Fig S12
\begin{figure}[htbp]
\begin{minipage}{\textwidth}
\centering
  \subfloat{\label{fig:prestretch}
    \begin{overpic}[width=0.40\textwidth]{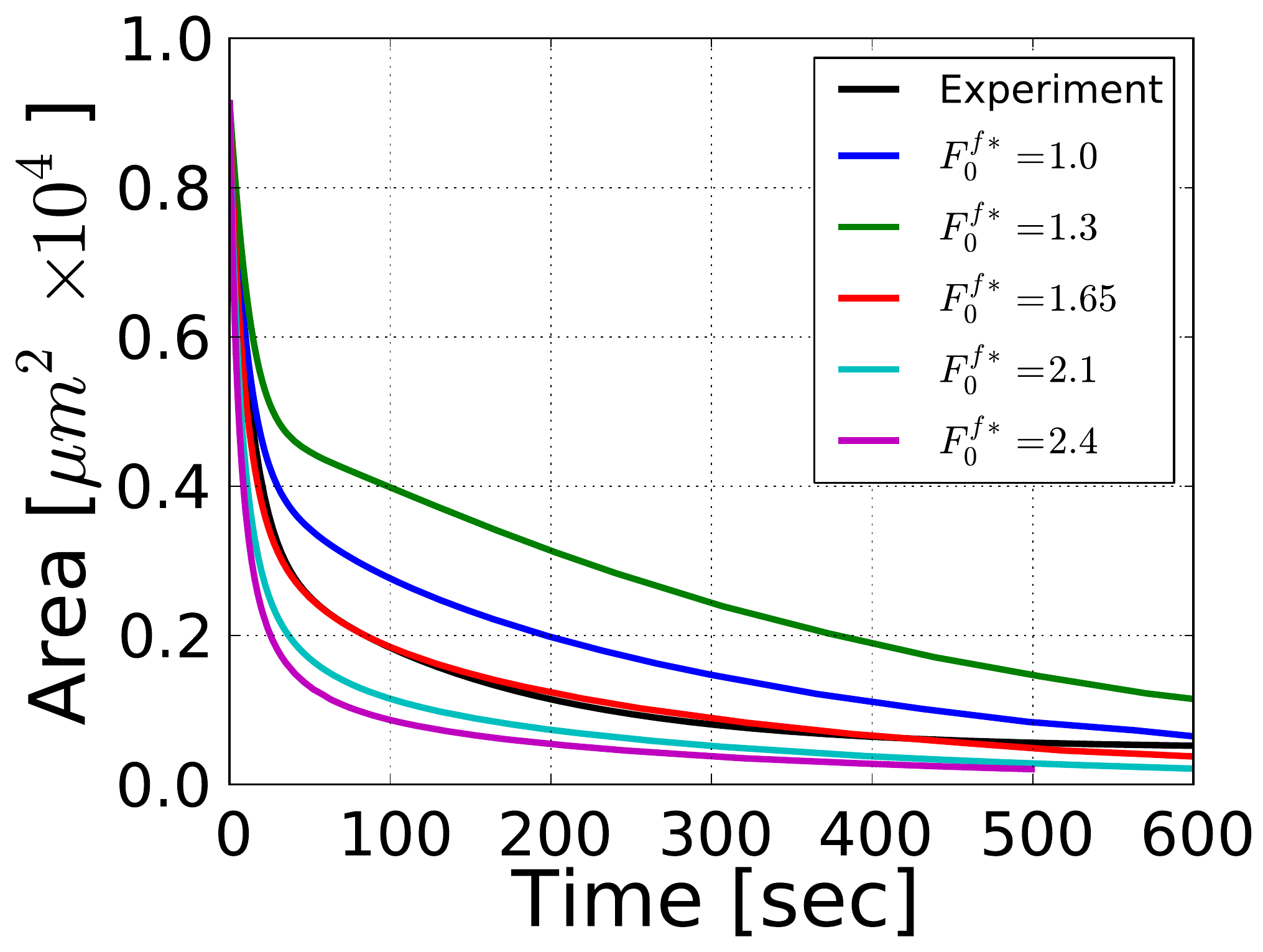}
        \put(0,0){\color{black}{\large (a)}} \end{overpic}}
  \subfloat{\label{fig:AltParams}
    \begin{overpic}[width=0.40\textwidth]{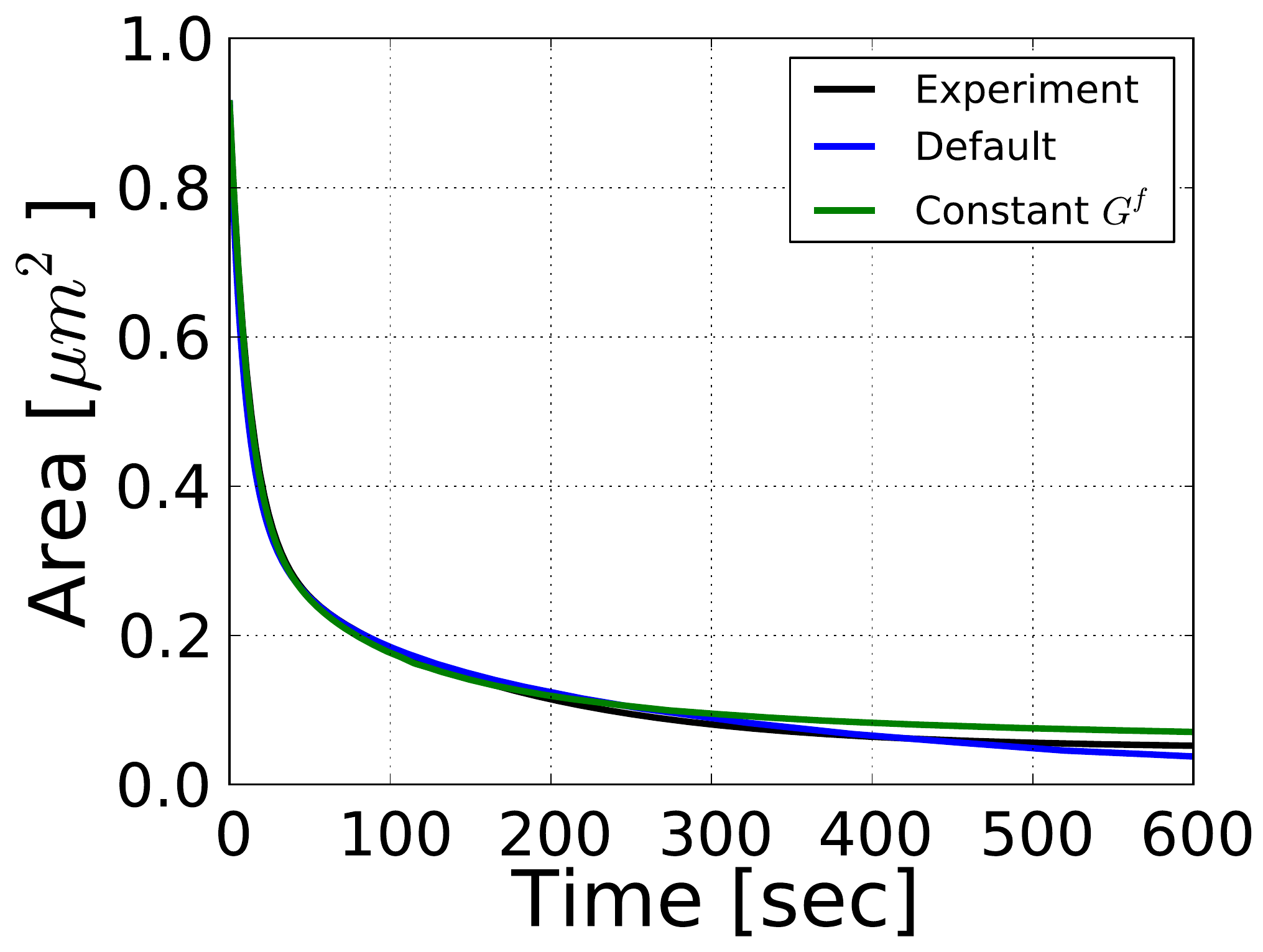}
        \put(0,0){\color{black}{\large (b)}} \end{overpic}}
\end{minipage}
\caption{
Effects of parameter variation on model behavior.
(a) Effect of prestretch $F^{f*}_0$ on circular wounds, with all other parameters as in Table~\ref{tab:Parameters}.
$F^{f*}_0$ has a strong effect on wound healing behavior.  
(b) Comparison circular wound experiment and model results for two different parameter sets.
``Default'' parameters correspond to those in Table~\ref{tab:Parameters}. For ``Constant $G^f$'' parameters,
$G^f_1=1$ (no fiber contraction) with $\tau_\phi=20$ min and $F^{f*}_0 = 3.1$.
Results are essentially indistinguishable given the natural variation in wound behavior.
}
\label{fig:AltParamsFig}
\end{figure}
}

%% file: Tables.tex
\definecolor{P1}{HTML}{0000FF}
\definecolor{P2}{HTML}{007F00}
\definecolor{P3}{HTML}{FF0000}
\definecolor{P4}{HTML}{000000}
\definecolor{P5}{HTML}{00BFBF}
\definecolor{P6}{HTML}{BF00BF}

\newcommand{\tableWoundFit}{
\begin{table}[!ht]
\centerline{
\small
\begin{tabular}{|r|c|c|c|c|c|c|c|}\hline
 & \textcolor{P1}{1} & \textcolor{P2}{2} & \textcolor{P3}{3} & \textcolor{P4}{4} & \textcolor{P5}{5} & \textcolor{P6}{6} & Mean $\pm$ SD\\
\hline \hline
$c_0$ [$\mu m^2$]& -160 & 60.2 & 459 & 2110 & 199 & 230 & 483 $\pm$ 751 \\
$c_1$ [$\mu m^2$]& 7090 & 5320 & 4170 & 5220 & 4690 & 8850 & 5890 $\pm$ 1600 \\
$c_2$ [$\mu m^2$]& 3000 & 2440 & 1780 & 3350 & 2430 & 3550 & 2760 $\pm$ 606 \\
$T_1$ [sec]& 13.5 & 8.83 & 9.62 & 16.1 & 12.4 & 13.7 & 12.4 $\pm$ 2.5 \\
$T_2$ [sec]& 186 & 82.7 & 111 & 131 & 134 & 195 & 140 $\pm$ 40 \\
\hline
\end{tabular}}
\caption
{
Parameters for best fit (Eq.~\ref{eqn:fit}) to experiment for six circular
wounds.  Numbers in column header indicate the wound in
Fig.~\ref{fig:circular_img}; colors match the plot in
Fig.~\ref{fig:CircularAreaFit} for each wound.
} \label{tab:WoundFitAll}
\end{table}
}

\newcommand{\tableGeometry}{
\begin{table}[!ht]
\centerline{
\small
\renewcommand{\arraystretch}{0.95}
\begin{tabular}{|c|c|c|}\hline
Geometric Parameters & $B_0$ & $b_0$ \\
\hline \hline
Circular wound radius & 38  $\mu$m ($R_w$) & 54 $\mu$m ($r_w$) \\
Elliptical wound major radius & 82  $\mu$m ($R_a$) & 91 $\mu$m \\
Elliptical wound minor radius & 5.4 $\mu$m ($R_b$) & 34 $\mu$m \\
Thick (cell) ring width & 41 $\mu$m ($\rho_c$) & 40 $\mu$m \\
Thin (fiber) ring width & 1.5 $\mu$m ($\rho_f$) & 1.3 $\mu$m \\
Membrane dimension & 1020 $\mu$m  & 1140 $\mu$m \\
\hline
\end{tabular}}
\caption
{
Wound geometries for the model in undeformed ($B_0$) and reference ($b_0$)
configurations (see Fig.~\ref{fig:Geometry3D}).  Membrane dimension is the
length of the square epithelial membrane quadrant
(Fig.~\ref{fig:GeometryMesh}).
} \label{tab:Geometry}
\end{table}
}

\newcommand{\tableParameters}{
\begin{table}[!ht]
\centerline{
\small
\renewcommand{\arraystretch}{0.95}
\begin{tabular}{|c|c|c|}\hline
\multicolumn{3}{|c|}{Material Parameters} \\
\hline \hline
\multicolumn{3}{|c|}{Fixed} \\ \hline
Cell shear modulus & $\mu_c$ & 40 Pa \\
Boundary stress & $\sigma_B$ & 42 Pa  \\
Poisson's ratio & $\nu$ & 0.45\\
Final fiber fraction & $\phi_1^f$ & 0.5 \\
Fiber shear modulus & $\mu_f$ & 3200 Pa  \\  % 80x mu_c
\hline \hline
\multicolumn{3}{|c|}{Free} \\ \hline
Cell contraction time & $\tau_c$ & 12 sec \\  % 0.02 * 600
Fiber contraction time & $\tau_f$ & 600  sec\\  % 1 * 600
Fiber formation time & $\tau_\phi$ & 180  sec\\  % 0.3 * 600
Cell final G& $G_1^c$ & 0.55 \\
Fiber prestretch & $F^{f*}_0$ & 1.65 \\  
Fiber final G& $G^f_1$ & 0.15 \\
\hline
\end{tabular}}
\caption
{
Cell and fiber parameters.  Fixed parameters have values assumed to be known
and were obtained as discussed in Section~\ref{sec:Parameters}.  Free parameter
values were found by an iterative procedure so that circular wound model area
trends reproduce those obtained by experiment
(Section~\ref{sec:ModelExperiment}).
} \label{tab:Parameters}
\end{table}
}

\newcommand{\tableTensors}{
\begin{table}[!ht]
\centerline{
\small
\begin{tabular}{|c|l|}\hline
$\boldG^i$ & Cell/fiber growth tensor\\
$\boldF^{i*}$ & Cell/fiber elastic deformation gradient tensor\\
$\boldF^{f*}_0$ & Fiber pre-stretch tensor\\
$\boldF$ & Total deformation gradient tensor of body (cell and fiber)\\
$\boldC^{i*}$ & Right Cauchy-Green elastic deformation tensor\\
$\boldsigma^i$ & Cell/fiber Cauchy stress tensor \\
$\lambda^{f*}$ & Fiber elastic stretch ratio \\
$\phi^i$ & Cell/fiber volume fraction \\
$J^{i*}$ & Cell/fiber elastic volume ratio \\
$W^i$ & Cell/fiber strain-energy density function \\
$\tau_j$ & Intermediate time $0 \leq \tau_j \leq t$ of virtual configuration \\
$I^*_1$, $I^*_3$ & Cell strain invariants \\
$G^i$ & Scalar cell/fiber growth measure\\
\hline
\end{tabular}}
\caption
{
Nomenclature definitions.  Superscript $i=c$ for cells, $i=f$ for fibers.
} \label{tab:Tensors}
\end{table}
}

%% file: Title.tex
\title{Computational and Experimental Study of the Mechanics of Embryonic Wound Healing}

\author[wu]{Matthew A. Wyczalkowski \corref{cor}\fnref{fn1}} \ead{wyczalkowskim@wustl.edu}
\author[pu]{Victor D. Varner } \ead{vdv@princeton.edu}
\author[wu]{Larry A. Taber } \ead{lat@wustl.edu}
\cortext[cor]{Corresponding Author}
\fntext[fn1]{Current address: The Genome Institute, Washington University, St. Louis, Missouri 63108, USA}

\address[wu]{Department of Biomedical Engineering, Washington University, St. Louis, MO 63130 }	
\address[pu]{Department of Chemical and Biological Engineering, Princeton University, Princeton, NJ 08544 }

%% file: 0_Abstract.tex
\begin{abstract}

% from LT email 7/9/13
Wounds in the embryo show a remarkable ability to heal quickly without leaving
a scar. Previous studies have found that an actomyosin ring (``purse string'')
forms around the wound perimeter and contracts to close the wound over the
course of several dozens of minutes. Here, we report experiments that reveal an
even faster mechanism which remarkably closes wounds by more than 50\% within
the first 30 seconds. Circular and elliptical wounds ($\sim$100 $\mu$m in size)
were made in the blastoderm of early chick embryos and allowed to heal, with
wound area and shape characterized as functions of time. The closure rate
displayed a biphasic behavior, with rapid constriction lasting about a minute,
followed by a period of more gradual closure to complete healing. Fluorescent
staining suggests that both healing phases are driven by actomyosin
contraction, with relatively rapid contraction of fibers at cell borders within
a relatively thick ring of tissue (several cells wide) around the wound
followed by slower contraction of a thin supracellular actomyosin ring along
the margin, consistent with a purse string mechanism. Finite-element modeling
showed that this idea is biophysically plausible, with relatively isotropic
contraction within the thick ring giving way to tangential contraction in the
thin ring. In addition, consistent with experimental results, simulated
elliptical wounds heal with little change in aspect ratio, and decreased
membrane tension can cause these wounds to open briefly before going on to
heal. These results provide new insight into the healing mechanism in embryonic
epithelia.

\begin{keyword}
biomechanics \sep chick embryo \sep epithelium \sep finite elements \sep epithelial morphogenesis \sep mechanobiology \sep growth
\end{keyword}

\end{abstract}
%\end{frontmatter}

%% file: 1_Introduction.tex
Embryonic epithelia display a remarkable ability to repair wounds, and use a
series of redundant mechanisms to do so.  Shortly after a wound is made, an
actomyosin cable forms at the wound edge and contracts, drawing the wound
closed by a purse string mechanism~\citep{Jacinto:2001,Martin:1992,Redd:2004}.
Filopodia and lamellipodia also frequently form to draw in apposing edges like
a zipper~\citep{Jacinto:2000, Wood:2002}.  Both these mechanisms require the
polymerization of new actin structures that can form relatively quickly, over a
period of minutes.  Here, we report on a cell contraction mechanism that
precedes these other mechanisms and operates on an even faster time scale.  
To our knowledge, this rapid healing phase has not been studied previously.
Together, these mechanisms (fast contraction, slow contraction, zippering)
constitute three phases for embryonic wound healing.

Much of the cellular machinery used in wound healing also is used during
morphogenesis~\citep{Martin:2004,Sonnemann:2011,Wood:2002}.  In {\em Drosophila}
dorsal closure, for example, two epithelial sheets are drawn together and fuse
in a process remarkably similar to the healing of embryonic wounds, with both
the purse string and zippering mechanisms playing a significant
role~\citep{Hutson:2003}.  The mechanisms of embryonic wound healing are
significantly different from adult wounds, which tend to involve crawling
fibroblasts and an inflammatory response that can lead to scarring.  Embryonic
wounds, by contrast, heal relatively quickly and do not form scars, and thus
may offer clinical insight to improving adult wound healing~\citep{Redd:2004,
Sonnemann:2011}.

The mechanics of wound healing are herein studied in the blastoderm of early
chick embryos.  We found that the rapid healing phase partly closes the wound
within tens of seconds, followed by the slower healing phase which continues over
the course of minutes.  Using fluorescence microscopy and finite-element
modeling to characterize the healing mechanisms, we found that the 
unexpectedly rapid and previously uncharacterized initial healing phase is driven
by an approximately isotropic contraction of cells within a relatively thick ring around the
wound.  Subsequent assembly and contraction of an actomyosin cable (``purse
string'') accounts for the second, slower phase, and these two mechanisms are
able to account for the observed behavior in the first ten minutes of wound
healing.  Filopodial zippering constitutes the third phase, which is not
incorporated into our model.

%% file: 2_Background.tex
%Chick embryos have been used in the study of embryonic development for well
%over a century~\cite{Waddington:1932,Rudnick:1944}, in large part due to their
%ability to heal quickly and completely after surgical manipulation.  

The chick embryo has been a popular model for studies of wound
healing~\citep{Bortier:1993, Brock:1996, England:1977b, Lawson:1998,
Mareel:1977, Martin:1992, Stanisstreet:1980}.  The early embryo consists of a
planar blastoderm, two to three cell layers thick, which is held under tension
and attached to a substrate (the vitelline membrane) only at its
periphery~\citep{ Bellairs:1967,Bortier:1993, New:1959}.  All forces required
to close a wound are generated within the blastoderm itself.  

A critical initial signal of a wounding event is the influx of extracellular
calcium (\Ca) into the perforated cells, as well as its release from
intracellular stores~\citep{ Benink:2005, Cordeiro:2013, Woolley:2000,
Xu:2011}.  Among other roles, \Ca triggers a signaling cascade involving small
GTPase  molecules which play a central and necessary role in the regulation of
wound healing, including recruitment of actin and myosin to the wound edge,
actomyosin assembly and contraction, and the formation of filopodia~\citep{
Bement:2006, Brock:1996, Clark:2009, Sonnemann:2011, Wood:2002}.  A number of
different wound healing mechanisms occur in various model organisms.
Actomyosin purse strings are widely observed in both single-cell and
multicellular wounds~\citep{ Clark:2009, Kiehart:1999, Martin:1992}.
Epithelial cells crawling over a mesenchymal substrate~\citep{Radice:1980}, as
well as contraction of the substrate itself~\citep{ Davidson:2002,
McCluskey:1995}, can also help close embryonic wounds. In the present study,
however, the wounds were cut completely through the blastoderm, and there is no
substrate to aid healing.

Published work on the healing of elliptical-shaped embryonic wounds is
inconclusive on how the shape of the wound changes as it heals.  Elongated
wounds in {\em Drosophila} embryos maintain an approximately constant aspect
ratio (AR) as they close~\citep{Hutson:2003}.  Small elliptical wounds in {\em
Xenopus} oocytes (frog eggs) round as they heal~\citep{Mandato:2001}.  Larger,
superficial rectangular wounds in multicellular {\em Xenopus} embryos with an
underlying mesenchyme, however, become more elongated during
healing~\citep{Davidson:2002}.

In contrast to wound healing in adult tissue, which operates by markedly
different processes, modeling of embryonic wound healing has received
relatively little attention~\citep{ Murray:2003, Olsen:1995,
Wyczalkowski:2012}.  Much of the published work is based on the mechanochemical
model of \citet{Murray:1984} for the morphogenesis of epithelial sheets.
\citet{Sherratt:1992} (see also~\citealt{Murray:2003}) considered the
quasi-static response of such a sheet to wounding, and \citet{Sadovsky:2007}
extended this model to incorporate wound closure.  Previously we presented a
model for a circular wound in which the healing response is governed by a
stretch-activated morphomechanical feedback law~\citep{Taber:2009}.
\citet{Nagai:2009} developed a vertex dynamics model where minimization of
interfacial energy leads to the closing of a wound.  Finally,
\citet{Hutson:2003} (see also~\citealt{Layton:2009}) developed a model
incorporating an actomyosin ring and filopodial zippering for the related
problem of dorsal closure in {\em Drosophila}. 

Tissue-scale deformations are driven by cell division, cell intercalation,
cytoskeletal contraction, or some combination of these and other processes. The
present analysis is based on our tissue-level theory for epithelial
morphogenesis~\citep{Taber:2009}, in which these processes are simulated by
active changes in the local zero-stress configuration.  The kinematic
equations of continuum mechanics are modified to include volumetric
growth \citep{Rodriguez:1994}, which has been used to effectively simulate a number of
morphogenetic processes \citep{Munoz:2007, Taber:2009, Varner:2010b,
Ambrosi:2011, Filas:2012}.  Here, we consider only
active contraction of tissue, as simulated by negative growth, and do not
consider the contributions of filopodia.

%% file: 3_ExperimentalMethods.tex
\subsection{Embryo Preparation and Culture}

Fertilized white Leghorn chicken eggs were incubated at 37\degree C for 20-22
hours (of a 21-day incubation period) in a humidified, forced draft incubator
to yield embryos at approximately Hamburger-Hamilton (HH) stage
4~\citep{Hamburger:1951}.  Whole embryos were harvested from the eggs using a
filter paper carrier method~\citep{Voronov:2002} which preserves the stresses
normally present in the tissue.

At this early stage of development, the embryonic blastoderm is organized as a
nearly flat sheet in a state of approximately uniform, isotropic
tension~\citep{Varner:2010b} (Fig.~\ref{fig:circular_img}).  The blastoderm is
connected to the vitelline membrane only at its outer periphery, and consists
of two epithelial cell layers, the endoderm and ectoderm.  A third cell layer,
the mesoderm, consists of migratory mesenchymal cells and spreads between the
endoderm and ectoderm, although at this stage of development its extent is
limited~\citep{Bortier:1993, England:1977a}.  
While the endoderm and ectoderm have important differences, wounds passing
through both layers heal in the same manner as ectoderm-only
wounds~\citep{Stanisstreet:1980}.  Insofar as we can tell the germ layers do
not move significantly with respect to one another during 
healing, and we do not distinguish between the layers.

\figureWoundImage

After harvesting, embryos were placed atop a 2\% agar/98\% phosphate buffered
saline (PBS) gel in Delta T Dishes (Bioptechs, Butler, PA), submerged under a
thin layer of liquid culture media~\citep{Voronov:2002}, and incubated for at
least one hour. Prior to wounding, the embryos were removed from the incubation
chamber and allowed to equilibrate in the Delta T4 Culture Dish Controller
under the same incubation conditions.

\subsection{Wounding and Imaging}\label{sec:Imaging}

Circular and linear incision wounds were cut completely through the blastoderm.
For dynamic imaging we created between 1 and 6 circular wounds per embryo, positioned as illustrated in
Fig.~\ref{fig:circular_img} for a representative embryo.  Elliptical wounds, as well as circular wounds
used for fluorescence imaging, were created in pairs and positioned anterior and lateral to Hensen's
node (as in Fig.~\ref{fig:elliptical_img}), where the mesoderm is
absent~\citep{England:1977a}.  The location of the wounds generally did not
affect their behavior.

% notes on making scalpels: 8/12/11, 11/13/11
To create circular wounds, a microforged glass micropipette (42 $\mu$m inner,
57 $\mu$m outer radii) held in a micromanipulator was used to punch through the
blastoderm and excise circular plugs of tissue of uniform
size~(Fig.~\ref{fig:circular_img})\citep{Varner:2010a}.  Linear incisions
approximately 200 $\mu$m in length were made by drawing a handheld
microscalpel~\citep{Conrad:1993} across the blastoderm surface.  Tension in the
membrane caused these wounds to open into an initial elliptical shape
(Fig.~\ref{fig:elliptical_img}).  In both cases, wounds were concealed by the
wounding apparatus while they were made, with the first unobscured frame
visible 2-4 seconds after wounding.

%The microscalpels were fashioned as described by Conrad {\em et
%al.}~\citep{Conrad:1993}.  A 1-inch length of tungsten wire (0.25mm diameter,
%Electron Microscopy Sciences, Cat. 73804) was held in a pin vice, bent into a
%``V" shape, and connected to the positive terminal of a 15V DC power supply.
%With a paper clip as the anode in a 1M NaOH solution, the wire was positioned
%with a micromanipulator such that the bottom of the V was in contact with the
%meniscus of the solution. The tungsten wire then eroded into a very sharp
%microscalpel (inset Fig.~\ref{fig:elliptical_img}).

% Below, as submitted.
%Wounds were created and allowed to heal under incubation conditions, and were
%observed for the entire duration of the experiment by a combination of video
%and still digital imaging.  Wound area was determined by thresholding each
%image to isolate the wound, with manual correction as necessary, using
%ImageJ~\citep{Abramoff:2004}.  The ratio of major to minor radius of elliptical wounds
%(the wound aspect ratio, AR) was obtained by fitting the isolated wound
%shape with an ellipse of equivalent area and second moment of inertia.  

Wounds were created and allowed to heal under incubation conditions, and were
observed for the entire duration of the experiment by a combination of video
and still digital imaging.  Wound area was determined by thresholding using
ImageJ~\citep{Abramoff:2004}, with the threshold value selected to isolate the
bright wound from the darker surrounding tissue.  
While somewhat subjective, variations in the threshold value resulted in relatively small changes in the 
calculated wound radii and do not significantly affect the results presented here.
Manual correction of wound shape was performed
in a few elliptical wound cases where debris temporarily impinged upon the
wound area.  The ratio of major to minor radius of elliptical wounds (the wound
aspect ratio, AR) was obtained by fitting the isolated wound shape with an
ellipse of equivalent area and second moment of inertia.  

Fluorescence staining was used to visualize the actin cytoskeleton and
phosphorylated myosin II light chain (pMLC), as described
previously~\citep{Filas:2012}.  Wounded embryos were fixed at a given time
after wounding by injecting 37\% formaldehyde into the media at a 1:10 volume
ratio.  F-actin was visualized with rhodamine phalloidin, while
immunofluorescence staining was used for pMLC.  Following staining, the
endoderm and ectoderm in the vicinity of the wound were separated using drawn
glass capillary tube needles.  Small pieces of both germ layers were then
excised and mounted for fluorescence imaging.

%The samples remained in the
%fixative for 1.5 hours and rinsed.  For actin staining, the embryos were soaked
%for two hours in a solution of 0.1\% Triton X-100 (Sigma) and 1\% bovine serum
%albumin (BSA) (Sigma) in phosphate buffered saline, after which rhodamine
%phalloidin (Molecular Probes) diluted 1:40 was added.  
%
%The immunofluorescence staining used to visualize phosphorylated myosin-II (pMLC)
%proceeded in the same manner as actin staining above until the first
%post-fixation rinse, at which point the endoderm and ectoderm layers around the
%wound were separated and excised.  The tissue samples were then soaked for two hours in
%a solution of 5\% normal goat serum (NGS) and 0.1\% Triton X-100, rinsed, and
%incubated overnight at 4$^\circ$C in a solution of PBS, 1.5\% NGS and 1:50
%dilution of primary p-MLC rabbit antibody (Cell Signaling, 3671S).  Samples
%were rinsed three times, one hour apiece in PBS, and incubated overnight at
%4$^\circ$C in a solution of 1.5\% NGS and 1:400 dilution of secondary goat
%anti-rabbit IgG conjugated to Alexa Fluor 488 (Molecular Probes, A-11070).
%After rinsing the samples were mounted as for actin staining.  

%% file: 4_ExperimentalResults.tex
% work on this - not entirely clear.
%Only wounds which closed completely, or nearly so, during the period
%of observation were included in this analysis.  

We analyzed 52 circular wounds from 15 embryos and 13 elliptical wounds from 10
embryos.  We discarded embryos due to blastoderm detachment, excessive floating
debris, or other experimental problems; we also discarded a minority of wounds which never healed.
Of those analyzed, nearly all (62/65) 
closed relatively quickly and typically healed fully within ten minutes to two
hours, depending on wound size as well as wound location and stage of embryonic
development.  Three circular wounds initially closed partly, then expanded for
several minutes before going on to heal.  
Relatively wide variations in healing time for such
wounds have been previously reported~\citep{Stanisstreet:1980}.  

Our analysis focuses on the first ten minutes after wounding.  Most of the
healing, as measured by wound area, took place during this time, and optical
characterization of wound area and shape is most reliable during this period,
as discussed below.  Subsequent modeling focuses on this time period as well.

\subsection{Circular Wound Area} \label{sec:CircularWoundArea}
The experimental wound area is shown as a function of time after wounding for
six circular wounds (Fig.~\ref{fig:CircularAreaFit} and {\em Supplemental Video
1}).  Immediately after formation, wound area decreased rapidly for about 30
seconds.  After this period the wounds continued to close, but significantly
more slowly.  To minimize variability from differing experimental conditions
and embryos 
and to facilitate quantitative statistics, 
we consider here only wounds from a single embryo (shown in
Fig.~\ref{fig:circular_img}).  However, the qualitative behavior shown in
Fig.~\ref{fig:CircularAreaFit} is representative of that observed in nearly all
embryos we analyzed; in particular, wounds broadly displayed distinct fast and
slow phases of healing.

\figureCircularAreaFit

Nonlinear regression indicated that a single exponential cannot adequately account
for the observed healing behavior, and
we approximate the area versus time relationship for each of the wounds in
Fig.~\ref{fig:CircularAreaFit} using a double exponential function of the form
\begin{equation}
a(t) = c_0 + c_1 \exp(-t / T_1) + c_2 \exp(-t / T_2), \label{eqn:fit}
\end{equation} 
which contains two characteristic time constants, $T_1$ and $T_2$.
Choosing $T_1 < T_2$, we identify $c_1$ and $c_2$ as the magnitudes of
the fast and slow phases of wound healing, respectively.  For each wound we
used nonlinear least squares fitting to obtain values for each of the five
parameters in Eq.~\eqref{eqn:fit} (listed in Table~\ref{tab:WoundFitAll}).  The
dashed curves in Fig.~\ref{fig:CircularAreaFit} show the resulting fit,
illustrating that the double exponential function reproduces trends in the data
quite well.

For all six wounds the fast phase is responsible for the bulk of the decrease
in wound area ($c_1 \sim 2 c_2$), and the associated time constant is $T_1 =
12.4 \pm 2.5$ sec.  The time constant of the slow phase is more variable, but
is generally ten or more times greater than $T_1$ ($T_2 = 140 \pm 40$
sec).  We used the mean parameters $T_i$ and $c_i$
(Table~\ref{tab:WoundFitAll}) to construct a mean wound response.  Not all
wounds closed within ten minutes, an aspect which is captured by the $c_0$
term, and which accounts for the large variation in $c_0$.  Evaluating
Eq.~\eqref{eqn:fit} for $t=0$ with the mean wound parameters yields the
reference wound area $a_0 = 9130~\mu$m$^2$, corresponding to a radius $r_w = 54~\mu$m.  
This initial wound size is intermediate between the pipette inner and outer
radii (42 and 57 $\mu$m, respectively).  This observation is consistent with
previous reports which indicate that the inner radius of the pipette is the
cutting edge \citet{Varner:2010a}.  Once cut, the wound edge then recoils due
to blastoderm tension.

The relatively large difference between the time constants $T_1$ and
$T_2$ suggests that there are (at least) two distinct physical mechanisms
operating to close the wound during the first ten minutes.  The initial phase,
associated with time constant $T_1$, results in the bulk of wound closing
within $\sim$20--30 seconds of wounding (area at 30 seconds is $35\%\,a_0$).
This is followed by a second phase, associated with $T_2$, which operates
more slowly to close the wound the rest of the way.  
The second healing phase appears consistent with actomyosin cable assembly and contraction,
with a time scale in line with published work on the dynamics of actin cable
formation~\citep{ AbreuBlanco:2011, Brock:1996, Clark:2009}.  As discussed
later, however, the rate of the initial mechanism appears too rapid for {\em de
novo} actomyosin assembly.  Filopodial zippering appears to play an insignificant role
at this time scale.

\subsection{Elliptical Wound Area and Shape} \label{sec:ResultsEllipticalAreaAR}
Following linear cuts, wounds opened into elliptical shapes as quickly as could
be observed.  The area and AR of five representative elliptical wounds, all
from different embryos, are plotted in Figure~\ref{fig:EllipseAreaAR}.  
The initial areas and aspect ratio trends of the elliptical wounds are more variable than the circular
ones, in part because they are from different embryos and wound size is somewhat less repeatable.
Consequently, we restrict ourselves here to a qualitative description of elliptical wound healing.

\figureEllipseAreaAR

Like circular lesions, elliptical wounds closed at a rapid initial rate, then
more gradually until fully healed.  For some wounds the area initially
increased for several seconds before decreasing, although this was not observed
in all elliptical wounds
(Figs.~\ref{fig:EllipseArea},\psubref*{fig:EllipseArea1min}).  No circular
wounds displayed such transient gaping behavior.

Figures~\ref{fig:EllipseAR},\psubref*{fig:EllipseAR1min} show the AR of healing
elliptical wounds.  The AR, obtained from the quotient of two fluctuating
quantities, is subject to large fluctuations, particularly as wounds become
small (see also Section~\ref{sec:Filopodia}).  Nevertheless, it is clear the AR
tends to stay relatively constant, indicating that the wounds neither round up
(AR$\rightarrow$ 1) nor become slit-like (AR $\rightarrow \infty$) as they
heal. Notably, in cases where the area of the wound transiently increases
immediately after wounding, there is a corresponding initial decrease in the
AR.

We reason that, since the same chemo-mechanical processes are involved for both
circular and elliptical wound healing, differences in their behavior (e.g., initial
increase in area) are likely caused by geometric effects, 
a topic we address in Section~\ref{sec:MechanicsWoundClosure}.

\subsection{Actin and Myosin Staining} \label{sec:Fluorescence}

To obtain clues concerning the origin of the forces that close the wounds, we
investigated the distributions of F-actin and activated myosin (pMLC) at
various times after wounding.  To visualize actin distribution more clearly, we
separated the two epithelial monolayers, endoderm and ectoderm, and imaged them
separately (Fig.~\ref{fig:WoundActin}).  These layers display different cell
morphologies, with the thicker ectoderm consisting of columnar epithelial cells
and the thinner endoderm of squamous cells~\citep{ Bellairs:2005, England:1993}.
Nevertheless, the conclusions drawn here regarding their healing response are
broadly similar. 

\figureWoundActin

Tens of seconds after wounding, we observed a broad, diffuse ring of enhanced
actin fluorescence near the wound in both the ectoderm
(Fig.~\ref{fig:ecto_20sec}) and the endoderm (Fig.~\ref{fig:endo_15sec}).  This
intensified fluorescence, an indicator of increased F-actin localization,
coincided mainly with cell borders.  A minute or so after wounding this ring had
largely dissipated, except at the wound border where the stain became more
intense~(Figs.~\ref{fig:ecto_1min},\psubref*{fig:endo_1min}, see also Fig.~\ref{fig:WoundActinElliptical}).  
By ten minutes
actin had formed a continuous, distinct supracellular structure along the wound
border, characteristic of an actomyosin cable
(Figs.~\ref{fig:ecto_11min},\psubref*{fig:endo_11min}).  For convenience, we
denote the initial and subsequent actin regions as the ``thick ring'' and
``thin ring,'' respectively.  In addition to the cable structure, finger-like
filopodia and leaf-like lamellipodia were seen in the endoderm
(Figs.~\ref{fig:endo_1min}, \ref{fig:endo_1min_s}), consistent with previous
reports (see Section~\ref{sec:Filopodia}).

While the boundary of the thick ring was not sharply defined and its precise size
and timing of disappearance were variable
(Figs.~\ref{fig:ecto_20sec}, \ref{fig:WoundActinElliptical}), it was roughly 40 $\mu$m in width and of
comparable size around both circular and elliptical wounds (data not shown).
The width of the thin ring was also variable but generally 1-3 $\mu$m wide,
consistent with published reports~\citep{Martin:1992}.  

To better establish the correlation between actin fluorescence intensity and
mechanical force production, we also stained circular wounds for phosphorylated
myosin II light chain (pMLC), which is the activated conformation of the myosin
molecule~\citep{Lecuit:2011}.  The results (Fig.~\ref{fig:WoundPMLC}) show
essentially the same features as actin staining, with a thick ring around the
wound immediately after wounding and a sharp, condensed cable structure ten
minutes later.  

These results suggest that the initial phase of wound healing, associated with
the time constant $T_1$, is driven by rapid cellular contraction within a
thick ring of cells around the wound.  The second phase, associated with the
time constant $T_2$, is in turn driven by slower contraction of a relatively
thin actomyosin cable.

%% file: 5_ModelMethods.tex
To investigate the plausibility of our proposed healing mechanisms, we
constructed a finite-element model that incorporates active tissue contraction.
The blastoderm is modeled as an initially homogeneous single-layered membrane
in plane stress under isotropic tension $\sigma_B$
(Fig.~\ref{fig:Geometry3Doverview}).   The material properties associated with
this epithelial membrane  capture the collective mechanical contributions of
both the endoderm and ectoderm, including cellular cytoplasm and cortical
actin.  From our experimental results, we propose that wounding triggers two
distinct mechanisms: (1) initial contraction of cells in a thick ring
surrounding the wound, and (2) subsequent assembly and contraction of
actomyosin fibers in a thin ring at the wound border.  

\figureGeometryTD

To incorporate these mechanisms into our model, the membrane is treated as a
constrained pseudo-elastic mixture consisting of two contractile components:
``cells'' and ``fibers''~\citep{Humphrey:2002}.  Immediately after wounding,
cells contract within the thick ring (red area, Fig.~\ref{fig:Geometry3D}),
while contractile fibers form more gradually in the thin ring, replacing cells
in the process (blue area, Fig.~\ref{fig:Geometry3D}).  The fibers are assumed
to be stiffer than cells, created in a prestretched configuration, and to
contract along their length.  Working together, cell contraction, assembly of
prestretched fibers, and subsequent fiber contraction generate the mechanical
forces that close the wound.  Both circular and elliptical wounds are
considered in our model (see \ref{sec:coordinates} for geometric
considerations).

% To satisfy comments 1.1.9, 1.2.7
In our model, active contraction is simulated as negative growth using the
theory of~\citet{Rodriguez:1994} for finite volumetric
growth~\citep{Ramasubramanian:2006, Ramasubramanian:2008b}.
As in theories for thermoelasticity and elastoplasticity~\citep{Lubarda:2004},
the total deformation gradient tensor is decomposed into elastic deformation and active
growth tensors. 
This theory is compatible with fundamental thermodynamic principles~\citep{Lubarda:2002, Menzel:2012}.
We consider herein two models for fiber formation.  
First, we discuss the ``multiple fiber'' model,
which is based on the constrained mixture and evolving natural configurations
theory of \citet{Humphrey:2002}.  Next, we describe a simpler ``single fiber''
model, which is a special case of the multiple fiber model and offers numerical
efficiency.

\subsection{Kinematic Equations}
\tableTensors

Applied equibiaxial tension $\sigma_B$ initially deforms the stress-free
membrane $B_0$ into the reference configuration $b_0$
(Fig.~\ref{fig:Geometry3Ddetail}).  With $\sigma_B$ held constant, contraction
is then specified as a function of position and time for $t>0$.  The theory
behind the present work is described below; further details can be found in
previous reports on the mechanics of growth and
morphogenesis~\citep{Rodriguez:1994,Taber:2001}.  

\figureGrowthSchematicMultiFiber

Following the creation of a wound, cells in the thick ring begin to contract
while actin and myosin are recruited to the wound
periphery~\citep{Sonnemann:2011}, where they assemble into a thin supracellular
cable that contracts over the course of minutes (Fig.~\ref{fig:WoundActin}).
As individual actomyosin fibers incorporate into the cable over time, they each
experience a different mechanical environment due to the ongoing deformation of
the tissue.  To model this behavior, each fiber is assumed to form at an
intermediate time $\tau$ and subsequently contract.

In the scheme illustrated in Figure~\ref{fig:GrowthSchematicMultiFiber}, the
first group of fibers is created at time $\tau_1$ (see Table~\ref{tab:Tensors}).  Before this time, each cell
in $B_0$ contracts to a new zero-stress state defined by the contraction
(growth) tensor $\boldG^c_{\tau_1/0}$, which generally varies with time and
space.  The cell then deforms through the elastic deformation gradient tensor
$\boldF^{c*}_{\tau_1/0}$.  This deformation is caused by surface loads, as well
as by geometric compatibility requirements when the cells are reassembled into
the intermediate configuration $b(\tau_1)$.  Fibers are created in $b(\tau_1)$
at a prestretch $\boldF^{f*}_0$ relative to their own zero-stress state.  Here,
we assume that $\boldF^{f*}_0$ is the same for all fibers.

Thereafter these newly created fibers are constrained to undergo the same total
deformation pointwise as the cells.  For the time scale considered here, we
assume that fiber degradation is negligible.  During the interval from $\tau_1$
until the next group of fibers form at $\tau_2$, the cells contract by
$\boldG^c_{\tau_2/\tau_1}$, while the fibers contract by
$\boldG^{f_1}_{\tau_2/\tau_1}$ relative to their individual zero-stress
configurations, with the label $f_1$ indicating the fiber under consideration.
The zero-stress configurations at $\tau_1$ are obtained by reversing the
elastic deformations in $b(\tau_1)$ of both cells ($\boldF^{c*}_{\tau_1/0}$)
and fibers ($\boldF^{f*}_0$).  The cells and fibers then undergo elastic
deformations $\boldF^{c*}_{\tau_2/\tau_1}$ and $\boldF^{f_1*}_{\tau_2/\tau_1}$,
respectively, to give the body $b(\tau_2)$ in which the second set of fibers
form.  This scheme continues until the current configuration $b(t)$.  
In general, we may write for cells ($i=c$) and fibers ($i=f$) $\bF^i_{t/0} =
\bF^i_{t/\tau} \cdot \bF^i_{\tau/0}$ and $\bG^i_{t/0} = \bG^i_{t/\tau} \cdot
\bG^i_{\tau/0}$. This decomposition does not hold, however, for the elastic deformation,
i.e., $\bF^{i*}_{t/0} \ne \bF^{i*}_{t/\tau} \cdot \bF^{i*}_{\tau/0}$.
For a constrained mixture, the deformation of the fibers, once formed, is identical
to the deformation of the cells and the composite body as a whole, i.e.,
$\bF^f_{t/\tau} = \bF^c_{t/\tau} = \bF_{t/\tau}$, where we write
$\bF^f_{t/\tau} = \bF^{f_i}_{t/\tau_i}$ as the total deformation at $t$ for
fibers created at time $\tau$.

The total deformation gradient tensor $\bF_{t/0}$ maps the undeformed
configuration $B_0$ into the current configuration $b(t)$.  For cells, the
sequence of transformations which together compose $\bF^c_{t/0} = \bF_{t/0}$ is
given by (Figure~\ref{fig:GrowthSchematicMultiFiber})
\begin{align}
\bF^c_{t/0} &= \bF^{c*}_{t/\tau_2} \cdot \bG^c_{t/\tau_2} \cdot (\bF^{c*}_{\tau_2/\tau_1})^{-1} 
    \cdot \bF^{c*}_{\tau_2/\tau_1} \cdot \bG^c_{\tau_2/\tau_1} \cdot (\bF^{c*}_{\tau_1/0})^{-1} 
    \cdot \bF^{c*}_{\tau_1/0} \cdot \bG^c_{\tau_1/0} \notag \\ 
&= \bF^{c*}_{t/\tau_2} \cdot \bG^c_{t/\tau_2} \cdot \bG^c_{\tau_2/\tau_1} \cdot \bG^c_{\tau_1/0} \notag \\ 
&= \bF^{c*}_{t/\tau_2} \cdot \bG^c_{t/0}, \label{eqn:Fc}
\end{align}
where $\bG^c_{t/0} \equiv \bG^c_{t/\tau_2} \cdot \bG^c_{\tau_2/\tau_1} \cdot \bG^c_{\tau_1/0}$.
This equation gives the elastic cell deformation as 
\begin{equation}
\bF^{c*}_{t/\tau_2} = \bF^c_{t/0} \cdot (\bG^c_{t/0})^{-1} = \bF^{c*}_{t}, \label{eqn:cell_elastic}
\end{equation}
where we write $\boldF^{c*}_{t/\tau_2} = \boldF^{c*}_{t}$ since the cell
elastic deformation is independent of the fiber creation time $\tau$.

For fibers created at $\tau_1$, the total deformation
$\bF^f_{t/\tau_1}=\bF_{t/\tau_1}$ is (see
Fig.~\ref{fig:GrowthSchematicMultiFiber})
\begin{align}
\bF^f_{t/\tau_1} &= \bF^{f_1*}_{t/\tau_2} \cdot \bG^{f_1}_{t/\tau_2} \cdot (\bF^{f_1*}_{\tau_2/\tau_1})^{-1} \cdot
    \bF^{f_1*}_{\tau_2/\tau_1} \cdot \bG^{f_1}_{\tau_2/\tau_1} \cdot (\bF^{f*}_0)^{-1} \notag \\
&= \bF^{f_1*}_{t/\tau_2} \cdot \bG^{f_1}_{t/\tau_2} \cdot 
    \bG^{f_1}_{\tau_2/\tau_1} \cdot (\bF^{f*}_0)^{-1} \notag \\
&= \bF^{f_1*}_{t/\tau_2} \cdot \bG^{f_1}_{t/\tau_1} \cdot (\bF^{f*}_0)^{-1}, \label{eqn:tot_Ff}
\end{align}
with $\bG^{f_1}_{t/\tau_1}=\bG^{f_1}_{t/\tau_2} \cdot \bG^{f_1}_{\tau_2/\tau_1}$. 
Solving for the elastic deformation of fibers created at $\tau_1$ gives
\begin{equation}
\bF^{f_1*}_{t/\tau_2} = \bF^f_{t/\tau_1} \cdot \bF_0^{f*} \cdot (\bG^{f_1}_{t/\tau_1})^{-1}, \label{eqn:fib1_elastic}
\end{equation}
and a similar expression can be found for fibers created at $\tau_2$.
In general, therefore, the elastic deformation for fibers created at time
$\tau$ is given as
\begin{equation}
\bF^{f*}_{t/\tau} = \bF^f_{t/\tau} \cdot \bF_0^{f*} \cdot (\bG^{f}_{t/\tau})^{-1}. \label{eqn:fib_elastic}
\end{equation}

In the current configuration $b(t)$, the volume fractions of the cells and
fibers are $\phi^c$ and $\phi^f$, respectively, with $\phi^c + \phi^f = 1$.
The fiber volume fraction in the current configuration is given by
\begin{equation}
\phi^f(t) = \int_0^t \phidot^f(\tau) \, J_{\tau/t} \, d\tau,
\end{equation}
where $\phidot^f(\tau)$ is the rate of fiber formation at time $\tau$ and
$J_{\tau/t} = J_\tau J^{-1}_t = \det[\boldF^f_{\tau/0} \cdot
(\boldF^f_{t/0})^{-1}]$ accounts for the change in fiber volume from $b(\tau)$
to $b(t)$.

\subsection{Stress and Equilibrium}

The total Cauchy stress tensor is given as~\citep{Humphrey:2002}
\begin{equation}
\boldsigma(t) = \phi^c(t) \boldsigma^c\left(\boldF^{c*}_{t}\right) + \int_0^t \phidot^f(\tau) \, \boldsigma^f\left(\boldF^{f*}_{t/\tau} \right)
J_{\tau/t} \, d\tau, \label{eqn:sigmf}
\end{equation}
where $\bF^{c*}_{t}$ and $\bF^{f*}_{t/\tau}$ are given by
Eqs.~\eqref{eqn:cell_elastic} and \eqref{eqn:fib_elastic}.  We assume that the
cells and fibers are nearly incompressible and only change shape as they
contract, so that $J_{\tau/t} \cong 1$.

With inertial effects being negligible, morphogenesis can be treated as
quasi-static, and the equilibrium equation is~\citep{Humphrey:2002} 
\begin{equation} 
\del\cdot\boldsigma = 0, \label{equil}
\end{equation}
where $\del$ is the gradient operator defined in $b(t)$.  

\subsection{Constitutive Relations} \label{sec:constitutive}
Because of their high water content, soft biological tissues often are treated
as incompressible materials.  Not only does this assumption often lead to
numerical challenges, but it really is not accurate.  During deformation, water
can enter or leave the tissue, as well as shift from one location to another
within the tissue.  For these reasons, we assume here that the blastoderm is
{\em nearly} incompressible.  In addition, to a first approximation, we assume
that it is pseudo-elastic~\citep{Fung:1993}.

For a compressible pseudo-elastic material, the constitutive relation can be written
in the form~\citep{Taber:2004} 
\begin{equation}
\boldsigma^{i} = \frac{2}{J^{i*}} \boldF^{i*} \cdot \frac{\partial W^i}{\partial \boldC^{i*}} 
    \cdot (\boldF^{i*})^T. \label{eqn:constitutive}
\end{equation}
Here, $W^i(\boldC^{i*})$ is the strain-energy density function for cells ($i=c$) or
fibers ($i=f$), $\boldC^{i*} = (\boldF^{i*})^ T \cdot \boldF^{i*}$ is the right
Cauchy-Green deformation tensor relative to the current zero-stress state, and
$J^{i*}=\det \boldF^{i*}\cong 1$ is the volume ratio~\citep{Taber:2004}.  In
the following, we drop subscripts on $\bF$ and related quantities with the
understanding that $\bF^{c*}$ depends on $t$ while $\bF^{f*}$ depends on $t$
and $\tau$.

Available experimental data suggest that tissues in the early embryo are
relatively linear and isotropic~\citep{ Xu:2010, Zamir:2004b}, especially
compared to mature tissues which have a more organized microstructure.  Hence,
the cells in the blastoderm are assumed to comprise a compressible isotropic
material whose strain-energy density function is
\begin{equation} 
W^c = \frac{\mu_c}{2} \left[ I_1^\ast -3 + \frac{1-2 \nu}{\nu} 
  \left(\left(I^\ast_3\right)^{\frac{\nu}{1-2\nu}}-1\right)\right], \label{eqn:W}
\end{equation}
which is the form for a Blatz-Ko material~\citep{Taber:2004}, where the
invariants are defined as $I_1^\ast = \tr \boldC^{c*}$ and $I_3^\ast =
(J^{c*})^2$.  In addition, $\mu_c$ and $\nu$ are the shear modulus and
Poisson's ratio, respectively, in the limit of small strain.

The fibers in the thin ring are modeled as a distinct transversely isotropic 
material.  By construction, such fibers lie parallel to the wound margin,
and the elastic stretch ratio of the fibers is given by
\begin{equation}
\lambda^{f*} = \sqrt{\bolde_\Theta \cdot \boldC^{f*} \cdot \bolde_\Theta}, \label{eqn:sigf}
\end{equation}
where $\bolde_\Theta$ is a unit vector parallel to the wound edge.
Analogously, $\bolde_R$ is a unit vector perpendicular to the wound edge, and
$\bolde_Z$ is normal to the plane of the membrane (see \ref{sec:coordinates}),
all in the undeformed ($B_0$) configuration.  The corresponding unit vectors 
($\bolde_r$, $\bolde_\theta$, $\bolde_z$) 
in the current configuration $b(t)$ are given by~\citep{Taber:2004}
\begin{equation}
\bolde_i = \frac{\boldF \cdot \bolde_I}{|\boldF \cdot \bolde_I|}. \label{eqn:unit}
\end{equation}
For convenience, we refer to the
$R$ and $\Theta$ directions as radial and circumferential, respectively, even
for elliptical wounds.  Similarly, we refer to elliptical semiaxes as major and
minor radii.  

Each fiber is taken as an incompressible neo-Hookean bar with~\citep{Taber:2004} 
\begin{equation}
W^f = \frac{\mu_f}{2} \left[\left(\lambda^{f*}\right)^2 + 2 \left(\lambda^{f*}\right)^{-1} - 3\right]. \label{eqn:Wf}
\end{equation}
The fiber stress is given by Eq.~\eqref{eqn:constitutive} with
\begin{equation}
\bF^{f*}=\lambda^{f*} \bolde_\theta \bolde_\Theta + \lambda^{T*}(\bolde_r \bolde_R + \bolde_z \bolde_Z)
\end{equation}
where the transverse fiber elastic stretch ratio
$\lambda^{T*}=1/\sqrt{\lambda^{f*}}$ satisfies the incompressibility condition
$\det \bF^{f*}=1$. 

\subsection{Single-Fiber Approximation}
Equation \eqref{eqn:sigmf} is a hereditary integral which depends on the
histories of deformation and fiber deposition.  
As such, it introduces
significant computational cost and complexity, and 
the time required to perform each
simulation would make it impractical to find parameter values iteratively. 
To simplify the calculation, we
take advantage of the separation in time scales of the healing mechanisms.  The
first phase of wound healing, driven by cellular contraction, is essentially
complete by the time fibers form in significant quantity, while actin staining
suggests that the bulk of the thin ring forms within the first few minutes
after wounding (Fig.~\ref{fig:WoundActin}).  Hence, cell contraction is
essentially constant as fibers form at a rate which is relatively fast compared
to subsequent contraction and healing.  As a first approximation, therefore, we
take the reference configuration for all fibers in the thin ring to be the
same, and set $\tau=0$. Hence, we assume that the fibers accumulate relatively
rapidly over the first few minutes of the healing process, while subsequent
contraction takes place over a longer time scale.  Then,
Eq.~\eqref{eqn:fib_elastic} becomes 
\begin{equation}
\bF^{f*}_{t/\tau} \cong \bF^{f*}_{t/0} = \bF^{f*}(t) = \bF^f_{t/0} \cdot \bF^{f*}_0 \cdot \left(\bG^f_{t/0}\right)^{-1}. \label{eqn:Gdef}
\end{equation}

With $\bF^{f*}$ no longer dependent on $\tau$, Eq.~\eqref{eqn:sigmf} reduces to
\begin{align}
\boldsigma(t) &= \phi^c(t) \boldsigma^c\left(\boldF^{c*}\right) + \boldsigma^f\left(\boldF^{f*} \right) \int_0^t \phidot^f(\tau) d\tau \notag \\
&= \phi^c \boldsigma^c + \phi^f \boldsigma^f, \label{eqn:sigtot_sf}
\end{align}
in which subscripts denoting time intervals have been dropped.
In this approximation, the total deformation gradient tensor for a constrained mixture
is given by $\bF = \bF^c = \bF^f$, and Eqs.~\eqref{eqn:Fc} and \eqref{eqn:Gdef} yield
\begin{equation}
\boldF = \boldF^{c*} \cdot \boldG^c = \boldF^{f*} \cdot \boldG^f \cdot \left( \boldF^{f*}_0\right)^{-1}.  \label{eqn:F} 
\end{equation}
This equation gives the elastic deformation gradient tensors $\boldF^{c*} = \bF
\cdot \left(\bG^c\right)^{-1}$ and ${\boldF^{f*} = \bF \cdot \bF^{f*}_0 \cdot
\left(\bG^f\right)^{-1}}$, which are used in Eq.~\eqref{eqn:constitutive}.
With $\bG^c(t)$, $\bG^f(t)$, and $\bF^{f*}_0$ specified, we employ the single
fiber approximation in the model described in the remainder of this section.
In \ref{sec:SingleMultiFiber} we show that this approximation is sufficiently
accurate for our purposes.

\subsection{Contraction Dynamics}
The cell contraction tensor $\boldG^c$ is taken in the form
\begin{equation}
\boldG^c = G^c_R~\bolde_R \bolde_R + G^c_\Theta \bolde_\Theta \bolde_\Theta + G^c_Z \bolde_Z \bolde_Z. \label{eqn:GcLoc}
\end{equation}

Since a cell contracts with little change in volume, we take
$\det \boldG^c = G^c_R G^c_\Theta G^c_Z = 1$.  Here, we consider three
contraction schemes which define $G^c_R$, $G^c_\Theta$, and $G^c_Z$ in terms of
a single specified cell contraction ``stretch ratio'' $G^c$:
\begin{equation}
G^c_R, G^c_\Theta, G^c_Z =
\begin{cases}
G^c, G^c, \frac{1}{(G^c)^2} & \text{isotropic contraction ($R$, $\Theta$ directions)} \\
1, G^c, \frac{1}{G^c} & \text{circumferential contraction ($\Theta$ direction)} \\
G^c, 1, \frac{1}{G^c} & \text{radial contraction ($R$ direction)}. 
\end{cases} \label{eqn:GcCases}
\end{equation}
We will consider behavior of all three of these schemes. % to point 1.1.7
As discussed below, $G^c$ is a specified function of time.  Other choices of
$(G^c_R, G^c_\Theta, G^c_Z)$ for circumferential and radial contraction schemes
that preserve $\det \bG=1$, including those where $G^c_Z=1$, do not
significantly alter the behavior of the model (not shown).

Since $\boldG^c$ is isochoric and the cells nearly incompressible, contraction
causes cells to thicken in the $Z$ direction, affecting cell stiffness (which
is proportional to thickness).  Because the epithelial membrane is restricted
to lie in the plane, such thickening is the only deformation in the $Z$
direction, and no out-of-plane bending is permitted.

% dropped mention of hypothesis, as per C.41
To investigate whether rapid cell contraction in the thick ring drives the first
phase of wound closure, we vary the degree of cell contraction $G^c$
with both position and time.  Outside of the thick ring, we set $G^c=1$
at all times, indicating passive cells.  Within the ring, $G^c$ decreases with
time from an initial value of 1 to the final value $G^c_1$.  To avoid
discontinuities, which can lead to numerical instabilities, the behavior of
$G^c$ transitions gradually at the ring boundary.  We take $G^c$ in the form
\begin{subequations}
\label{eqn:Gc}
\begin{equation}
G^c(t,\rho) = 1 - \left[1-g^c(t)\right] f^c(\rho), \label{eqn:Gca}
\end{equation}
where $f^c$ and $g^c$, respectively, govern the spatial and temporal aspects of $G^c$: 
\begin{align}
f^c(\rho) &= \left[1+\exp\left(\frac{\rho-\rho_c}{k_c}\right)\right]^{-1}  \\
g^c(t) &= (1 - G_1^c) \exp(-t/\tau_c) + G_1^c. \label{eqn:gct}
\end{align}
\end{subequations}
Here, $\tau_c$ is the characteristic cell contraction time (within the thick
ring), $\rho_c$ is the width of the thick ring, and $k_c$ determines the width
of the transition at the ring boundary.  The distance from the wound edge
$\rho$ is given in \ref{sec:coordinates} by Eqs.~\eqref{eqn:rhoCylindrical} and
\eqref{eqn:rhoElliptical} for circular and elliptical wounds, respectively.
Figure~\ref{fig:Gc_Plot} shows $G^c(\rho)$ for various times.

\figureGplots

Consistent with the single fiber approximation, the thin contractile ring forms
with a prestretch $F^{f*}_0$ and subsequently contracts ($G^f$ decreases) while
additional assembly occurs ($\phi^f$ increases).  Fibers are assumed to
contract only along their lengths, and the fiber prestretch and contraction
tensors, respectively, are taken as
\begin{subequations}
\begin{align}
\boldF^{f*}_0 &= 1/\sqrt{F^{f*}_0}~\bolde_r \bolde_R + F^{f*}_0 \bolde_\theta \bolde_\Theta + 1/\sqrt{F^{f*}_0}\bolde_z \bolde_Z \\
\boldG^f &= 1/\sqrt{G^f}~\bolde_R \bolde_R + G^f \bolde_\Theta \bolde_\Theta + 1/\sqrt{G^f}\bolde_Z \bolde_Z,  \label{eqn:GfLoc}
\end{align}
\end{subequations}
which satisfies $\det \boldF^{f*}_0=\det \boldG^f=1$, and both are symmetric
relative to the cross-fiber direction.  To define the dynamics of the fiber
contraction ratio $G^f(t)$, we assume that $G^f$ decreases exponentially with
time from $G^f=1$ at $t=0$ to a final value $G_1^f$ at $t=\infty$, with the
rate of decrease given by the characteristic time constant $\tau_f$
(Fig.~\ref{fig:Gf_Plot}):
\begin{equation}
G^f(t) = (1 - G^f_1) \exp(-t/\tau_f) + G^f_1  \label{eqn:G_f}.
\end{equation}

The spatial distribution of the thin ring is defined through $\phi^f$, which
varies both spatially and temporally.  The value of $\phi^f$ is uniformly zero
at $t=0$, increases with time in the vicinity of the wound at a rate
characterized by $\tau_\phi$, and approaches asymptotically the maximum value
$\phi_1^f$ at the wound edge.  After wounding ($t>0$), the spatial distribution
of $\phi^f$ takes the form of a decaying exponential away from the wound edge
whose characteristic width $\rho_f$ corresponds to the nominal thin ring width.
Mathematically,
\begin{equation}
\phi^f(t,\rho) = 
\phi_1^f \left[ 1-\exp(-t/\tau_\phi) \right]  \exp(-\rho/\rho_f) \label{eqn:phi_f}\\ 
\end{equation}
which is plotted in Fig.~\ref{fig:Phi_Plot}.

\subsection{Solution Procedure}
\begin{comment}
% This text heaviy shortened to satisfy 1.2.10
All model simulations were performed using COMSOL Multiphysics (v 3.5a; Comsol,
Inc.), with plane stress conditions assumed; see~\citet{Taber:2008} for more
details about the implementation of growth.  The simulations take place in two
stages: starting with the undeformed configuration $B_0$, the boundary stress
is increased from zero to the prescribed value $\sigma_B$, where it is
subsequently held fixed for the remainder of the simulation.  There is no
contraction or fiber assembly during this preparatory stage (i.e., $G^i=1$ and
$\phi^f=0$ everywhere).  For convenience in comparing to experiment, where the
wound is created at $t=0$, we take the preparatory stage to begin at $t=0^-$ in
the undeformed configuration $B_0$, and conclude at $t=0$ with the system in
the reference configuration $b_0$.  The simulation stage begins at $t=0$, where
$G^c$, $G^f$, and $\phi^f$ vary with time as previously described.  All results
presented here are for $t\geq0$.
\end{comment}

All model simulations were performed using COMSOL Multiphysics (v 3.5a; Comsol,
Inc.), with plane stress conditions assumed; see~\citet{Taber:2008} as well as ~\ref{sec:matrix} for more
details about the implementation of growth.  
To establish initial conditions which correspond to the reference configuration
$b_0$ at $t=0$, the boundary stress is increased from zero in the undeformed
configuration $B_0$ to the prescribed value $\sigma_B$, where it is
subsequently held fixed for the remainder of the simulation.  
Because of the symmetry of the system, we simulate one quadrant of the
epithelial membrane.  We approximate an incision wound by an ellipse with a
large AR, and use a triangular mesh of third-order Lagrange elements with a
progressively finer mesh in the thick and thin ring regions, resulting in 4474
and 4419 elements in the circular and elliptical models, respectively.
We perform a quasi-static analysis, with the
direct UMFPACK spatial solver and BDM time stepper.  Relative and absolute tolerances
are 0.001, 0.0001, respectively.  We found that the model converges successfully for 
the time period of interest (first 10 minutes) and that the results are not
significantly different with the use of a finer mesh size.  Furthermore, we
found no evidence of numerical instabilities (e.g., extremely large stresses,
deformations, or ``checker-boarding'').
Figure~\ref{fig:GeometryMesh} illustrates the mesh geometry and boundary
conditions of the finite-element model.

\subsection{Parameter Values} \label{sec:Parameters}
The model parameters can be divided into three groups.  Parameters defining the
geometry of representative circular and elliptical wounds were obtained from
our experimental data.  A second set of parameters were obtained from
experiments or deduced, and are considered known.  A final set of parameters
was determined by iteratively fitting the circular wound model to experimental
results.  These parameters were then used also for the elliptical wound model.
In this section, we discuss the choice of the first two types of parameters,
which are then considered ``fixed.''  The selection of the third type,
considered ``free'' parameters, is discussed later.

\subsubsection{Geometric Parameters} \label{sec:geometric_parameters}
In constructing the model geometry we distinguish between the stress-free
undeformed configuration $B_0$ and the reference configuration $b_0$, which is
deformed by the boundary stress $\sigma_B$ but not fiber or cell contraction
(Fig.~\ref{fig:Geometry3Ddetail}).  The geometry of the model is defined in the
$B_0$ configuration, but is chosen such that dimensions in $b_0$ match
experimental measurements.  

\tableGeometry

The circular wound radius $r_w$ in $b_0$ was obtained from measurements
immediately after wounding (see Section~\ref{sec:CircularWoundArea}).  The
elliptical wound dimensions were chosen such that in $b_0$ the elliptical and
circular wound areas are approximately equal, and the AR (=2.67) is consistent
with experiment (Section~\ref{sec:ResultsEllipticalAreaAR}).  
We approximate linear incisions with an elliptical
wound of a relatively large aspect ratio (AR=15)
for numerical reasons, since sharp corners introduce stress concentrations and lead to numerical
instabilities (see also \ref{sec:param_sup}). 
Thick and thin
ring widths ($\rho_c$, $\rho_f$) were estimated from fluorescence microscopy
images of circular wounds (Fig.~\ref{fig:WoundActin}), and $k_c$ is taken as
$\rho_c/10$.  The same parameters are used for elliptical wounds. 

A circular hole in a plate under tension
perturbs the state of stress only locally, and its effect decays quickly with distance from the hole.
At a distance of four wound diameters, the stress 
differs from the far field value by just 6\% according to linear plate theory~\citep{Timoshenko:1951}.  
Stress in our membrane model, although subject to
relatively large deformations, is in generally good agreement with these
theoretical predictions, and we choose the dimensions of the membrane to be at
least an order of magnitude greater than the wound to eliminate
far-field boundary effects.  For computational efficiency we simulate a quarter of the entire
domain, as illustrated in Fig.~\ref{fig:GeometryMesh}.

Table~\ref{tab:Geometry}
summarizes the geometric parameters, and \ref{sec:coordinates} provides details
of ring geometry specification.

\subsubsection{Other Fixed Parameters}

As discussed in \ref{sec:varner}, we estimate the cell shear modulus to be
$\mu_c \simeq 40$~Pa.  Literature suggests a wide range of fiber stiffness
values, from $\sim 10$ kPa~\citep{Lu:2008} to well over 1
MPa~\citep{Deguchi:2006}, with fibers much stiffer than cells ($\mu_f \gg
\mu_c$)~\citep{Rauzi:2011}.  We find that model sensitivity to $\mu_f$
decreases for $\mu_f \gtrsim 100 \, \mu_c $ as fiber dynamics alone dominate
wound closure (data not shown), and thus choose $\mu_f = 80 \mu_c = 3.2$~kPa.
We further chose $\phi_f$ to have a maximum value of $0.5$, and note that model
behavior depends on the product $\phi_f \mu_f$ (analysis not shown.)

\tableParameters

The boundary tensile stress $\sigma_B$ was estimated from previous
work~\citep{Varner:2010a,Varner:2010b}.  Immediately after wounding, circular
epithelial wounds open to a diameter approximately 1.2 times that of the punch
used to make them~\citep{Varner:2010b}, a result that corresponds to an
equibiaxial membrane stretch ratio of 1.1~\citep{Varner:2010a}.  We find that a
boundary stress $\sigma_B=1.05 \times \mu_c = 42$~Pa reproduces this strain.
Model results are relatively insensitive to Poisson's ratio in the range $0.4
\leq \nu \le 0.49$, and we take $\nu=0.45$.  Note that nearly incompressible
models for anisotropic materials can be susceptible to errors for $\nu
\rightarrow 0.5$~\citep{NiAnnaidh:2012}.  All parameters used in the cell and
fiber model, including those obtained from a fit to experiment (see below) are
listed in Table~\ref{tab:Parameters}.

%% file: 6_ModelResults.tex
% dropped reference to hypothesis
We now determine the remaining free parameters and examine whether the model
captures the main features of the measured wound healing response.  

\subsection{Cellular Contraction is Relatively Isotropic in Thick Ring} \label{sec:CellRingContraction}

Fluorescence microscopy images (Section~\ref{sec:Fluorescence}) suggest that
contraction at cell borders occurs within a thick ring around the wound, but do
not indicate its nature.  For example, is the contraction circumferential
around the wound (like the contraction of an actomyosin cable) or isotropic?
Other types of contraction anisotropy also may
occur~(Fig.~\ref{fig:ModelCompare_schematic}).  Models of circular wounds are
unable to discriminate between these possibilities, since multiple schemes can
close a wound.  Elliptical wound models, however, predict distinct temporal 
trends in the AR as the wound closes, and allow for competing contraction schemes to be
critically evaluated.

\figureModelCompare

Consider the area and AR of an elliptical wound for three different cell
contraction schemes: isotropic, circumferential, and radial~(see
Eq.~\eqref{eqn:GcCases} and Fig.~\ref{fig:ModelCompare}).  As expected, radial
contraction causes the wound to open further, allowing that scheme to be
rejected.  Both isotropic and circumferential contraction close the wound, but
the AR trends differ qualitatively. For the circumferential case, the AR
increases without bound, with the wound becoming increasingly slit-like as it
heals. This behavior is at odds with experimental observations
(Fig.~\ref{fig:EllipseAreaAR}).  Isotropic contraction, by contrast, yields an
AR that stays relatively constant as the wound closes,  behavior generally
consistent with experiment (Fig.~\ref{fig:EllipseAreaAR}).  

We conclude, therefore, that cellular contraction in the thick ring is
relatively isotropic, with cells on average contracting in both the radial and
circumferential directions simultaneously.  Future data may indicate that the
contraction is actually anisotropic, but here we take contraction in the thick
ring as isotropic to a first approximation.  This scheme will be used for the
thick ring in all subsequent models.

\subsection{Model-Predicted Changes in Wound Geometry Agree with Experimental Results} \label{sec:ModelExperiment}

With the geometric and material parameters established in
Section~\ref{sec:Parameters}, six free model parameters remain: three that
quantify the degree of contraction ($F^{f*}_0$, $G^c_1$, $G_1^f$) and three
that characterize rates ($\tau_c$, $\tau_\phi$, $\tau_f$) (see
Table~\ref{tab:Parameters}).  The goal is to find a set of biologically
plausible parameter values that reproduce the average experimental circular
wound area trends, i.e., the area given by Eq.~\eqref{eqn:fit}.  Multiple
circular wound simulations were performed with systematically varying parameter
values, which were adjusted iteratively to obtain an area vs. time curve that
reasonably matches experiment.

\figureModelFitWound

The healing response  during the first 20-30 seconds is dominated primarily by
cell contraction, as the fibers have not yet formed in significant quantity.
The wound behavior during this time is controlled by the rate of cell
contraction $\tau_c$ and the final cell contraction parameter $G^c_1$ (see
Eq.~\ref{eqn:gct}).  The values $G_1^c=0.55$ and $\tau_c=12$ sec reproduce this
initial stage of contraction (Fig.~\ref{fig:ModelFitCircle}).  

The parameters corresponding to the fiber prestretch ($F^{f*}_0$) and
contraction ($G_1^f$) govern the remainder of wound closure.  We find that
fibers forming rapidly ($\tau_\phi=180$ sec) in a state of significant
prestretch ($F_0^{f*}=1.65$) and thereafter contracting more slowly ($\tau_f =
600$ sec, $G_1^f=0.15$), together with the other parameters in
Table~\ref{tab:Parameters}, reproduce the experimentally observed area versus
time curves quite well~(Figure~\ref{fig:ModelFitCircle}).  The uniqueness and
biological plausibility of these values are discussed later.

Once obtained for circular wounds, these same parameters were then used in the
elliptical wound model, with the resulting area and AR plotted in
Figures~\ref{fig:ModelFitEllipseArea}, \psubref*{fig:ModelFitEllipseAR}.  From
an experimental perspective, elliptical wounds are less reproducible, and we
restrict ourselves to a qualitative comparison of wound trends.  As in the case
of circular wounds, and consistent with experimental trends, the elliptical
wounds in the model close in two phases -- a rapid initial closing followed by
slower healing (Fig.~\ref{fig:ModelFitEllipseArea}).  The AR displays a brief
decrease followed by a rebound and relatively constant value
(Fig.~\ref{fig:ModelFitEllipseAR}).  Both features are broadly consistent with
our experimental observations.

\subsection{Peak Stresses Increase as Wounds Close} \label{sec:CircularStress}

Stress distributions near the wound boundary are shown for circular wounds,
with radial ($\sigma_r$) and circumferential ($\sigma_\theta$) stress
components plotted for four illustrative time points
(Fig.~\ref{fig:ModelStressCircular}).  At $t=0$ (i.e., $b_0$ configuration),
the membrane is stretched passively by $\sigma_B$.  Significant cell
contraction has occurred by $t=12$ sec, but fiber formation is negligible.  At
$t=300$ and 600~sec, cell contraction is fully developed, with changes in
stress driven by ongoing fiber formation and contraction.

\figureModelStressCircular

For a circular wound, $\sigma_r$ develops two peaks as the wound closes
(Fig.~\ref{fig:sig_rad_R}).  One peak occurs near the wound edge, while the
other develops just outside the thick ring.  The distribution of
$\sigma_\theta$ is dominated by a strong peak at the edge of the wound,
increasing from about $2\,\sigma_B$ at $t=0$ to $40\,\sigma_B$ at $t=300$ sec.
Thereafter $\sigma_\theta$ decreases at the wound edge and becomes slightly
compressive immediately outside the cell ring; the reason for this effect is
discussed later.  In all cases, both stresses approach the far-field value
$\sigma_B$ away from the wound.

Stress distributions near an elliptical wound exhibit similar characteristics
(Fig.~\ref{fig:ModelStressElliptical}). For an ellipse, the strongest stress
concentrations occur near the wound edge along the major axis, while the
stresses are reduced relative to those for a circular wound along the minor
axis.

\subsection{Elliptical Wounds Briefly Open at Low Tension} \label{sec:EllipticalGape}

The principal effect of varying $\sigma_B$ is to change the initial size of the
wound. Also, as $\sigma_B$ increases, elliptical wounds become initially
rounder (Fig.~\ref{fig:sigb_compare}).

For a circular wound, area decreases in a similar manner regardless of
$\sigma_B$, with wounds under higher tension closing more rapidly
(Fig.~\ref{fig:Fbnd_area_1min}).  Elliptical wounds at high tension likewise
begin to close immediately after contraction begins, but for $\sigma_B \lesssim
15$ Pa the wound area increases momentarily before closing
(Fig.~\ref{fig:Fbnd_area_1min}).  This initial ``gaping'' of elliptical wounds
at low $\sigma_B$ is associated with a rapid decrease in the wound AR
(Fig.~\ref{fig:Fbnd_AR_1min}).  Similar behavior was observed experimentally,
where some elliptical wounds initially increased in area while becoming more
round (Fig.~\ref{fig:EllipseAreaAR}), 
but it is important to note tissue tension $\sigma_B$ was not determined experimentally.

\figureModelFitSigB

This transient area increase is driven primarily by a brief expansion of the
minor radius of the wound (Fig.~\ref{fig:ExptRadiusCompare}).  This effect is
captured by the elliptical wound model when the initial tension is reduced.
With $\sigma_B$ set to the estimated physiological value $\sigma_B = 42$ Pa
(Table~\ref{tab:Parameters}), the area decreases immediately after wounding
(Fig.~\ref{fig:Fbnd_area_1min},\psubref*{fig:Fbnd_AR_1min}) and both the major
and minor axes decrease (Fig.~\ref{fig:ModelRadiusCompare}), with the AR
remaining relatively constant (Fig.~\ref{fig:Fbnd_AR_1min}).  For low
$\sigma_B$ wound area increases, the minor axis increases, and the AR decreases
briefly just after contraction begins
(Fig.~\ref{fig:Fbnd_area_1min},\psubref*{fig:Fbnd_AR_1min},
\ref{fig:ModelRadiusCompare}).  The reason behind this behavior is discussed
later.  The qualitative aspects of wound closure are relatively insensitive to
wound size and shape (see Supplemental \ref{sec:param_sup}).

%% file: 7_Discussion.tex
Our results suggest that wounds in the early chick embryo heal by three
distinct and essentially sequential mechanisms.  First, during about the first
30 seconds, a ring of cells (3-4 cells deep) contracts to quickly close the
wound area by more than 50\%.  Second, a relatively thin contractile ring forms
at the wound edge and contracts more slowly to close the wound nearly
completely over a period of several minutes.  Finally, filopodia pull and zip
the edges of the wound together to complete the healing process.  The last two
phases have been known for some time~\citep{ Jacinto:2001, Woolley:2000}, and
the initial rapid healing phase was observed in chick embryos by
\citet{Bortier:1993}, but they did not investigate it in detail.  Notably, this
phase seems to be absent in {\em Drosophila} embryonic
wounds~\citep{AbreuBlanco:2011, Hutson:2009, Ma:2009}, so it may not be a
universal phenomenon.  

As discussed below, our model captures the fundamental
behavior of the first two phases of wound healing and yields insight into the
detailed mechanics.
Because it does not include filopodial zippering, however, the utility of our
model is restricted to wounds relatively large compared to cell size.  At time
scales on the order of hours the assumption that the blastoderm can be treated
as a single-layered membrane also breaks down, as germ layers develop and move
with respect to one another.  As a result, our model is valid from the first
few seconds to ten minutes or so after wounding.

\subsection{Parameter Values for Contraction Dynamics}

For our wound healing model to be plausible, it is important that the values of
the free parameters are consistent with those reported in the literature for
related systems.  Our model incorporates three separate time constants
(Table~\ref{tab:Parameters}): $\tau_c$ for cell contraction, $\tau_\phi$ for
fiber formation, and $\tau_f$ for fiber contraction. Three additional
parameters specify the degree of initial and final contraction: final cell
contraction is given by $G^c_1$, and initial (prestretch) and final fiber
contractions are given by $(F^{f*}_0)^{-1}$ and $G^f_1$, respectively.

The rate of cell contraction $\tau_c$ (12 sec) matches the time constant
$T_1$ (12.4 sec) for the first phase of healing circular wounds
(Section~\ref{sec:CircularWoundArea}, Table~\ref{tab:WoundFitAll}).  Both the
speed and magnitude of cell contraction ($G^c_1=0.55$) are in line with
observations for smooth muscle cells~\citep{An:2007}.  The rate of fiber
formation $\tau_\phi$ (3 min) is consistent with our observed time rate of
fiber formation (see Fig.~\ref{fig:WoundActin} and Sec.~\ref{sec:phase1}), and
the fiber contraction rate $\tau_f$ (20 min) is consistent with actomyosin
fiber dynamics continuing over tens of minutes following a stimulus~\citep{
An:2007, Brock:1996}.  Embryonic tissues can shorten by 70\% or more within a
relatively short time period~\citep{Varner:2012b}, supporting the value
$G^f_1=0.15$.  Such a large contraction likely involves significant remodeling
of actomyosin fibers, similar to that observed in smooth muscle
cells~\citep{Matsumoto:2012} or the contractile ratcheting mechanism described
by \citet{Martin:2009}.

Finally, the dynamics of wound closure are quite sensitive to the value of the
prestretch $F^{f*}_0$ (Fig~\ref{fig:prestretch}). Our value ($F^{f*}_0=1.65$)
is somewhat larger than that reported in the literature (1.10 to 1.35;
\citealt{Kaunas:2011}).  This is discussed further in
Section~\ref{sec:thin_ring}. 

The parameters in Table~\ref{tab:Parameters} are not unique.  For instance, a
set of parameters where fibers form slowly ($\tau_f=20$ min) with a larger
prestretch $F^{f*}_0 = 3.1$ and without further contraction ($G^f_1 = 1$) can
also reproduce our wound closing results (Fig.~\ref{fig:AltParams}).  Our
experimental data are insufficient to discriminate between these and other
possible parameter sets, and this certainly warrants further study.

\subsection{Phases of Embryonic Wound Healing}
Unlike adult wounds, embryonic wounds generally heal quickly without leaving a
scar~\citep{Redd:2004}. The healing process involves many of the same
mechanisms used for morphogenesis~\citep{Wood:2002}, which often produce
dramatic and rapid changes in tissue shape driven by contractile
forces~\citep{Davies:2005}.  Fusion of epithelia is also common, e.g., during
neurulation and heart tube formation~\citep{Colas:2001, MorenoRodriguez:2006,
Ray:2012}.  The three healing phases close wounds efficiently and robustly.

\subsubsection{Phase 1: Rapid Cellular Contraction (Thick Ring)}\label{sec:phase1}
Activation of upstream GTPase regulators of actin and myosin occurs 10-20
seconds after wounding \citep{Bement:2006, Clark:2009}, and wound-induced
actomyosin structures are first observed 1-3 minutes following wounding in a
variety of organisms~\citep{Brock:1996, Kasza:2011}. These characteristics are
consistent with our observations on the formation and contraction of the thin
contractile ring (see Sec.~\ref{sec:thin_ring} and
Figs.~\ref{fig:ecto_1min},\psubref*{fig:endo_1min}).  However, we have found
that healing actually begins within seconds of wounding, ruling out the
possibility of {\em de novo} actomyosin assembly in the first phase.

Presumably, the initial rapid-healing phase is instead powered by pre-existing
cellular cytoskeletal structures.  Epithelial cells generally contain a ring of
actin and myosin associated with adherens junctions around cell
borders~\citep{Lodish:2004}
(Figs.~\ref{fig:ecto_20sec},\psubref*{fig:endo_15sec}), and contraction of
these fibers plays a role in many morphogenetic processes~\citep{Martin:2010,
Rauzi:2011}.  Shortening of these relatively randomly oriented fibers would
result in tissue-level isotropic contraction, consistent with our model.  A
possible signal for the contractile response is calcium (\Ca), which can induce
swift but short-lived actomyosin
contraction~\citep{Cordeiro:2013,Tomasek:2002}, is rapidly
modulated~\citep{Clark:2009,Xu:2011}, can be activated several cell layers away
from the wound~\citep{Klepeis:2001, Woolley:2000, Xu:2011}, and has been
implicated in a variety of wound healing processes~\citep{Benink:2005,
Sonnemann:2011}.  Thus, the proposed cell contraction mechanism is reasonable
from a cellular biological perspective.

Most previous studies of embryonic wound healing have focused on time
scales of a minute or longer, but \citet{Bortier:1993} mention in passing rapid
dynamics of chick epithelial wounds similar to those we report here. 
The authors note that ``the wounds
showed a viscoelastic reaction upon wounding: they enlarged within one second,
immediately followed by a narrowing, until wound diameters were between the
original and largest size.  After these reactions the wound submarginal region
was thickened..."  No further discussion or analysis of this healing mechanism 
is provided and, to our knowledge, other investigators have not addressed this observation.
It seems likely that the submarginal region, reported as
75-150 $\mu$m wide, corresponds to the thick ring where isotropic contraction
occurs.  The authors also observed blebbing cells in this region, while
cell bulging and ``mounding'', as well as tissue thickening, has been reported
near the margin of embryonic wounds in several organisms~\citep{England:1977b,
Jacinto:2001, Smedley:1984,Stanisstreet:1980}.  Both cell bulging and blebbing
suggest high intracellular pressure~\citep{Charras:2008}, which is consistent
with strong isotropic contraction.

\subsubsection{Phase 2: Slow Fiber Contraction (Thin Ring)} \label{sec:thin_ring}

Following the creation of a multicellular wound, actin and myosin are recruited
to the wound margin where they assemble a continuous supracellular cable which
spans cells through adherens junctions.  The cable forms over the course of
minutes and contracts to generate tension which helps close the
wound~\citep{Brock:1996,Clark:2009,Kiehart:1999,Martin:1992,Sonnemann:2011}.
Contractile actomyosin cables are used by the embryo in a variety of contexts,
for example to divide cells and close the dorsal ectoderm in {\em Drosophila},
as well as to heal
wounds~\citep{Davies:2005,RodriguezDiaz:2008,Sonnemann:2011}.  Studies of
actomyosin dynamics at high temporal and spatial resolution have shown
contraction to be pulsatile, with periods of active shortening occurring
between periods of quiescent stabilization~\citep{ Gorfinkiel:2011,
Levayer:2012, Martin:2010}.  We did not observe such pulsatility, as localized
pulses may be averaged out over the perimeter of a multi-cellular wound, and we
did not include this in the model.  Being relatively stiff, the fiber ring can
generate the force needed to overcome the high circumferential stresses that
develop as the wound closes
(Fig.~\ref{fig:ModelStressCircular})~\citep{Stricker:2010}. 

The remodeling theory used in our model was originally developed to describe
the turnover of passive tissue constituents, e.g., elastin and collagen, which
are thought to be created with significant prestretch~\citep{Humphrey:2002}.
It is not clear, however, that actomyosin fiber assembly involves an initial
stretch.  Hence, we interpret $F^{f*}_0$ as the result of a rapid contraction
of the initially unloaded fiber (time constant $\tau_{un} \ll \tau_f$) that
slows ($\tau_{un} \rightarrow \tau_f$) as tension increases, with the observed
prestretch $F^{f*}_0=(G^f)^{-1}$ resulting from the contribution of many such
fibers.  This Hill-like behavior is consistent with some recent models for
stress fibers~\citep{Deshpande:2007, Stachowiak:2008}, and an ability of
actomyosin fibers to rapidly contract by large amounts has been observed in
smooth muscle cells~\citep{An:2007}.

\subsubsection{Phase 3: Filopodial Zippering} \label{sec:Filopodia}

Filopodia-mediated zippering is also an important wound healing mechanism in
early chick embryos.  Although cell protrusions have not been reported during
wound healing in older chick embryos~\citep{ Brock:1996, Martin:1992},
filopodia, lamellipodia, and microvilli have been seen at early (HH 3-5)
embryonic stages~\citep{England:1977b,Mareel:1977,Stanisstreet:1980}, and were
apparent here as well (Fig.~\ref{fig:endo_11min}, \ref{fig:endo_1min_s}).  Such
protrusions can contact and pull cells at opposite sides of the wound together,
closing it by a zippering mechanism~\citep{Jacinto:2000,Wood:2002}.  Because of
their short length ($\sim$5~$\mu$m~\citep{Wood:2002}), filopodia and related
structures are unlikely to play a significant role until the wound opening is
relatively small. Such effects are not captured by our model.

The three active processes -- isotropic cell contraction, actomyosin ring
contraction, and filopodial zippering -- work together to close a wound, a
redundancy which has been observed in other morphogenetic processes~\citep{
Davidson:2002, Jacinto:2001, Kiehart:2000}.  Very small wounds tend to close
quickly (data not shown), with isotropic contraction likely playing the
dominant role.  The capacity of this mechanism is limited, however, and large
wounds require a contractile ring.  In some cases wounds never heal, or heal so
slowly that closing does not occur during the period of observation.  The
reason may be that they are created in areas of older embryos where blastoderm
tension is high~\citep{Varner:2010b}, or they excessively perturb the tissue
past the point of viability.  Nevertheless, the capacity of the early chick
embryo to heal is remarkable, and we find that nearly all wounds, if given
enough time, do close.

\subsection{Mechanics of Wound Closure} \label{sec:MechanicsWoundClosure}

The way circumferential contraction closes a wound is consistent with physical
intuition, i.e., by shortening the perimeter.  The mechanics of embryonic
wound healing, however, contain some subtleties that warrant discussion.  
In particular, geometric effects, coupled closely to mechanics, can lead to 
counter-intuitive behavior in the elliptical wound (e.g., gaping under low tension)
which is not observed in circular wounds, even though the underlying contractile
mechanisms and model parameters are the same.

For example, it is important to note that cellular contraction occurs with
little or no change in cell volume.  Hence, circumferential contraction is
accompanied by radial expansion, which helps healing by pushing the edges of
the wound inward.  Because of structural stiffening in regions of high
curvature, these radial stresses deform edges along the minor axis of
elliptical wounds more easily than those along the major axis, causing the
wound to become more slit-like.  This explains in part why the AR increases for
the case of circumferential contraction only (Fig.~\ref{fig:ModelCompare_AR}).
In contrast, radial contraction pulls the long edges outward, increasing the AR
(Fig.~\ref{fig:ModelCompare_AR}).

Since radial and circumferential contraction in the thick ring have opposite
effects on the AR, these two effects essentially cancel out for isotropic
contraction, and the wound shape remains relatively constant as it closes
(Fig.~\ref{fig:ModelCompare_AR}).  The interaction of these effects with
stress-induced elastic deformation underlies the initial opening of the
elliptical wounds in membranes subjected to relatively small tension
(Fig.~\ref{fig:Fbnd_area_1min}).  As the applied tension increases, the stress
concentrations near the edge of the wound increase.  Since radial stress is
considerably smaller than circumferential stress near the wound edge
(Fig.~\ref{fig:ModelStressElliptical}), radial contraction is more effective
initially than circumferential contraction, and the wound first opens before
starting to close.  Hence, the detailed wound geometry depends on a balance
between radial and circumferential contraction, as well as the tissue stresses
developed in these directions.  

Another item of note is that inward radial movement of the membrane reduces its
perimeter and tends to decrease circumferential stress.  Unless this ``slack"
is taken up by circumferential contraction, compression can result.  This
effect is clearly seen in plots of $\sigma_\theta$
(Fig.~\ref{fig:ModelStressCircular}), where the circumferential stress
decreases, and in some cases becomes negative, immediately outside of the thick
ring, as well as inside the ring toward the end of the simulation.  In a
physical system such compression, if large enough, would manifest as buckling
or wrinkling of tissue.  Careful observation with a stereomicroscope
revealed no such out-of-plane folding or buckling.

The stress concentrations predicted by our model, particularly around
elliptical wounds, are of a magnitude consistent with fiber stress exerted by
smooth muscle cells~\citep{An:2007}.  Viscoelastic effects, not incorporated in
the model, are likely to reduce the peak stress considerably.  Overall, the
coupling between passive stretch, active contraction, and nonlinear geometric
effects driven by large tissue deformation resist simple explanation and
highlight the utility of computational models.

\subsection{Conclusions and Future Work}
% From B
The results of our study suggest that wound healing in embryonic epithelia
consists of three main sequential phases: (1) rapid contraction of actomyosin
fibers at cell borders within a relatively thick zone near the wound
(approximately isotropic at the tissue scale); (2) slower contraction of a thin
supracellular actomyosin ring at the wound margin; and (3) filopodia-mediated
zippering of the wound edges.  While the latter two phases have been studied in
considerable detail~\citep{Wood:2002}, the first phase has apparently remained
unknown or underappreciated by most previous investigators.

Several aspects of this problem warrant future investigation.  For example,
work is needed to determine the molecular mechanisms that regulate the
formation and activity of the contractile fibers, as well as how this activity
is integrated to produce tissue-level forces (see related work by
\cite{Hutson:2003}).  Also, mechanical signals govern the spatiotemporal
contractility of tissue in a variety of ways~\citep{ Gorfinkiel:2011,
Hutson:2008, Levayer:2012, Vogel:2006, Wozniak:2009}, and mechanical feedback
has long been suggested to play a role in wound healing~\citep{Martin:1992}.
While we have recently proposed a mechanical feedback model for embryonic wound
healing~\citep{Taber:2009}, this is not included in the current model.

Finally, it is important to note that other mechanisms may be involved in
embryonic wound healing.  Some of these may be redundant mechanisms that
activate when contraction fails.  For example, increased adhesion affinity
between cells can cause them to elongate radially and shorten
circumferentially, creating tension that helps close the wound.  Cell
rearrangement may also play a role.  Ultimately, a complete understanding of
embryonic wound healing will require integrating experiments and computational
models across multiple disciplines as well as multiple scales.

%% file: A_Coordinate_Systems.tex
For clarity, in this appendix we use the notation ($R$, $\Theta$, $Z$) for
cylindrical polar and ($\xi$, $\eta$, $Z$) for elliptical cylindrical
coordinates, respectively.  In the body of the manuscript, for convenience we
use ($R$, $\Theta$, $Z$) to refer to coordinates in both coordinate systems,
with the understanding that $\xi=R$ and $\eta = \Theta$ in the context of an
elliptical geometry.  Lower and upper cases refer to current ($b$) and
undeformed ($B_0$) configurations, respectively.

\subsection{Coordinate Systems} 

Because the contractile fibers are oriented parallel to the wound edge, we
construct coordinate systems with a unit vector lying parallel to the fibers in
the undeformed state $B_0$.  For circular wounds we use a cylindrical polar
coordinate system, which relates the coordinates $(R,\Theta,Z)$ to the
Cartesian $(X,Y,Z)$ coordinates as (Fig.~\ref{fig:GeometryCylindrical})
\label{eqn:polar_coords}
\begin{equation}
X = R \cos \Theta, ~~~ Y = R \sin \Theta, ~~~ Z = Z ~~~\text{(polar)}.
\end{equation}
For an elliptical wound centered at the origin with foci at positions $\pm \alpha$ on
the $X$ axis, the coordinates $(X,Y,Z)$ relate to the elliptical cylindrical
coordinates $(\xi,\eta,Z)$ as \citep{Arfken:2005} (Fig.~\ref{fig:GeometryElliptical})
\begin{equation}
X = \alpha \cosh \xi \cos \eta, ~~~ Y = \alpha \sinh \xi \sin \eta, ~~~ Z = Z ~~~\text{(elliptical)}.  \label{eqn:elliptical_coords}
\end{equation}
The unit vectors $\bolde_R$ and $\bolde_\xi$ are normal to the wound edge in
the polar and elliptical undeformed coordinate systems, respectively, while
$\bolde_\Theta$ and $\bolde_\eta$ are tangent to the wound edge.  In both
cases $\bolde_Z$ is normal to the plane of the membrane.

\figureGeometryCoord 

\subsection{Contractile Ring Width}\label{sec:rho}
To construct rings of uniform width around a wound in the undeformed state
$B_0$, we calculate the distance from any given point $\boldQ$ to the wound
edge.  
For a circular wound this task is straightforward.
We define $\boldR = R_R \bolde_R $ as
the nearest point on the wound edge to $\boldQ = R_Q \bolde_R $ (Fig.~\ref{fig:GeometryCylindrical}). 
Thus, the magnitude of
the vector $\boldrho = \boldQ - \boldR = \rho \, \bolde_R$ is given simply as,
\begin{equation}
\rho = R_Q - R_R    ~~~ \text{(polar)}. \label{eqn:rhoCylindrical}
\end{equation}

An analogous approach fails in elliptical coordinates because the vector
$(\xi_Q - \xi_R) \bolde_\xi$ does not have a uniform length around the perimeter
of the wound (Fig.~\ref{fig:GeometryElliptical}).  
Instead, we define $\boldR$
as the point on the wound edge with the same $\eta$ coordinate as $\boldQ$, and
expand $\boldQ$ as a Taylor series in $d\xi = \xi_Q - \xi_R$ about $\boldR$.
Retaining terms to $O(d\xi^2)$, we write
\begin{align}
\boldrho &= \boldQ - \boldR, \nonumber \\
&\simeq \left(\boldR + \left.\frac{\partial \boldR}{\partial \xi}\right|_\boldR d\xi + 
    \frac{1}{2} \left.\frac{\partial^2 \boldR}{\partial \xi^2}\right|_\boldR d\xi^2\right) - \boldR, \nonumber \\
&= \left.\frac{\partial \boldR}{\partial \xi}\right|_\boldR d\xi + \frac{1}{2} \left.\frac{\partial^2 \boldR}{\partial \xi^2}\right|_\boldR d\xi^2. \label{eqn:rhoTSE}
\end{align}
Using Eq.~\eqref{eqn:elliptical_coords}, we can write the derivatives of $\boldR = X\, \bolde_X + Y\, \bolde_Y$ 
as
\begin{align}
\left.\frac{\partial \boldR}{\partial \xi}\right|_\boldR &= \alpha \sinh \xi_R \cos \eta_Q \bolde_X + \alpha \cosh \xi_R \sin \eta_Q \bolde_Y, \nonumber \\
\left.\frac{\partial^2 \boldR}{\partial \xi^2}\right|_\boldR &= \alpha \cosh \xi_R \cos \eta_Q \bolde_X + \alpha \sinh \xi_R \sin \eta_Q \bolde_Y .
\end{align}
Finally, from  Eq.\eqref{eqn:rhoTSE} we find the magnitude of $\boldrho$ for the elliptical wound as 
\begin{align}
\rho &= \frac{\alpha}{2} \left[(2 \sinh\xi_R d\xi + \cosh\xi_R d\xi^2)^2 \cos^2\eta_Q \right. + \notag \\ 
& \left. (2 \cosh\xi_R d\xi + \sinh\xi_R d\xi^2)^2 \sin^2\eta_Q\right]^{1/2} \label{eqn:rhoElliptical}
\text{(elliptical)}, 
\end{align}
where $\xi_R$ is constant along the wound edge and $\eta_Q$ is given by
$\boldQ$.  While this approximation strictly holds only for $\boldQ$ in the
immediate neighborhood of $\boldR$, it serves as an adequate approximation over
the domain of both contractile regions. 

Given the model geometry (Table~\ref{tab:Geometry}), we find the focus position
$\alpha=82.8~\mu$m and $\xi$ at wound boundary $\xi_R=0.0663$.  The contours of
$\rho$ equal to $\rho_f$ and $\rho_c$, which demarcate the borders of the fiber
and cell contractile regions, respectively, are plotted in
Fig.~\ref{fig:GeometryElliptical}, and we find that $\rho$ in the elliptical
coordinates evaluated along the X and Y axes in the $B_0$ configuration varies
less than 5\%.

%% file: B_SingleMultiFiber.tex
To evaluate the accuracy of the single-fiber approximation, we consider the
simple case of an unstretched homogeneous bar fixed at both ends
(Fig.~\ref{fig:FiberStress} inset), and neglect the contribution of cells by
setting $\mu_c=0$.  In this one-dimensional approximation,
$F^{f*}=\lambda^{f*}$, $F^{f*}_0=\lambda^{f*}_0$, $C^{f*}={\lambda^{f*}}^2$,
and $J^{f*}=1$. For simplicity we set $\mu_f=1$ and $\lambda^{f*}_0 = 1$, and
write the stress as a linear function of the elastic stretch ratio,
\begin{equation}
\sigma^f = \lambda^{f*} - 1 .
\end{equation}

For the multiple-fiber model, the fiber stretch ratio is given by Eq.~\eqref{eqn:fib_elastic} as
\begin{equation}
\lambda^{f*}_{t/\tau} = \frac{\lambda^f_{t/\tau}}{  G^f_{t/\tau} },
\end{equation}
and for the single-fiber model, Eq.~\eqref{eqn:F} gives
\begin{equation}
\lambda^{f*}=\frac{\lambda}{G^f}.
\end{equation}
In the present isometric problem, the total stretch ratios are $\lambda^f_{t/\tau}=\lambda=1$.  

With the above equations, the fiber stress, given by the 1-D forms of Eqs.~\eqref{eqn:sigmf} and \eqref{eqn:sigtot_sf}, is
\begin{equation}
\sigma_{MF}(t) = \int_0^t \phidot^f(\tau) \, \left( \frac{1}{G^f_{t/\tau}} - 1 \right) d\tau \label{eqn:sigmfB}
\end{equation}
for the multiple-fiber model and
\begin{equation}
\sigma_{SF}(t) = \phi^f(t) \, \left( \frac{1}{G^f} - 1 \right)  \label{eqn:sigsfB}
\end{equation}
for the single-fiber model.

For illustrative purposes, we take $G^f(t) = \exp(-t/\tau_f)$ for the single-fiber model.  This gives
\begin{equation}
G^f_{t/\tau} = G^f(t) / G^f(\tau) = \exp[-(t-\tau)/\tau_f] \\
\end{equation}
for the multi-fiber model.  The fiber fraction, from Eq.~\eqref{eqn:phi_f}, is
\begin{equation}
\phi^f(t) = 1 - \exp(-t / \tau_\phi).
\end{equation}

We also nondimensionalize time by the fiber contraction rate and define
\begin{align}
t' &\equiv t / \tau_f,  \nonumber \\
\alpha &\equiv \tau_f / \tau_\phi.
\end{align}
Finally, we can write Eqs.~\eqref{eqn:sigmfB} and \eqref{eqn:sigsfB} as
\begin{align}
\sigma_{MF}(t') &= \frac{1}{1+\alpha} \exp(-\alpha \, t') + \frac{\alpha}{1+\alpha} \exp(t') - 1 \nonumber \\
\sigma_{SF}(t') &= [1 - \exp(-\alpha \, t')]\,[\exp(t') - 1] 
\end{align}
where the multi-fiber stress was integrated analytically.  Both stresses are parameterized by $\alpha$,
which is large when fibers form quickly and contract slowly, and small when the opposite is true. 

\figureFiberStress

The stresses $\sigma_{MF}$ and $\sigma_{SF}$ are plotted for a range of
$\alpha$ (Fig.~\ref{fig:FiberStress}).  Differences between the single and
multi-fiber models decrease with increasing $\alpha$. As $\alpha \rightarrow
\infty$ all fibers appear instantaneously at $t=0$, so that the two models
yield identical results.  The parameters used in the body of the manuscript
correspond to $\alpha = 3.3$ (Table~\ref{tab:Parameters}).  We conclude that
the single fiber model, while a simplification of a more biologically realistic
multi-fiber formulation, is sufficiently accurate for the chosen parameters to
justify its significant computational advantages.

%% file: C_Properties.tex
To estimate the mechanical properties of the early chick blastoderm (i.e., a
value for $\mu_c$), we constructed a simple linear model for indentation of a
plate under in-plane tension, and used it to simulate microindentation
experiments of HH stage 5 embryos previously conducted in our laboratory 
\citep{Zamir:2003,Varner:2010b,Varner:2012a}. Briefly, following the framework employed by
\citep{Zamir:2004a}, we model the blastoderm near the indenter as a thin,
isotropic annular plate under in-plane tension. For small deflection, the
governing differential equation is given by~\citep{Szilard:1974}
\begin{equation}
{D_p}{\nabla ^4}w - {T_r}{\nabla ^2}w = p, \label{eqn:plate}
\end{equation}
where $w$ is the transverse deflection, $\nabla$ is the gradient operator,
$D_p=Eh^3/12(1-\nu)^2$ is the flexural rigidity, $T_r$ is the radial in-plane
force per unit length, $p$ is the applied surface pressure, $E$ is the Young's
modulus, $\nu$ is Poisson's ratio, and $h$ is the plate thickness.

Here, we consider the case of a circular plate clamped at the outer radius $b$, which
represents the distance at which the local disturbance from the indenter is
effectively zero.  The plate also is taken as fixed at the inner radius to a
rigid cylindrical indenter of radius $a$ that exerts
a force P.  The boundary conditions take the form 
\begin{subequations}
\label{eqn:bc}
\begin{align}
V(a) &=  -\dfrac{P}{2\pi a},  \nonumber \\
\Theta(a) &=  0,  \nonumber \\
w(b) &= 0,  \nonumber \\
\Theta(b) &= 0,
\end{align}
\end{subequations}
where $V$ is the transverse shear per unit length and $\Theta$ is the rotation.
Note that, in this case, no surface pressure is applied to the plate (i.e.,
$p=0$), so the applied load enters not through any specific term in equation
\eqref{eqn:plate}, but rather through enforcement of the boundary condition
\eqref{eqn:bc}.

Hole punching experiments in HH stage 5 embryos have indicated that the
blastoderm is initially in an approximately uniform, equibiaxial state of tension
\citep{Varner:2010a,Varner:2010b}. In addition, comparing the geometry of the
wounds to the dimensions of the punching pipette have revealed that the
initial tissue strains are $\epsilon_r = \epsilon_\theta \equiv \epsilon \simeq 0.1$.
For plane stress, the tension is then given by 
\label{eqn:ps}
\begin{align}
\sigma_r = \sigma_\theta &= \frac{E}{1-\nu^2}(\epsilon+\nu\epsilon),\nonumber\\
&= \frac{E \epsilon}{1-\nu}.
\end{align}
If we assume material incompressibility (i.e., $\nu=0.5$) and $\epsilon=0.1$, then
\begin{equation}
T_r = \sigma h=0.2Eh.  \label{eqn:inplane}
\end{equation}

Microindentation experiments performed in our laboratory have indicated an
approximate blastodermal stiffness of 0.2-0.6 mdyne/$\mu$m using an indenter
with a radius $a=10$ $\mu$m \citep{Varner:2010b,Varner:2012a}. Moreover, optical
coherence tomography reconstructions of early chick embryos have shown the
thickness of the blastoderm to be approximately $h=50$--90 $\mu$m
\citep{Varner:2010b}.

If we use these values in combination with equation \eqref{eqn:inplane} and take
the distance from the embryonic midline to the boundary of the area pellucida
($\sim$1--2 mm) as the radius of the circular plate $b$, we can solve equations
\eqref{eqn:plate} and \eqref{eqn:bc} for different values of $E$ and compare the computed
stiffness values, given by $P/w(a)$, to those reported experimentally \citep{Varner:2012a}. Doing so
indicates an approximate Young's modulus of 120 Pa for the blastoderm, which
(given material incompressibility) yields the shear modulus $\mu_c=E/3= 40$ Pa.

%% file: X_Acknowledgements.tex
This work was supported by grants F32 GM093396 (MAW), R01 GM075200 and R01
N5070918 (LAT) from the National Institutes of Health, as well as grant
09PRE2060795 from the American Heart Association (VDV).  We thank Elliot Elson,
Benjamen Filas, and Yunfei Shi for helpful discussions.

%% file: S_Supplemental.tex
\subsection{Parameters for Experimental Wound Area}

\tableWoundFit

Table~\ref{tab:WoundFitAll} lists the parameters obtained from a best fit of
Eq.~\eqref{eqn:fit} to closing trends of six circular wounds
(Figs.~\ref{fig:woundImage} and ~\ref{fig:CircularAreaFit}).  Mean parameter
values yield area versus time curve against which model results are compared.

\subsection{Implementation Details} \label{sec:matrix}

%Writing out key tensors in matrix form.  Copied from notebook V.vi p.29.
We provide here details of coordinate transformations as well as key tensors in
matrix form as an aid to understanding and implementation (see also~\citep{Taber:2008}).  Our goal is to
calculate the stress carried by cells and fibers, $\boldsigma^c$ and $\boldsigma^f$ (Eq.~\ref{eqn:constitutive}),
given the total deformation $\boldF$ and growth tensors $\boldG^c$ and
$\boldG^f$ (Eqs.~\ref{eqn:GcLoc}, \ref{eqn:GfLoc}).

\subsubsection{Coordinate Transforms}
Given, for instance, the tensor $\bG^c$ whose components are provided in local (i.e., polar or elliptical)
coordinates as in Eq.~\eqref{eqn:GcLoc}, we can obtain the components in Cartesian coordinates
using the relation \citep{Taber:2004}
\begin{equation}
G^c_{IJ} = \bolde_I \cdot \boldG^c \cdot \bolde_J.
\end{equation}
The quantities $R$ and $\Theta$, as well as the
dot product between local ($\bolde_\alpha$) and Cartesian ($\bolde_A$) unit vectors, $e_{A \alpha} \equiv \bolde_A \cdot \bolde_\alpha$,
can be calculated as,
\begin{equation}
\begin{matrix}
R = \sqrt{X^2 + Y^2}, & \Theta = \tan^{-1} Y / X, \\
e_{XR} = \cos(\Theta), & e_{YR} = \sin(\Theta), \\
e_{X\Theta} = -\sin(\Theta), & e_{Y\Theta} = \cos(\Theta), \\
\end{matrix}
\end{equation}
for polar coordinates and
\begin{equation}
\renewcommand\arraystretch{1.5}
\begin{matrix}
R = \operatorname{Re}(\cosh^{-1}\frac{X+ i\,Y}{\alpha}), & \Theta = \operatorname{Im}(\cosh^{-1}\frac{X+ i\,Y}{\alpha}), \\
e_{XR} = \frac{\sinh(R)}{g} \cos(\Theta), & e_{YR} = \frac{\cosh(R)}{g} \sin(\Theta), \\
e_{X\Theta} = -\frac{\cosh(R)}{g} \sin(\Theta), & e_{Y\Theta} = \frac{\sinh(R)}{g} \cos(\Theta),  \\
\multicolumn{2}{c}{g =\sqrt{\cosh^2 R \sin^2 \Theta + \sinh^2 R \cos^2 \Theta}}, \\
\end{matrix}
\end{equation}
for elliptical coordinates, where $\alpha$ is a real number defined in
\ref{sec:coordinates}, and with the understanding that for elliptical coordinates
$R=\xi$ and $\Theta=\eta$ (see \ref{sec:coordinates}).  We can thus write
$\bG^c$ in Cartesian coordinates as,
\begin{equation}
\begin{bmatrix}
G^c_{XX} & G^c_{XY} & 0 \\ 
G^c_{YX} & G^c_{YY} & 0 \\ 
0 & 0 & G^c_{ZZ}
\end{bmatrix}  =  \begin{bmatrix}
G^c_R e^2_{XR} + G^c_\theta e^2_{X \Theta} & G^c_R e_{XR} e_{YR} + G^c_\Theta e_{X\Theta} e_{Y\Theta}& 0 \\
G^c_R e_{XR} e_{YR} + G^c_\Theta e_{X\Theta} e_{Y\Theta}   & G^c_R e^2_{YR} + G^c_\theta e^2_{Y \Theta} & 0 \\
0 & 0 & G^c_Z \end{bmatrix}.
\end{equation}

\subsubsection{Cell Stress}

With $\bG^c$ expressed in Cartesian coordinates, we can write the elastic
deformation gradient tensor $\boldF^{c*}$ in Cartesian components as,
\begin{equation}
\bF^{c*} = \bF \cdot \left(\bG^c\right)^{-1} = D^{-1} \begin{bmatrix}
F_{xX} G^c_{YY} - F_{xY} G^c_{XY} & F_{xY} G^c_{XX} - F_{xX} G^c_{XY} & 0 \\
F_{yX} G^c_{YY} - F_{yY} G^c_{XY} & F_{yY} G^c_{XX} - F_{yX} G^c_{XY} & 0 \\
0 & 0 & D^2\,F_{zZ} \end{bmatrix},
\end{equation}
with 
\begin{equation}
D = G^c_{XX} G^c_{YY} - (G^c_{XY})^2,
\end{equation}
and find the cell strain invariants as
\begin{align}
I^*_1 = \tr \left[(\bF^{c*})^T\cdot \bF^{c*}\right] &= \left(F^{c*}_{xX}\right)^2 + \left(F^{c*}_{xY}\right)^2 + 
    \left(F^{c*}_{yX}\right)^2 + \left(F^{c*}_{yY}\right)^2 + \left(F^{c*}_{zZ}\right)^2, \\
I^*_3 = (\det \bF^{c*})^2 &= \left[ \left( F^{c*}_{xX} F^{c*}_{yY}  - F^{c*}_{xY} F^{c*}_{yX}\right) F^{c*}_{zZ}\right]^2.
\end{align}
We calculate $W^c$ from Eq.~\eqref{eqn:W} and obtain the 
first Piola-Kirchoff stress tensor $\boldP^c$ by numerical differentiation as,
\begin{equation}
\renewcommand\arraystretch{1.25}
\bold{P^c} = \frac{J}{J^{c*}} \frac{\partial W^c}{\partial \bF^T} = \frac{J}{J^{c*}}\begin{bmatrix}
\frac{\partial W^c}{\partial F_{xX}} & \frac{\partial W^c}{\partial F_{yX}} & 0 \\
\frac{\partial W^c}{\partial F_{xY}} & \frac{\partial W^c}{\partial F_{yY}} & 0 \\
0 & 0 & \frac{\partial W^c}{\partial F_{zZ}} 
\end{bmatrix}.
\end{equation}
Finally, the cell Cauchy stress is obtained as,
\begin{equation}
\boldsigma^c = J^{-1} \bF \cdot \bold{P}^c.
\end{equation}

\subsubsection{Fiber Stress}
The fiber elastic stretch ratio $\lambda^{f*}$, given by Eq.~\eqref{eqn:sigf}, can be calculated
(using Eqs.~\ref{eqn:F}, \ref{eqn:GfLoc}) as,
\begin{equation}
\lambda^{f*} = \frac{\sqrt{C_{\Theta\Theta}}}{G^f},
\end{equation}
where $C_{\Theta\Theta}$ is the $\bolde_\Theta \bolde_\Theta$ component of $\boldC = \bF^T\cdot\bF$.
We calculate fiber stress by analytical differentiation
and write the first Piola-Kirchoff fiber stress components in local coordinates as
\begin{equation}
\boldP^f = \frac{2 \mu_f J}{G^f J^{f*}} \left(1-(\lambda^{f*})^{-3}\right) 
    \left(F^{f*}_{r\Theta} \bolde_\Theta \bolde_r + F^{f*}_{\theta \Theta} \bolde_\Theta \bolde_\theta \right).
\end{equation}
This quantity is then transformed into Cartesian Cauchy stress components as
described previously, and the total stress $\boldsigma$ carried by the tissue
is given by Eq.~\eqref{eqn:sigtot_sf}.

\subsection{Actin and Myosin Staining}

\figureWoundColor
\figureWoundpMLC

\figureWoundActinElliptical

Fluorescent staining for actin at one minute after wounding
(Fig.~\ref{fig:endo_1min_s}) illustrates the formation of the actin purse
string and the presence of lamellipodia.  Staining for phosphorylated myosin
light chain (pMLC) at different times after wounding (Fig.~\ref{fig:WoundPMLC})
yields distributions similar to those for actin (Fig.~\ref{fig:WoundActin}).
These results indicate that pMLC co-localizes with actin in the thick and thin
rings, and demonstrate that these structures are actively contractile.  

Cell and fiber rings around elliptical wounds are similar to those around
circular wounds.  Figure ~\ref{fig:WoundActinElliptical}, shows actin distribution around
elliptical wounds at 3 and 15 minutes after wounding.  Three minutes after
wounding both cell and fiber rings are visible, and at 15 minutes a well
developed fiber ring is visible in both germ layers.

\subsection{Stress and Displacement}

\figureModelStressElliptical

%The principal difference between circular and elliptical wounds is more intense
%stress concentrations along the major axis of the ellipse

Circular and elliptical wounds differ primarily in that stress concentrations are 
more intense along the major axis of the ellipse
(Fig.~\ref{fig:ModelStressElliptical}, \ref{fig:logmisesCs}, \ref{fig:logmisesEs}). Along both axes, at t = 0 the radial
stress $\sigma_r$ is zero at the wound boundary and increases with distance,
converging to $\sigma_B$ some distance away. However, the increase is very
sharp along the major axis and much more gradual along the minor one.
Similarly, the circumferential stress at t = 0 has a value about 70 times
greater than $\sigma_B$ at the wound border along the major axis, while it
decreases to less than half of $\sigma_B$ along the minor axis. Otherwise,
contraction of cells and fibers changes the stress distribution in a manner
broadly similar to that observed for the circular wound
(Sec.~\ref{sec:CircularStress}).  
Figures~\ref{fig:dispCs} and \ref{fig:dispEs} plot the tissue displacement relative to the reference ($b_0$) 
configuration for circular and elliptical wounds, respectively.  In both cases the largest displacement
occurs inside of the cell region.

\figureLogMisesCs
\figureDispCs
\figureLogMisesEs
\figureDispEs

\subsection{Parameter Sensitivity Studies} \label{sec:param_sup}

\figureRadiusCompare

Differences in membrane pre-stress can explain the initial variations in the
healing behavior observed in elliptical wounds.  Fig.~\ref{fig:RadiusCompare}
plots the major and minor radii of both experimental and model elliptical
wounds.  The initial opening of elliptical wounds is driven by an increase in
the minor radius, a behavior which is reproduced by the model for low values of
$\sigma_B$ (see Sec.~\ref{sec:EllipticalGape}).

\figureModelFitR
\figureModelFitAR

To evaluate the role of wound size and shape on the rate of wound closure, we
varied the initial wound size and aspect ratio.  With other parameters
unchanged, a smaller wound implies proportionately wider cell and fiber
contractile rings.  The model predicts that smaller wounds close more quickly,
particularly in the initial phase of healing where cell contraction dominates
(Fig.~\ref{fig:rab_area}).  The aspect ratios of smaller wounds increase
somewhat more than those of larger wounds in the second, fiber-driven phase of
wound healing (Fig.~\ref{fig:rab_AR}), an effect which is analogous with the
result of Fig.~\ref{fig:ModelCompare_AR}.  In general, however, qualitative
aspects of wound closure tend to be relatively insensitive to wound size.  This
is broadly consistent with experimental observations, where we find that wounds
much larger than those considered here will typically heal, albeit much more
slowly.

Changing the elliptical wound geometry by modifying the minor radius yields
negligible changes in the relative rate of wound closure, and predictable
variation in the aspect ratio (Fig.~\ref{fig:ModelFitAR}).  Thus, neither the
wound size nor shape has a strong effect on wound behavior, and at least for
the value of $\sigma_B$ under consideration, circular and elliptical wounds
behave similarly.  
This result also supports our approximation of a linear incision, which can be considered an elliptical wound of infinite aspect ratio,
by an elliptical wound of an aspect ratio AR=15 (Section~\ref{sec:geometric_parameters}).

\figureAltParams

The dynamics of wound closure are quite sensitive to the value of the
prestretch $F^{f*}_0$, as illustrated in Fig~\ref{fig:prestretch}.  Parameter
sets other than those listed in Table~\ref{tab:Parameters} can reproduce the
experimental circular wound trends, as shown in Fig.~\ref{fig:AltParams}